\documentclass[14pt]{article}
\pdfoutput=1
\usepackage{amsmath,amssymb,amsfonts,amsthm,bm,bbm,cancel,wasysym}
\usepackage{epsfig,graphics,graphicx,epstopdf,caption,subcaption}
\graphicspath{{}}
\usepackage{array,booktabs,colortbl,colordvi,multirow}
\usepackage{colordvi,color,xcolor}
\usepackage{hyperref}
\usepackage{rotating}
\usepackage{comment}
\usepackage{feynmf}
\usepackage{tikz}
\usetikzlibrary{calc, shapes, arrows, decorations.pathmorphing, decorations.pathreplacing, 3d, shapes.multipart}

\tikzstyle{particle} = [diamond, draw, text centered, rounded corners, minimum height = 2em]
\tikzstyle{coupling} = [circle, draw, text centered, rounded corners, minimum height = 2em]
\tikzstyle{action} = [rectangle, draw, text centered, rounded corners, minimum height = 2em]
\tikzstyle{input} = [circle, draw, text centered, minimum height = 2em]
\tikzstyle{graph} = [rectangle, draw, rounded corners, text centered, minimum height = 2em]
\tikzstyle{mlp} = [rectangle, draw, text centered, minimum height = 2em]
\tikzstyle{pooling} = [diamond, draw, text centered, minimum height = 2em]

\usepackage{verbatim}
\usepackage{cite}
\usepackage{amsmath}
\usepackage{physics}
\usepackage{mathtools}
\usepackage{amsthm}
\usepackage{algorithm2e}
\usepackage{setspace}
\usepackage{url}
\usepackage[percent]{overpic}
\usepackage{slashed}
\usepackage{authblk}
\usepackage{xspace}
\usepackage{fullpage}
\usepackage[numbers,sort&compress]{natbib}
\usepackage{multirow}
\usepackage{array}
\newcolumntype{P}[1]{>{\centering\arraybackslash}p{#1}}

\def\eg{{\it e.g.}}

\newskip\zatskip \zatskip=0pt plus0pt minus0pt
\def\matth{\mathsurround=0pt}

\def\atversim#1#2{\lower0.7ex\vbox{\baselineskip\zatskip\lineskip\zatskip
  \lineskiplimit 0pt\ialign{$\matth#1\hfil##\hfil$\crcr#2\crcr\sim\crcr}}}

\newif\ifdiagrams
\diagramstrue
\ifdiagrams

\else
  \excludecomment{fmffile}
\fi

\begin{document}

%----------------------------------- TITLE AND AUTHORS -----------------------------------------%

%Preprint numbers
\begin{flushright}
%SLAC-PUB-17609\\
\today
\end{flushright}
\vspace*{5mm}

\renewcommand{\thefootnote}{\fnsymbol{footnote}}
\setcounter{footnote}{1}
\begin{center}

{\Large {\bf Graph Reinforcement Learning for Exploring BSM Model Spaces}}\\
%\vspace*{0.15cm}

\vspace*{0.75cm}

{\bf George N. Wojcik~\footnote{gwojcik@wisc.edu}, Shu Tian Eu~\footnote{eu@wisc.edu}, and Lisa L. Everett~\footnote{leverett@wisc.edu}}

\vspace{0.5cm}

{Department of Physics, University of Wisconsin-Madison, Madison, WI 53706, USA}

\end{center}
\vspace{.5cm}

%--------------------------------------------- ABSTRACT ---------------------------------------------%
\begin{abstract}
 
\noindent  
We present a methodology for performing scans of beyond Standard Model (BSM) parameter spaces with reinforcement learning. We identify a novel procedure using graph neural networks that is capable of exploring spaces of models without the user specifying a fixed particle content, allowing broad classes of BSM models to be explored -- in theory, the technique is applicable to nearly any model space with a pre-specified gauge group. We provide a generic procedure by which a suitable graph grammar can be developed for any BSM model which features user-specified symmetry groups and a finite number of different possible particle species, the use of which is applicable to a variety of machine learning tasks over the actions of BSM theories beyond our particular reinforcement learning use case. As a proof of concept, we construct the graph grammar for theories with vector-like leptons that may or may not be charged under a dark $U(1)$ group, inspired by portal matter extensions of the sub-GeV vector portal/kinetic mixing simplified dark matter models. We then use this graph grammar to create a reinforcement learning environment tasked with creating models with these vector-like leptons that are consistent with a list of a variety of precision observables. The reinforcement learning agent succeeds in developing models that can address the observed muon anomalous magnetic moment discrepancy while remaining consistent with flavor violation and electroweak precision observables, including both constructions that have previously been studied as well as new models which have not, to our knowledge, previously been identified. By inspecting the resulting ensembles of models that the agent produces and experimenting with different configurations for our reinforcement learning environment and graph grammar, we also infer various lessons about the development of these environments that can be transferable to reinforcement learning scans of more complicated model spaces, and comment on future directions for the development of this technique into a more mature tool.
\end{abstract}

\renewcommand{\thefootnote}{\arabic{footnote}}
\setcounter{footnote}{0}
\thispagestyle{empty}
\vfill
\newpage
\setcounter{page}{1}

%-------------------------------- DOCUMENT: INTRODUCTION ---------------------------------%

\section{Introduction}\label{sec:intro}

As machine learning techniques become more powerful and accessible, their utility to scientific inquiry expands. High energy physics is no exception, and a variety of recent studies have explored the utility of machine learning techniques, particularly those based on neural networks, to problems in the field \cite{Halverson:2019tkf,Krippendorf:2021uxu,Nishimura:2020nre,Dersy:2022bym,Wojcik:2023usm,Goodsell:2022beo,Matchev:2024ash,Cole:2021nnt,Li:2024uju,hepmllivingreview,Larkoski:2017jix,Dersy:2023job,Guest:2018yhq,Karagiorgi:2021ngt,Schwartz:2021ftp}. While much of this work has focused on experimental problems, where large and noisy data sets present difficult problems in simulation and data analysis well-suited to machine learning techniques, some significant work has also been done in applying these techniques to theoretical problems \cite{Halverson:2019tkf,Krippendorf:2021uxu,Nishimura:2020nre,Dersy:2022bym,Wojcik:2023usm,Dersy:2023job,Matchev:2024ash,Cole:2021nnt}.

Reinforcement learning (RL) has demonstrated particular promise in the field of model building beyond the Standard Model (BSM).  As the space of possible BSM extensions is enormous (encompassing all theories which contain the Standard Model effective field theory and possess only additions which avoid present experimental constraints), a conventional automated exploration of this space is infeasible. Reinforcement learning offers a method by which automated exploration of the space might be achieved, as in such a scan the reinforcement learning agent is trained to recommend actions to modify a model in ways which that maximize its expected reward, which in this case will be designed to be correlated to some metric of the empirical or theoretical viability of the model, rather than sampling actions from a pre-specified random distribution. Most notably, in contrast to supervised or unsupervised learning, it is not necessary to train the RL agent with a large dataset as it learns solely through its interactions with theorist-specified environment. This technique has previously enjoyed success in probing large spaces in string theory \cite{Halverson:2019tkf,Krippendorf:2021uxu,Cole:2021nnt}, and has been used in \cite{Harvey:2021oue,Nishimura:2020nre} to the problem of identifying theoretically viable fermion charges in the Frogatt-Nielsen model \cite{Froggatt:1978nt}. Both of these studies demonstrated the efficacy of reinforcement learning in identifying new points in the nontrivially distributed viable regions of an extremely large space of discrete values (for example, those of the discrete Frogatt-Nielsen charges); however, because both relied on learning over a fixed-dimensional discrete parameter space, the applicability of the technique to more general BSM model building tasks remained undetermined. If reinforcement learning scans are to be generalized effectively to a broader class of BSM model building problems, it is crucial that the procedure be adapted to scanning over spaces of models where the BSM particle content, and therefore the feature dimensionality of the subspace, is variable.

In pursuit of this goal, we identify the utility of graph neural networks. As data structures, graphs are represented by an indeterminate number of nodes and connections between them. By applying learnable transformations that are generalizable across arbitrary graph topologies, a graph neural network is capable of effective learning tasks across graphs with varying size, which may in turn represent theories with varying particle content. We develop a generic and systematic recipe by which a graph grammar may be developed to represent any class of 4-dimensional BSM theories with a finite number of fields, subject only to the restriction that a theorist specifies the theory's symmetry group and which representations particles might appear in.

In principle, this system of graph grammars has potential applications for a broad array of learning tasks over BSM actions. However, as a demonstration of its utility toward our original goal of a reinforcement learning scan, we apply it to create an agent capable of performing a scan over a space of models in which the particle content is undetermined. For a proof-of-concept example, in this work and in our companion short paper \cite{Wojcik:2024azu}, we consider a subclass of BSM models with vector-like leptons and a dark $U(1)$ gauge symmetry, inspired by portal matter constructions \cite{Rizzo:2018vlb,Rizzo:2022lpm}, in which heavy fermions charged under both the SM and dark gauge groups will generate a kinetic mixing between the dark photon and the electroweak gauge bosons \cite{Holdom:1985ag,Holdom:1986eq} of a magnitude which is consistent with the viable parameter space for a sub-GeV vector portal/kinetic mixing simplified dark matter model of the type described in, \eg, \cite{Pospelov:2007mp,Izaguirre:2015yja,Essig:2013lka,Curtin:2014cca}. Rewarding the agent for producing models which maximize the difference in log-likelihood with the SM based on a collection of precision observables, combined with a penalty for each BSM particle introduced to incentivize simplicity, we find that our agent successfully identifies models which address the discrepancy between the SM prediction and the experimental observation of the muon anomalous magnetic dipole moment \cite{Muong-2:2006rrc,Muong-2:2021ojo,Muong-2:2023cdq,Aoyama:2020ynm,Aoyama:2012wk,Aoyama:2019ryr,Czarnecki:2002nt,Gnendiger:2013pva,Davier:2017zfy,Keshavarzi:2018mgv,Colangelo:2018mtw,Hoferichter:2019mqg,Davier:2019can,Keshavarzi:2019abf,Kurz:2014wya,Melnikov:2003xd,Masjuan:2017tvw,Colangelo:2017fiz,Hoferichter:2018kwz,Gerardin:2019vio,Bijnens:2019ghy,Colangelo:2019uex,Blum:2019ugy,Colangelo:2014qya}. In particular, in addition to identifying constructions of a type previously identified in \cite{Wojcik:2022woa}, the agent also produces simple alternatives which suggest different collider phenomenology and UV completions (see e.g.~\cite{Crivellin:2018qmi,Crivellin:2021rbq} for a detailed analysis of general model classes that can yield viable theories that can accommodate the muon anomalous magnetic moment). 

Our paper is laid out as follows. In Section \ref{sec:graph}, we describe a general procedure by which a suitable graph grammar might be constructed for representing different classes of BSM models with a fixed gauge group. In Section \ref{sec:reinforcement}, we provide a brief introduction to reinforcement learning, focusing on the concepts which are relevant to the remainder of the work. In Section \ref{sec:model}, we describe the class of vector-like lepton models that we use in our case study in detail, including identifying the observables which we task the reinforcement learning agent with fitting and specifying the graph grammar that we use to represent our models to the agent. In Section \ref{sec:experiment-setup}, we describe the reinforcement learning environment that we have developed to perform our scan, including identifying the structure of rewards and the actions that the agent is permitted to perform on a model. In Section \ref{sec:experiments}, we present the results of reinforcement learning scans across a variety of configurations for the environment and agent, and comment on the performance of the agent across various metrics. Finally, in Section \ref{sec:conclusion}, we discuss our results, including inferences that we may make about the best practices for similar scans over more complex model spaces, and identify directions for future work both in reinforcement learning and in analysis of BSM actions with graph neural networks more broadly.

\section{Expressing BSM Theories as Graphs}\label{sec:graph}

A crucial component of our work is highlighting the utility of the mathematical graphs in representing arbitrary BSM theories, with learning tasks accomplished via a graph neural network. In the case of our reinforcement learning study, we leverage this graph structure to allow the agent to consider models over a space of BSM extensions with varying particle content, and therefore an uncertain number of discrete and continuous parameters. In this Section, we shall provide a brief pedagogical review of the relevant characteristics and functioning of graphs and graph neural networks, as well as argue that graph structures are extremely well-suited to expressing BSM actions for a broad variety of learning tasks. Finally, we shall present a general recipe for creating graph grammars for different classes of BSM theories, demonstrating the wide applicability of this technique for expressing theories in a machine-learnable format.

Mathematically, a graph consists of two sets $\{ V, E \}$ of nodes $V$, and edges $E$ which connect them. The elements of the set $V$ are various feature vectors that describe each node, while $E$ consists simply of a list of the pairs of connected nodes in $V$, as well as corresponding feature vectors for each edge (in the event that only one type of edge is required in the graph, the graph may not include an edge feature vector). A graph neural network then operates on a graph via message-passing layers. Message-passing layers will transform a node's feature vectors via a trainable function of the node's feature vector itself and some aggregation of its neighboring nodes (that is, the nodes connected to it by an edge). The operation of a simple linear message-passing layer, translating a node feature vector $\mathbf{x}_i$ to an output feature vector $\mathbf{x}'_i$ might be described by the function
\begin{align}\label{eq:message-passing}
    \mathbf{x}'_i = \mathbf{W}_1 \cdot \mathbf{x}_i + \mathbf{W}_2 \cdot \sum_{j \in \mathcal{N}(i)} (\mathbf{x}_j + \mathbf{W}_3 \cdot \mathbf{e}_{j,i}),
\end{align}
where $\mathcal{N}(i)$ represents the neighbors of the node $i$, $\mathbf{e}_{j,i}$ represents the edge feature vector (if necessary) of an edge connecting the node $j$ to the node $i$, and $\mathbf{W_{1,2,3}}$ are trainable weight matrices. We stress that Eq.(\ref{eq:message-passing}) is simply an example of the action of a message-passing layer, and should not be considered a general depiction of all possible, often sophisticated, message-passing operations that have been developed in the literature. As message-passing layers (interspersed with nonlinearities or other operations) are applied in succession, information about arbitrarily distant nodes in the graph will be incorporated into each output node. Graph neural networks have demonstrated the capacity to perform learning on graph-structured data sets \cite{2023AIChE..69E7938S,2015arXiv150909292D,2022arXiv220314132B,khemani:2024,electronics12040825,2019arXiv190206667H,2021arXiv210809568S}, and demonstrate many characteristics which shall be useful in a learning task over a subspace of BSM actions. First, because each message-passing layer applies the same trainable function to all nodes, graph neural networks can learn non-trivial behavior across different graphs of differing topologies, or in practical terms, data of the same type with a differing number of input parameters-- since we do not wish to pre-specify a model's field content, this is critical to our reinforcement learning task. Furthermore, the uniformity of message-passing functions allows learned knowledge about various nodes representing a part of a BSM action to be readily transferred to other nodes with similar feature vectors-- that is, the neural network that has learned a nontrivial constraint on, for example, a vector-like electroweak triplet fermion will automatically apply similar constraints to an arbitrary number of additional such fields in the theory. Finally, a graph neural network has no {\it a priori} notion of the ordering of its inputs-- all nodes are processed simultaneously. Because a BSM action consists of an arbitrary sum of new fields' kinetic and interaction terms, introducing a learner which did not automatically recognize additive commutativity in these actions would increase the possible space of inputs by combinatoric factors, which will explode as more fields are introduced to the theory.

Having argued for the utility of a graph to represent a BSM action in our learning task (and, indeed, in learning tasks featuring such models as inputs in general), we must now determine how to develop a suitable graph grammar for these actions, and establish under what conditions we can do so. First, we shall do so in the abstract, to illustrate our procedure's general applicability, and then for clarity, we shall provide some toy examples. In principal, virtually any data may be represented as a graph, as long as each edge connects only two nodes, the space of possible nodes and edge types are finite-dimensional, and each datum to be represented as a graph has a finite number of nodes.\footnote{An alternative generalized structure, hypergraphs, describe structures in which edges can connect more than two nodes. We do not find its application necessary for expressing BSM actions.} We can address these conditions by carefully considering the characteristics of a general 4-dimensional BSM action:

\begin{enumerate}
    \item A BSM action consists of an arbitrary number of different fields, which in turn interact with one another via different coupling terms. Each of these fields has a representation under the spacetime symmetry group (the Poincare group in 4D), as well as a representation under whatever internal symmetry(s) that the model proposes. Excluding theories with infinite towers of states, such as emerge from dimensional reduction of higher-dimensional theories, the total number of fields in any action will be finite.
    \item Each interaction term includes a coupling constant and some number of fields, contracted together to form a singlet under the theory's preserved symmetries.
    \item Due to constraints in the forms of interactions from the internal and spacetime symmetries of the action, there are a finite number of possible distinct interaction operators that can be written in the action, as long as two principles hold: (i) there are a finite number of \emph{distinct} group representations appearing in the model, and (ii) there is a maximum mass dimension for the operators that are written down.\footnote{The first of these conditions is almost trivially satisfied: It is difficult to imagine a theory in which there are even countably infinite \emph{different} group representations appearing within its elementary field content. The second is in practice satisfied by nearly any action that can be written down: A renormalizable theory will have no interaction terms with mass dimension greater than 4 (at least in 4 spacetime dimensions), and effective field theory analyses must truncate their analysis at some finite order (which, due to the decoupling of UV physics from the IR, is computationally justified).}
\end{enumerate}

Characteristics 1 and 2 naturally suggest a basic graph structure for a BSM action graph grammar: Each node in a graph represents either a field or an interaction term. Edges can then connect field nodes to coupling terms, allowing interaction terms with arbitrary numbers of different fields to be represented with edges that connect only pairs of particles. Identifying conditions under which the space of graph and edge features is finite-dimensional is straightforward. We see in characteristic 3 that under very broadly applicable assumptions, we might describe a class of BSM theories with $N$ different possible field representations under the Poincare and internal symmetry groups and $M$ different possible forms of operators, with $N$ and $M$ being finite numbers. If we then describe each possible field representation $i$ with $n_i$ parameters (which may be discrete or continuous, for example $U(1)$ charges or particle masses) and each type of interaction term $j$ with $n_j$ (similarly discrete or continuous) parameters, we might craft our node representations in one of two ways. First, using a \emph{heterogeneous} graph architecture, we can construct a graph which supports multiple different node types with potentially different feature vectors, as long as the total number of types of nodes is finite -- message-passing is then done for each node and edge type individually and aggregated according to a pre-specified operation. This leads us to having $N+M$ different node types, each with finite feature dimensionalities given by $n_i$ and $n_j$. Alternatively, we can use a homogeneous graph architecture with a single feature vector, and specify different node types with discrete components of the feature vector -- for example, we might specify different charges under several $U(1)$ groups using a vector of charge values for each $U(1)$. At worst, we can always represent the different node types with $N+M$ one-hot encodings and  $\sum_i n_i + \sum_j n_j$ elements to represent possible numerical field and coupling parameters.\footnote{The  parameters that aren't applicable to a particular node, such as particle mass for a node representing a coupling term, will be trivially set to zero.} The difference between the two embedding strategies will depend on the precise mathematical form of the message-passing layers used in the graph neural network, and in practice both embedding strategies can be combined, with any discrete parameter in a feature vector substituted for differing node types for each of its values. At the end of this Section, we shall discuss the broad considerations that can influence a choice between these paradigms, and we shall explore the differing merits of both strategies in our particular reinforcement learning task in more detail in Section \ref{sec:results}.

In either case, it is clear that, as long as our individual node types are specified by a finite number of parameters (which in practice is almost trivially true), we have crafted a finite-dimensional space of possible nodes, specifically one of dimension $N+M+\sum_i n_i + \sum_j n_j$. Our sole remaining task is to establish that the space of edge features in a theory is also finite-dimensional. To that end, we must consider the form of a generic interaction term in a quantum field theory. Each such term will feature a finite number of different fields, but these fields are not necessarily treated equivalently by the contraction-- a trivial example would be that, for two Dirac fermions $\psi$ and $\chi$ (and the usual left-handed chiral projection operator $P_L$), we see that
\begin{align}\label{eq:dirac-fermion}
    \overline{\psi} P_L \chi \neq \overline{\chi} P_L \psi.
\end{align}
As we can see in the form of the simple message-passing layer depicted in Eq.(\ref{eq:message-passing}), in the absence of edge feature vectors message-passing layers will apply the same operation to \emph{all} a node's neighbors-- even for more complicated message-passing layers, this universality must hold because aggregation of nodes' neighbors must be well-defined for any number of neighbors. If no edge feature vectors are employed, then, a coupling node representing a Dirac mass term of the sort depicted in Eq.(\ref{eq:dirac-fermion}) could not be adequately represented in a graph, since a coupling node for this interaction connected to $\psi$ and $\chi$ would not uniquely represent one of these operators. We can readily address this issue with edge features, however-- in our example operation we could include a 2-dimensional feature vector of one-hot encodings, so that the $\psi$ node was connected to the coupling term with a ``barred" edge (with, \eg, feature vector [1,0]) and the $\chi$ node was connected to the coupling term with an ``unbarred" edge (feature vector [0,1]) in order to represent the operator on the left side of Eq.(\ref{eq:dirac-fermion}), and vice versa for the operator on the right.

Generally, each of a model's $M$ different types of coupling terms will feature a finite number of distinct fields, which we can describe as coupling type $j$ being a contraction of $f_j$ fields. Then, there are at most $f_j$ different edge types necessary to uniquely specify an interaction term, assuming that each field in $j$ appears differently in the contraction. This in turn means that the edge feature space will have dimensionality of \emph{at most} $\sum_j f_j$, which is finite. Therefore, we have demonstrated that in addition to the node feature space being finite-dimensional, we have a finite-dimensional edge feature space for a generic BSM theory. In the heterogeneous graph language, instead of encoding these edge features in vectors, we can simply use explicitly different edge types. Message-passing is done for each edge type individually, with different trainable weights, and aggregated.

To get a sense of how our node embedding procedure works in practice, we present graph grammars for two toy models of increasing complexity. The first of these is a theory with $N$ real scalar singlets $\phi_{1,...,n}$ as a graph, subject to a $Z_2^N$ global discrete symmetry so that the mass and interaction terms for the scalars are of the form
\begin{align}\label{eq:simple-scalar}
    \mathcal{L}_{\textrm{scal},Z_2} \supset -\sum_i \bigg( \frac{m_i^2}{2} \phi_i^2 + \lambda_i \phi_i^4 + \sum_{j > i} \lambda_{ij} \phi_i^2 \phi_j^2  \bigg) 
\end{align}
Together with kinetic terms, Eq.~(\ref{eq:simple-scalar}) depicts the entire action for the toy theory. Working in a canonically normalized field basis without loss of generality, the kinetic terms are entirely determined by the fields' representation under the Poincare group (in this case, singlets), so we note that the interaction terms depicted in Eq.~(\ref{eq:simple-scalar}) contain all the information to uniquely specify a theory of this class, and we merely need to devise a graph grammar to represent these expressions. Following our outlined strategy, we have exactly one field representation in our model (namely, scalar singlets), which can have two internal features: A mass $m_i^2$ and a quartic self-coupling $\lambda_i$. We furthermore have a single type of coupling node, representing the $\lambda_{ij}$ interactions, which have the numerical value of the coupling constant $\lambda_{ij}$ as their sole features. As each coupling term $\lambda_{ij}$ treats both scalar fields appearing in it identically, we require no edge features (or in the heterogeneous setup, only one edge type) to represent all the multiparticle interactions in the theory. We have summarized this graph grammar, including sample representation vectors for either a homogeneous or heterogeneous graph, in Table \ref{tab:simple-scalar}. To illustrate the grammar's application, in Figure \ref{fig:simple-scalar}, we show the graph representing a theory in the class of models of Eq.(\ref{eq:simple-scalar}) with three total scalars.

\begin{table}[]
    \centering
    \begin{tabular}{| c | c | c | c |}
    \hline
    & Name & Homogeneous Graph & Heterogeneous Graph\\
    \hline
    \multirow{2}{*}{Nodes} & Scalar $\phi_i$ & $\{1, 0, m_i^2, \lambda_i, 0 \}$ & $\{ m_i^2, \lambda_i \}$\\
    & Coupling $\lambda_{ij}$ & $\{0, 1, 0, 0, \lambda_{ij} \}$ & $\{ \lambda_{ij} \}$\\
    \hline
    Edges & $\phi_i^2$ factor & N/A & N/A\\
    \hline
    \end{tabular}
    \caption{A summary of the graph grammar used to represent a model of the class of Eq.(\ref{eq:simple-scalar}), featuring sample feature vector representations of both scalar and interaction nodes using either a traditional homogeneous graph (with one-hot encodings for different node types) or a heterogeneous graph. Only a single edge type, which will link scalar nodes to coupling nodes, is required, so the edge representation will be trivial in either the homogeneous or heterogeneous graph paradigms.}\label{tab:simple-scalar}
\end{table}
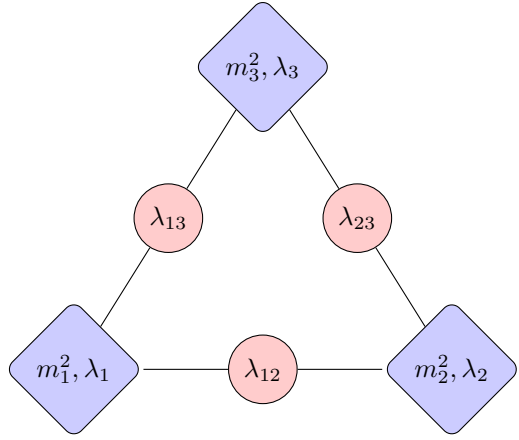
\begin{figure}
    \centering
    \begin{tikzpicture}
        \node at (0,0) [particle, fill=blue!20] (p1-id) {$m_{1}^2, \lambda_1$};
        \node at (5,0) [particle, fill=blue!20] (p2-id) {$m_{2}^2, \lambda_2$};
        \node at (2.5,4) [particle, fill=blue!20] (p3-id) {$m_{3}^2, \lambda_3$};
        \node at (2.5, 0) [coupling, fill=red!20] (c12-id) {$\lambda_{12}$};
        \node at (1.25, 2) [coupling, fill=red!20] (c13-id) {$\lambda_{13}$};
        \node at (3.75, 2) [coupling, fill=red!20] (c23-id) {$\lambda_{23}$};
        \draw[-] (p1-id) -- (c12-id);
        \draw[-] (p2-id) -- (c12-id);
        \draw[-] (p1-id) -- (c13-id);
        \draw[-] (p3-id) -- (c13-id);
        \draw[-] (p2-id) -- (c23-id);
        \draw[-] (p3-id) -- (c23-id);
    \end{tikzpicture}
    \caption{A visual depiction of a graph in the class of theories given by the action in Eq.(\ref{eq:simple-scalar}), assuming there are three scalars in the model. Diamonds denote particle nodes, which contain the mass squared and quartic self-couplings as features. Circles denote quartic couplings, with their coupling constants following the notation conventions of Eq.(\ref{eq:simple-scalar}).}
    \label{fig:simple-scalar}
\end{figure}
We can already see patterns which will characterize our general approach to constructing graph grammars: Particle nodes will generally include self-interactions as features in their feature vectors, and will be connected by edges to interaction nodes that represent different couplings between particles. The models of Eq.(\ref{eq:simple-scalar}), however, are almost trivial-- in particular, we have intentionally selected a construction which requires no edge features. To illustrate our graph grammar procedure for a theory in which edge features are required, we relax the $Z_2^N$ symmetry of Eq.(\ref{eq:simple-scalar}) to a single $Z_2$ symmetry, so that all terms which are of even order in the fields can appear in the action. In this case, the action of Eq.(\ref{eq:simple-scalar}) becomes
\begin{align}\label{eq:scalar-2}
    \mathcal{L}_{\textrm{scal}} \supset \mathcal{L}_{\textrm{scal},Z_2} -\sum_i \bigg( \sum_{j \neq i} \tilde{\lambda}_{i j} \phi_i \phi_j^3 + \sum_{j \neq i} \sum_{k \neq i, k > j} \lambda_{ijk} \phi_i^2 \phi_j \phi_k + \sum_{j > i} \sum_{k > j} \sum_{l > k} \lambda_{ijkl} \phi_i \phi_j \phi_k \phi_l \bigg).
\end{align}
We have elected to work (without loss of generality) in a field basis in which the mass matrix for the scalars is diagonal, so that there are no terms of the form $m_{ij}^2 \phi_i \phi_j$ in the action. Because this theory's particles have the same self-interactions as our previous construction, our field nodes are the same. However, instead of one type of coupling ($\lambda_{ij}$), we now have three additional coupling types, $\tilde{\lambda}_{ij}$, $\lambda_{ijk}$, and $\lambda_{ijkl}$. Furthermore, these coupling types do \emph{not} treat the fields which appear in them equivalently, and therefore different edge types are required as well. In theory, because we have 4 total coupling types which each contain 4 fields, the maximum possible number of edge types we require according to our earlier analysis would be 16-- 4 for each coupling variety. Of course, given the commutativity of the scalar fields, we note that we can dramatically improve on this upper bound by simply creating edges representing the inclusion of different powers of a field $\phi_i$ in the coupling term: One edge type for a single factor, another for a factor of $\phi_i^2$, and a third for a factor of $\phi_i^3$. We have summarized the graph grammar for the theories in the class of Eq.(\ref{eq:scalar-2}) in Table \ref{tab:scalar-2}, and depicted a sample graph with three scalar fields in Figure \ref{fig:scalar-2}.
\begin{table}[]
    \centering
    \begin{tabular}{| c | c | c | c |}
    \hline
    & Name & Homogeneous Graph & Heterogeneous Graph\\
    \hline
    \multirow{5}{*}{Nodes} & Scalar $\phi_i$ & $\{1, 0, 0, 0, 0, m_i^2, \lambda_i, 0, 0, 0, 0 \}$ & $\{ m_i^2, \lambda_i \}$\\
    & Coupling $\lambda_{ij}$ & $\{0, 1, 0, 0, 0, 0, 0, \lambda_{ij}, 0, 0, 0 \}$ & $\{ \lambda_{ij} \}$\\
    & Coupling $\tilde{\lambda}_{ij}$ & $\{0, 0, 1, 0, 0, 0, 0, 0, \tilde{\lambda}_{ij}, 0, 0 \}$ & $\{ \tilde{\lambda}_{ij} \}$\\
    & Coupling $\lambda_{ijk}$ & $\{0, 0, 0, 1, 0, 0, 0, 0, 0, \lambda_{ijk}, 0 \}$ & $\{ \lambda_{ijk} \}$\\
    & Coupling $\lambda_{ijkl}$ & $\{0, 0, 0, 0, 1, 0, 0, 0, 0, 0, \lambda_{ijkl} \}$ & $\{ \lambda_{ijkl} \}$\\
    \hline
    \multirow{3}{*}{Edges} & $\phi_i$ factor & $\{ 1, 0, 0 \}$ & N/A\\
    & $\phi_i^2$ factor & $\{ 0, 1, 0 \}$ & N/A\\
    & $\phi_i^3$ factor & $\{ 0, 0, 1 \}$ & N/A\\
    \hline
    \end{tabular}
    \caption{A summary of the graph grammar used to represent a model of the class of Eq.(\ref{eq:scalar-2}), featuring sample feature vector representations of both scalar and interaction nodes using either a traditional homogeneous graph (with one-hot encodings for different node types) or a heterogeneous graph, as well as homogeneous graph representations of edge feature vectors. Different edge types in the hetereogeneous graph representation are not represented by edge feature vectors, but rather are simply stored as different edge types in the graph data structure.}\label{tab:scalar-2}
\end{table}
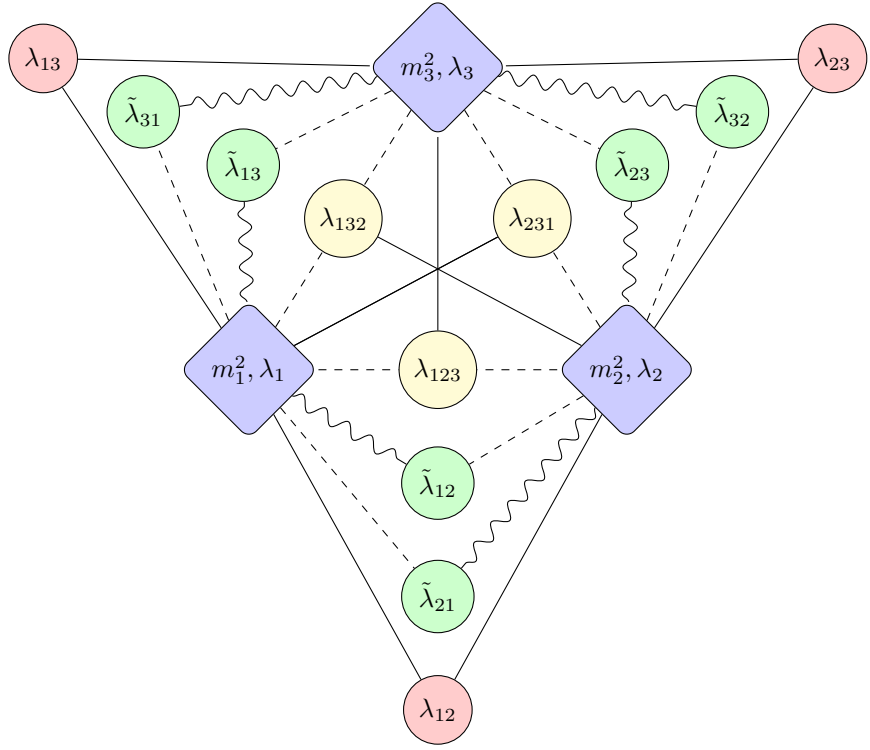
\begin{figure}
    \centering
    \begin{tikzpicture}
        \node at (0,0) [particle, fill=blue!20] (p1-id) {$m_{1}^2, \lambda_1$};
        \node at (5,0) [particle, fill=blue!20] (p2-id) {$m_{2}^2, \lambda_2$};
        \node at (2.5,4) [particle, fill=blue!20] (p3-id) {$m_{3}^2, \lambda_3$};
        \node at (2.5,0) [coupling, fill=yellow!20] (c123-id) {$\lambda_{123}$};
        \node at (2.5,-1.5) [coupling, fill=green!20] (cc12-id) {$\tilde{\lambda}_{12}$};
        \node at (2.5,-3) [coupling, fill=green!20] (cc21-id) {$\tilde{\lambda}_{21}$};
        \node at (2.5,-4.5) [coupling, fill=red!20] (c12-id) {$\lambda_{12}$};
        \node at (1.25, 2) [coupling, fill=yellow!20] (c132-id) {$\lambda_{132}$};
        \node at (3.75, 2) [coupling, fill=yellow!20] (c231-id) {$\lambda_{231}$};
        \node at (5.074, 2.706) [coupling, fill=green!20] (cc23-id) {$\tilde{\lambda}_{23}$};
        \node at (6.397, 3.412) [coupling, fill=green!20] (cc32-id) {$\tilde{\lambda}_{32}$};
        \node at (7.721, 4.118) [coupling, fill=red!20] (c23-id) {$\lambda_{23}$};
        \node at (-0.074, 2.706) [coupling, fill=green!20] (cc13-id) {$\tilde{\lambda}_{13}$};
        \node at (-1.397, 3.412) [coupling, fill=green!20] (cc31-id) {$\tilde{\lambda}_{31}$};
        \node at (-2.721, 4.118) [coupling, fill=red!20] (c13-id) {$\lambda_{13}$};
        \draw[dashed] (p1-id) -- (c123-id);
        \draw[dashed] (p2-id) -- (c123-id);
        \draw[-] (p3-id) -- (c123-id);
        \draw[dashed] (p1-id) -- (c132-id);
        \draw[dashed] (p3-id) -- (c132-id);
        \draw[-] (p2-id) -- (c132-id);
        \draw[dashed] (p2-id) -- (c231-id);
        \draw[dashed] (p3-id) -- (c231-id);
        \draw[-] (p1-id) -- (c231-id);
        \draw[-, decorate,decoration=snake] (p1-id) -- (cc12-id);
        \draw[dashed] (p2-id) -- (cc12-id);
        \draw[-, decorate,decoration=snake] (p2-id) -- (cc21-id);
        \draw[dashed] (p1-id) -- (cc21-id);
        \draw[-] (p1-id) -- (c12-id);
        \draw[-] (p2-id) -- (c12-id);
        \draw[-, decorate,decoration=snake] (p2-id) -- (cc23-id);
        \draw[dashed] (p3-id) -- (cc23-id);
        \draw[-, decorate,decoration=snake] (p3-id) -- (cc32-id);
        \draw[dashed] (p2-id) -- (cc32-id);
        \draw[-] (p2-id) -- (c23-id);
        \draw[-] (p3-id) -- (c23-id);
        \draw[-] (p1-id) -- (c231-id);
        \draw[-, decorate,decoration=snake] (p1-id) -- (cc13-id);
        \draw[dashed] (p3-id) -- (cc13-id);
        \draw[-, decorate,decoration=snake] (p3-id) -- (cc31-id);
        \draw[dashed] (p1-id) -- (cc31-id);
        \draw[-] (p1-id) -- (c13-id);
        \draw[-] (p3-id) -- (c13-id);
    \end{tikzpicture}
    \caption{A visual depiction of a graph in the class of theories given by the action in Eq.(\ref{eq:scalar-2}), assuming there are three scalars in the model. Diamonds denote particle nodes, which contain the mass squared and quartic self-couplings as features. Circles denote quartic couplings, with their coupling constants following the notation conventions of Eq.(\ref{eq:scalar-2}). Different coupling types are denoted using different colors, while different edge types are denoted with different line styles: A dashed line denotes a single factor of $\phi_i$ in the interaction term, a solid line denotes two factors, and a wavy line denotes three.}
    \label{fig:scalar-2}
\end{figure}

We note that our graph grammar as shown in Table \ref{tab:scalar-2} illustrates another characteristic of this procedure: An efficient graph grammar for a given class of theories is not unique. For example, all coupling terms appearing in the action of Eq.(\ref{eq:scalar-2}) are simply specialized forms of $\lambda_{ijkl}$, and thus in practice we might streamline our representation further by keeping only a single type of coupling node. Furthermore, since all of the edge types that we have enumerated are composed of one or more single factors of $\phi_i$, we might reduce our graph grammar even more by eliminating all edge types except the one corresponding to the inclusion of a single scalar factor in the coupling term. The other edge types could be represented by redundant instances of this edge, so that, for example, the coupling $\lambda_{ij} \phi_i^2 \phi_j^2$ would feature two edges between the $\lambda_{ij}$ node and the $\phi_i$ node, and two between the $\lambda_{ij}$ node and the $\phi_j$ node. While this graph grammar is more economical, the purpose of our example is to demonstrate the representation of a theory with multiple coupling and edge types, which for more general theories are necessary, and therefore we have not developed this simpler representation here.

The graph grammars of Tables \ref{tab:simple-scalar} and \ref{tab:scalar-2} illustrate the basic strategy for depicting \emph{any} BSM model (or, indeed, any quantum field theory action in general) subject to extremely weak assumptions: Namely, that consideration of only fields in a pre-specified set of allowed representations under the spacetime and internal symmetry groups of some class of models be allowed, that only operators up to a pre-specified mass dimension be considered, and every theory in the class of models have only a finite number of fields. This encompasses very nearly the entire space of possible BSM models, with the exception of those theories with infinite towers of states stemming from, \eg, Kaluza-Klein reduction. Additional types of fields (potentially with differing spins or nontrivial representations under some symmetry group), as well as additional interaction terms, can be readily introduced to these graph grammars by simply extending the feature vectors or node types. We can summarize our suggested procedure as follows:

\begin{enumerate}
    \item Specify the internal and spacetime symmetry group $\mathcal{G}$ of a class of models that are desired to be automatically included in the graph grammar. Notably, theories with symmetries beyond those specified here can still be expressed in the graph, with the additional symmetries reflected in the graph topology at the price of increased graph complexity and diminished human-readability of the data structure. For example, by specifying $\mathcal{G}$ as solely consisting of the 4-dimensional Poincare group, and leaving all internal gauge groups unstated, one can represent a 4-dimensional quantum field theory in flat spacetime with arbitrary gauge symmetries, or none at all, with the same graph data structure.
    \item Identify a finite number $N$ of different field representations under $\mathcal{G}$ that will be considered in this learning task. These different field representations will correspond to different particle node types encoded in feature vectors. The node feature vectors can encode these different node types as different node classes in a heterogeneous graph, as a discrete vector with $N$ total possible values (at worst, an $N$-dimensional one-hot encoding), or a combination of these two strategies. To complete a representation of the different fields in the model, append these node feature vectors with numerical parameters specifying interactions that do not involve multiple fields in the graph. These can either be self-interactions or interactions with fields that might be left implicit in a graph grammar, such as the SM field content in a BSM model building learning task.
    \item Identify all possible forms for interaction terms that can appear in the model, up to a finite mass order. As long as the truncation at some mass order is observed, this will be a finite number, $M$. Different interaction types can be encoded as either up to $M$ different node types in a heterogeneous graph, or parameterized as a discrete vector (at worst, with $M$ different one-hot encodings),  and include numerical parameters giving the magnitudes of each coupling as additional node features.
    \item In the homogeneous graph paradigm, the concatenation of the vector representations of the field and interaction nodes will create the node feature vectors of the graph. In the heterogeneous graph paradigm, these two classes of node feature vectors are considered separately.
    \item For each interaction term, identify the number of distinct edge types that are required to uniquely specify an operator. In a homogeneous graph, these edge types will be embedded in edge feature vectors with one-hot encodings. In the heterogeneous graph, they will be simply included as different edge types without the need for any edge feature vector.
\end{enumerate}

We shall follow this general outline when we encode BSM actions for different models as graphs for the reinforcement learning study we embark on in this work. As we have observed here, there exists some nontrivial ambiguity regarding how one might construct such a graph grammar, in particular considering the encoding of different particle representations or coupling forms using discrete feature vector components or different node types in a heterogeneous graph. Ultimately, the choice amounts to a decision regarding the degree of transferability of knowledge about nodes of one type compared to another. If separate node types are used, then the neural network will use entirely independent sets of weights to operate on each. If, instead, discrete feature vector components are used, similar node types (which might share certain feature parameters) will share many of the same weights, allowing for knowledge gained about one node type to be applied to the other. The efficacy of either strategy will depend on the parameterization of various nodes (in particular, the number of features which multiple node types might share) as well as the physics of the problem in question and the architecture of the neural network training on the graphs. As an example, in this work we shall consider two graph representations of the same class of theories in which we make different choices regarding the use of heterogeneous node types versus discrete features to encode the models.

Beyond the reinforcement learning use case we explore in this work, we emphasize that our graph procedure has potentially broad applications for a variety of learning tasks. For example, such a graph encoding could allow for machine learning of computationally intensive likelihood calculations that are applicable to theories with differing particle contents, supervised classification tasks for such theories, or even the use of generative models to produce new theories. By including only the Poincare group as our underlying symmetry group $\mathcal{G}$, we might also represent more or less arbitrary theories in a graph grammar over which a single graph neural network can learn, permitting learning tasks to be performed readily using data sets drawn from nearly the entire space of quantum field theory actions that can be written down.

\section{Reinforcement Learning Basics}\label{sec:reinforcement}

As discussed in the Introduction, a particular use case for the graph technology that we have developed in Section \ref{sec:graph} lies in a reinforcement learning scan of the space of a particular class of BSM models. Reinforcement learning is particularly well-suited to exploring large spaces in search of narrow acceptable regions, and the technique has been successfully employed to identify promising regions in Froggatt-Nielsen parameter space \cite{Harvey:2021oue,Nishimura:2020nre}. In this section, we provide a brief review of some essential concepts in reinforcement learning that we have employed in our current analysis, which seeks to broaden the applicability of these search techniques to model spaces with variable BSM particle content, and hence variable numbers of parameters.

Reinforcement learning is modelled as a Markov decision process (MDP). This consists of:
\begin{itemize}
    \item A state space $\mathcal{S}$, which in our case is some theorist-defined subspace of the space of possible BSM models, which we shall describe in greater detail in Section \ref{sec:model}.
    \item A space of actions $\mathcal{A}$, which in our case is the space of possible changes to a BSM Lagrangian density that we permit our agent to take.
    \item A policy $\pi_\theta$, which given a state $s \in S$, will assign probabilities that the agent will take various actions in $\mathcal{A}$.
\end{itemize}

At a given step in training, the reinforcement learning agent is passed a state $s_t \in \mathcal{S}$, and then samples an action $a_t \in \mathcal{A}$ based on probabilities given by $\pi_\theta$. This produces a new state, $s_{t+1}$, and assigns a numerical reward, $r_t$, to the agent's action based on a user-specified reward function. The goal of the reinforcement learning agent is to learn parameters $\theta$ for the policy $\pi_\theta$ that maximize the expected accumulated rewards over some collection of steps, called an \emph{episode}.

A number of algorithms exist to train the policy $\pi_\theta$. For this work, we select Proximal Policy Optimization (PPO) \cite{2017arXiv170706347S}. In the interest of accessibility, we shall briefly outline the relevant core concepts of the PPO algorithm in this section, but for a detailed (and more mathematically precise) discussion, we refer the reader to Appendix \ref{appendix:PPO}. Conceptually, PPO represents a policy gradient strategy, in which an objective function that can be stochastically estimated by sampling over some collection of reinforcement learning steps is maximized over the course of training by gradient ascent. By a policy gradient theorem, it can be demonstrated that this stochastically estimated function should share the same gradient with respect to the policy parameters (and therefore the same local extrema) as the accumulated rewards over the course of an episode, which we wish to optimize. PPO is also an example of an \emph{actor-critic} architecture, in which in order to learn tasks with long time horizons, the agent's policy includes not just an actor which recommends action probabilities, but also a critic which learns to estimate the expected long-term rewards that the policy will receive, if starting from a given input state. This value function in turn is inserted into the algorithm's objective function, allowing the objective function to reflect long-term expected returns even if training samples only contain fractions of full episodes. Schematically, the program flow for training with PPO (or, indeed, any actor-critic architecture) follows these simple steps:

\begin{enumerate}
    \item Sample a trajectory (or parallel trajectories for many environments) through state space $\mathcal{S}$ by acting according to the actor policy $\pi_\theta$.
    \item For episodes that have not terminated (according to some user-defined conditions), estimate the long-term value of their current states using the critic network.
    \item Use the sampled data (trajectories and values) to construct the PPO objective function, $L_{PPO}$, to be optimized (see Appendix \ref{appendix:PPO} for the form of the objective function in PPO).
    \item Optimize the parameters of the neural network to maximize the loss function, $L = L_{PPO} - L_{\textrm{critic}} + \beta S[\pi_\theta]$ using gradient ascent. In addition to the $L_{PPO}$ objective function, the loss is supplemented by $L_{\textrm{critic}}$ (the mean squared error of the critic network outputs compared to the observed values of the states along the trajectory) and $S[\pi_\theta]$, the mean entropy of the policy's output distributions over the sample trajectory, scaled by a hyperparameter $\beta$. The hyperparameter $\beta$ plays a crucial role in policy training, both as a regularizer during training and as a tunable method of encouraging exploration of the state space $\mathcal{S}$ rather than simple exploitation of a single optimal strategy.
    \item Repeat the steps 1-4 for some user-specified number of rounds.
\end{enumerate}

Since our task involves both continuously parameterized actions (such as adjusting particle parameters) and discrete actions (such as removing particles from the model), we employ a variant of the original PPO algorithm for hybrid spaces, first introduced in \cite{2019arXiv190301344F}, which the authors have called H-PPO. H-PPO in our setting functions virtually identically to conventional PPO, with the only difference being that instead of one actor, there are now an entire collection of actors representing policies for various discrete and continuous actions. This multi-actor architecture can also accommodate hierarchical actions, in which one action (for example, the discrete choice of which particle to modify) might necessarily be accompanied by other actions which parameterize it (for example, the continuous set of modifications to make to the chosen particle). The PPO loop described above functions as before, however the objective function $L_{PPO}$ is modified to treat each actor separately, and only to include a given actor's contribution to it only for steps in which that actor's output is used to modify the state (so for example, the actor which selects a specific particle for removal will not be updated by data from steps in which the actor which selects high-level actions did not choose to remove a particle from the model). 

\section{Model Building with Vector-Like Leptons}\label{sec:model}

Having outlined the tools we will use to explore a BSM parameter space, we can now begin to specify the precise space we shall be exploring. Because the space of all possible BSM models is enormous, for the purposes of this exploration it behooves us to limit our consideration to a particular class of such models. Nonetheless, we wish to explore the ability of the graph architecture that we propose to identify compelling models in a complicated, large parameter space with rich phenomenology. A reasonable space in which to work might be inspired by the portal matter paradigm of \cite{Rizzo:2018vlb}, and in particular the lepton-focused setup of \cite{Wojcik:2022woa,Wojcik:2023ggt} which has indicated some promise for these models to explain the observed anomaly in the anomalous magnetic moment of the muon. In this setup, the SM gauge symmetry is augmented with a dark Abelian gauge group, $U(1)_D$, under which all SM particles are uncharged, but an SM singlet dark matter candidate will be charged. The gauge boson for $U(1)_D$ (the dark photon) then couples feebly to SM fields via kinetic mixing with the SM hypercharge, which is achieved via one-loop exchange of new heavy ``portal matter'' particles that are charged under both the SM and $U(1)_D$ groups. If both the dark photon and the dark matter candidate possess approximately $O(0.1-1 \; \textrm{GeV})$ masses, then the portal matter-induced kinetic mixing will be of approximately the correct magnitude to produce the observed dark matter relic abundance via freeze-out. In \cite{Rizzo:2018vlb}, the author argued that a natural construction for such portal matter fields would be vector-like fermions which possess the same SM quantum numbers as existing SM fermions, and are mixed with SM fermions via Yukawa couplings to a sub-GeV dark Higgs field, which shall also break the $U(1)_D$ symmetry and impart mass to the dark photon. Assuming a fixed sub-GeV field content of a dark Higgs boson and a dark photon, we then allow our agent to consider the class of models with an arbitrary content of vector-like leptons, either in the form of portal matter or more conventional, $U(1)_D$-uncharged species-- the specific varieties of BSM vector-like fermions our model can consider are listed in Table \ref{tab:BSM-particles}.

\begin{table}[]
    \centering
    \begin{tabular}{| c | c | c |}
    \hline
    Field & $SU(2)_L \times U(1)_Y$ & $U(1)_D$\\
    \hline
    $L_0$ & $\mathbf{2}_{-\frac{1}{2}}$ & $0$ \\
    $L_{\pm 1}$ & $\mathbf{2}_{-\frac{1}{2}}$ & $\pm 1$\\
    $E_0$ & $\mathbf{1}_{+1}$ & $0$\\
    $E_{\pm 1}$ & $\mathbf{1}_{+1}$ & $\pm 1$\\
    \hline
    \end{tabular}
    \caption{The different vector-like leptons that the agent will attempt to introduce to the SM. These fermions may be mixed with SM fermions and each other via Yukawa couplings to the SM Higgs boson and an SM singlet dark Higgs boson with $U(1)_D$ charge $+1$.}
    \label{tab:BSM-particles}
\end{table}

To facilitate mixing between the vector-like leptons and the SM (and therefore facilitate the former's prompt decay into the latter), it is necessary that we assume that our dark Higgs field has a $U(1)_D$ charge of $+1$\footnote{We may also specify a charge of $-1$, however this choice is phenomenologically inconsequential}. Then, given our proposed particle content, the action of the model is specified by the Lagrangian density (after spontaneous breaking of $U(1)_D$ by the dark Higgs field taking on a vev of $v_D / \sqrt{2} \lesssim \textrm{GeV}$, and the usual spontaneous breaking of the electroweak symmetry by the SM Higgs mechanism):
\begin{align}\label{eq:model-action}
    \mathcal{L} &= \mathcal{L}_{\textrm{SM}} + \mathcal{L}_{\textrm{Dark}} + \mathcal{L}_{\textrm{VLL}} \nonumber \\
    \mathcal{L}_{\textrm{Dark}} &\equiv -\frac{1}{4} (F^D_{\mu \nu})^2 + \frac{1}{2} m_{A_D}^2 + \frac{1}{2} (\partial_{\mu} h_D)^2 - \frac{1}{2} m_{h_D}^2 + ... \nonumber\\
    \mathcal{L}_{\textrm{VLL}} &\equiv \sum_{q = 0, + -} \big( \sum_{k} \bar{L}_q^k (i \slashed{D} - M^{L,q}_{k}) L_q^k + \sum_{\alpha} \bar{E}_q^\alpha (i \slashed{D} - M^{E,q}_{\alpha}) E_q^\alpha \big) - \mathcal{L}_{y,6} - \mathcal{L}_{\lambda, 6} - \mathcal{L}_{y,8} - \mathcal{L}_{\lambda, 8L} - \mathcal{L}_{\lambda, 8E} \nonumber\\
    \mathcal{L}_{y,6} &\equiv \frac{(h + v)}{\sqrt{2}} \sum_{f = e, \mu, \tau} \bigg[ \sum_{i} y_{L, if}^0 (\overline{L}_0^i)_L e^f_R + \sum_{\alpha} y_{E, \alpha f}^0 \overline{e}^f_L (E_0^\alpha)_R \bigg] + h.c. \\
    \mathcal{L}_{y,8} &\equiv \frac{(h + v)}{\sqrt{2}} \sum_{q= 0, +, -}
    \sum_{i, \alpha} \bigg[ y_{LE,i \alpha}^q (\overline{L}_q^i)_L (E_q^\alpha)_R + y_{EL,i \alpha}^q (\overline{E}_q^\alpha)_L (L_q^i)_R \bigg] + h.c. \nonumber\\
    \mathcal{L}_{\lambda,6} &\equiv \frac{(h_D + v_D)}{\sqrt{2}} \sum_{q = +, -} \sum_{f = e, \mu, \tau} \bigg[ \sum_{\alpha} \lambda_{E,\alpha f}^q (\overline{E}^\alpha_q)_L e^f_R + \sum_{i} \lambda_{L,i f}^q  \overline{e}^f_L (L^i_q)_R\bigg] + h.c. \nonumber\\
    \mathcal{L}_{\lambda,8L} &\equiv \frac{(h_D + v_D)}{\sqrt{2}} \sum_{q = +, -} \sum_{i,j} \bigg[ \lambda^{0q}_{L,i j} (\overline{L}_0^i)_L (L_q^j)_R + \lambda^{q0}_{L,ij} (\overline{L}_q^j)_L (L_0^i)_R \bigg] + h.c. \nonumber \\
    \mathcal{L}_{\lambda,8E} &\equiv \frac{(h_D + v_D)}{\sqrt{2}} \sum_{q = +, -} \sum_{\alpha, \beta}\bigg[ \lambda^{0q}_{E,\alpha \beta} (\overline{E}_0^\alpha)_L (E_q^\beta)_R + \lambda^{q0}_{E,\alpha \beta} (\overline{E}_q^\beta)_L (E_0^\alpha)_R \bigg] + h.c. \nonumber
\end{align}
The notation of Eq.(\ref{eq:model-action}) bears some explanation. The $\mathcal{L}_{\textrm{SM}}$ term represents the SM Lagrangian and requires no further elaboration. $\mathcal{L}_{\textrm{Dark}}$ encapsulates the theory's sub-GeV dark sector. We only explicitly include the terms which are relevant to the portal matter-induced physics that we shall observe here, which are the existence of a massive sub-GeV dark Higgs that imparts a sub-GeV mass to the dark photon. Other elements of this sector, such as an actual dark matter candidate or dark Higgs interaction terms with the dark photon, do not significantly affect any of the observables that we consider in this work, and so we remain agnostic about them. The central focus of our model building lies in $\mathcal{L}_{\textrm{VLL}}$, which describes the action of the vector-like lepton sector. Here, we have an unspecified set of isospin doublet (singlet) vector-like leptons, enumerated with the lower case Roman (Greek) indices, which may have a $U(1)_D$ charge of $0$, $+1$, or $-1$. $\mathcal{L}_{y(\lambda),6}$ contains SM (dark) Higgs Yukawa coupling terms between vector-like leptons and SM fermions, which after integrating out the vector-like leptons appear in dimension-6 SMEFT operators. $\mathcal{L}_{y,8}$ ($\mathcal{L}_{\lambda,8L}$, and $\mathcal{L}_{\lambda,8E}$), meanwhile, contains SM (dark) Higgs Yukawa interaction terms between different vector-like leptons and first enter a low-energy effective field theory at the dimension-8 level. In spite of their subleading EFT contributions, we note that these dimension-8 operators tend to dominate BSM contributions to anomalous lepton magnetic dipole moments in these theories, due to chiral enhancements to their contributions relative to dimension-6 operators.

The class of models that we are considering here possesses some attractive characteristics as a demonstration space for our reinforcement learning methodology. First, we note that two features of this model class ease our method's implementation in this context: Because the only BSM gauge group is a dark sector $U(1)$, there is no need to explicitly modify any SM particles' quantum numbers or parameters (which in turn means that we can limit the information passed to our reinforcement learning agent only to BSM particles and couplings, rather than explicitly representing the SM). Furthermore, the present experimental constraints on these models are well-understood, having been explored in, \eg, \cite{Wojcik:2023ggt, Wojcik:2022woa,Ishiwata:2013gma,CarcamoHernandez:2019ydc,Lavoura:2003xp,Leveille:1977rc,Crivellin:2020ebi}, and are principally dominated by precision observables, contributions to which can be rapidly computed for given numerical points in theory space with modest computing power. We emphasize that these characteristics simply ease implementation, and are by no means necessary preconditions for applicability of the analysis methodology that we shall present here-- as we have seen in Section \ref{sec:graph}, the graph representation of a quantum field theory action, upon which our reinforcement learning agent is built, is extremely general, and so our techniques can theoretically be applied to any BSM model with finite field content.

While comparatively simple to implement, the general leptonic portal matter models we consider also exhibit a rich and complex phenomenological structure which our agent can be tasked to learn. In particular, because the flavor structure of the theory is kept entirely general, each BSM particle is capable of having three independent couplings to the SM, one per generation. In practice, two of these must be highly suppressed in order to avoid severe constraints from measurements of lepton-flavor-violating processes. A natural challenge to our reinforcement learning agent is to then produce flavor-conserving (at least to within experimental tolerance) models of new physics in a regime which requires extreme fine-tuning; we shall later argue that random sampling-based methods of a parameter space scan are poorly-suited to this task without being supplied considerably more domain knowledge by a human theorist, while our reinforcement learning agent is not. A further intriguing aspect of this model space lies in the potential phenomenological significance of dimension-8 operators in important experimental constraints. As discussed earlier, these Yukawa couplings between different vector-like species can provide a chirally-enhanced contribution to anomalous lepton magnetic dipole moments. Given that the muon anomalous magnetic dipole moment currently exhibits significant tension with the SM calculation,\footnote{Lattice calculations of the hadronic vacuum polarization (HVP) contribution to the SM value of the anomalous magnetic moment of the muon \cite{Borsanyi:2020mff,Ce:2022kxy,Alexandrou:2022amy,Colangelo:2022vok} have suggested that the currently observed tension may be attributable to this contribution, which is the dominant source of theoretical uncertainty in the SM prediction. As the purpose of this paper is to explore the utility of reinforcement learning as a method to explore BSM model spaces, and as discussed in this Section the muon $g-2$ anomaly evinces several attractive characteristics as a benchmark measurement to fit for this task, we have elected to take the current tension at face value as evidence of BSM physics for purposes of this study.} a reinforcement agent designed to produce models which diminish tensions between SM predictions and experiments will be richly rewarded for crafting a model which addresses this discrepancy. In turn this means that a successful reinforcement learning agent must learn not just individual particle species and parameters which can fit observational data well, it also must learn models holistically and consider the interactions \emph{between} different new physics particles, and learn to appropriately tune the Yukawa interactions $y_{LE}^{0,+,-}$, $\lambda_{L,ij}^{0\pm}$, $\lambda_{L,ij}^{\pm0}$, $\lambda_{E,\alpha \beta}^{0\pm}$, and $\lambda_{E,\alpha \beta}^{\pm 0}$ for given particle ensembles.

\subsection{Observables}\label{sec:observables}

In order for our reinforcement learning agent to learn to find promising models, we must have a reasonable definition of what constitutes ``promising'' in this context. Perhaps the most natural choice for a metric by which we can evaluate models is with a $\chi^2$ likelihood computation, using relevant physical observables. Our task then becomes identifying which observables to use in our fit. Broadly speaking, a model with vector-like leptons is constrained by both direct collider searches and low-energy precision observables. The likelihood, or rather log-likelihood, stemming from precision observables can be readily computed using conventional techniques, and treating the measurements as Gaussian (or half-Gaussian) distributions.\footnote{For this work we compute the likelihoods (or rather log-likelihoods) while ignoring correlations between observables. For the purpose of generating plausible models and testing our reinforcement learning methodology, this is sufficient.} Performing a robust computation of a model likelihood from aggregating collider searches is non-trivial, and therefore beyond the scope of this work (see, however, \cite{Waltenberger:2020ygp}). Fortunately, we can easily sidestep this issue for the purposes of this demonstration study by simply providing a lower bound on the vector-like lepton masses which lies well in excess of existing searches' bounds \cite{OsmanAcar:2021plv,ATLAS:2015qoy,ATLAS:2023sbu,Guedes:2021oqx,Kawamura:2023zuo,Bhattiprolu:2019vdu}.

The precision observables which we compute for our agent's suggested models are summarized in Table \ref{tab:observables}. Apart from the classic electroweak precision observables, we note that allowing models' vector-like lepton content to couple to multiple generations of SM leptons will lead to nontrivial lepton flavor violating processes. To estimate whether such effects would be disqualifying to our models, we incorporate limits on three lepton-violating decay processes, $\mu \rightarrow e \gamma$, $\tau \rightarrow \mu \gamma$, and $\tau \rightarrow e \gamma$, which are induced at the one-loop level from couplings between SM and vector-like leptons.\footnote{A reader may be concerned that $\mu \rightarrow 3 e$ constraints, such as those arising in \cite{Ishiwata:2013gma}, may dominate over $\mu \rightarrow e \gamma$, since the former arises at tree-level and the latter at the loop level. In practice, however, because we are specifically selecting for models which have a large BSM contribution to the muon anomalous magnetic dipole moment, many of the contributing diagrams of which have immediate analogues in the one-loop $\mu \rightarrow e \gamma$ process, while our favored parameter space also has small mixing between the vector-like and SM leptons relative to the parameter space considered in \cite{Ishiwata:2013gma}, we find that the $\mu \rightarrow e \gamma$ constraints will dominate (or at worst be comparable to) lepton flavor conservation constraints from $\mu \rightarrow 3e$ throughout our model space. As such, we use the one-loop $\mu \rightarrow e \gamma$ constraints, and analogous decays for other generations, as a reasonable approximation for constraints extracted when the $\mu \rightarrow 3 e$ and other 3-lepton LFV decays are included.} We further include the effect of $\mu-e$ conversion in gold nuclei, which can be induced by electroweak flavor-changing neutral currents at tree level.
% \footnote{Note that because the dark photon possesses only extremely feeble couplings to the quark sector stemming from kinetic mixing with the SM gauge bosons, while the dark Higgs is entirely decoupled to the quark sector absent mixing with the SM Higgs, the dark photon and dark Higgs do not contribute significantly to the $\mu-e$ conversion.}
Finally, as previously mentioned, models in the class we consider can produce significant corrections to the anomalous magnetic moment of SM leptons, so we include the BSM corrections to both the muon and electron anomalous magnetic moments. The former of these represents the most significant tension with the SM of all measurements that we consider, and therefore the anomaly which the reinforcement agent is most incentivized to fit.

\begin{table}[]
    \centering
    \begin{tabular}{| c | c | c | c |}
    \hline
    Observable & Definition & Value & Source \\
    \hline
    $M_W$ & $W$ boson mass & $80.377 \pm 0.012 \; \textrm{GeV}$ & \cite{Workman:2022ynf,LHCb:2021bjt,CDF:2022hxs,ATLAS:2017rzl,CDF:2012gpf,D0:2012kms,DELPHI:2008avl,OPAL:2005rdt,L3:2005fft,ALEPH:2006cdc,D0:2002fhu,CDF:2000gwd}\\
    \hline
    $BR(W \rightarrow e \nu)$ & \multirow{3}{*}{$W$ partial width} & $0.1071 \pm 0.0016$ & \multirow{3}{*}{\cite{Workman:2022ynf,CMS:2022mhs,OPAL:2007ytu,DELPHI:2003ftu,L3:2004lwm,ALEPH:2004dmh}}\\
    $BR(W \rightarrow \mu \nu)$ & & $0.1063 \pm 0.0015$ & \\
    $BR(W \rightarrow \tau \nu)$ & & $0.1138 \pm 0.0021$ & \\
    \hline
    $R_e$ & \multirow{3}{*}{$Z$ partial width ratio (hadrons to leptons)} & $20.804 \pm 0.050$ & \multirow{3}{*}{\cite{Workman:2022ynf,OPAL:2000ufp,DELPHI:2000wje,L3:2000vgx,ALEPH:1999smx}}\\
    $R_\mu$ & & $20.784 \pm 0.034$ & \\
    $R_\tau$ & & $20.764 \pm 0.045$ & \\
    \hline
    $A_e$ & \multirow{3}{*}{$Z$ pole electron asymmetry parameter} & $0.1515 \pm 0.0019$ & \cite{Workman:2022ynf,OPAL:2001brm,SLD:2000ujp,ALEPH:2001uca,DELPHI:1999yne,L3:1998oan,SLD:1996gjt,SLD:1994kvj}\\
    $A_\mu$ & & $0.142 \pm 0.015$ & \cite{Workman:2022ynf,SLD:2000ujp}\\
    $A_\tau$ & & $0.143 \pm 0.004$ & \cite{Workman:2022ynf,OPAL:2001brm,SLD:2000ujp,ALEPH:2001uca,DELPHI:1999yne,L3:1998oan}\\
    \hline
    $A_{FB}^{(0,e)}$ & \multirow{3}{*}{$Z$ pole forward-backward asymmetry} & $0.0145 \pm 0.0025$ & \multirow{3}{*}{\cite{Workman:2022ynf,OPAL:2000ufp,DELPHI:2000wje,L3:2000vgx,ALEPH:1999smx}}\\
    $A_{FB}^{(0,\mu)}$ & & $0.0169 \pm 0.0013$ & \\
    $A_{FB}^{0,\tau}$ & & $0.0188 \pm 0.0017$ & \\
    \hline
    $\Delta a_e$ & \multirow{2}{*}{lepton anomalous magnetic moment\footnote{We are using a somewhat outdated value of the observed muon $g-2$ anomaly, including only the experimental results up through the 2021 FNAL measurement \cite{Muong-2:2021ojo} rather than the updated results from 2023 \cite{Muong-2:2023cdq}. This is to highlight the utility of our approach even in the absence of an overwhelming (that is, $> 5 \sigma$) signal. In practice our results should be extremely similar if we use the latest results for $\Delta a_{\mu}$, with only a mild change in the log-likelihoods of models that our agent produces.}}& $(-8.8 \pm 3.6) \times 10^{-13}$ & \cite{2018Sci...360..191P}\\
    $\Delta a_\mu$ & & $(2.51 \pm 0.59) \times 10^{-9}$ & \cite{Muong-2:2006rrc,Muong-2:2021ojo,Aoyama:2020ynm,Aoyama:2012wk,Aoyama:2019ryr,Czarnecki:2002nt,Gnendiger:2013pva,Davier:2017zfy,Keshavarzi:2018mgv,Colangelo:2018mtw,Hoferichter:2019mqg,Davier:2019can,Keshavarzi:2019abf,Kurz:2014wya,Melnikov:2003xd,Masjuan:2017tvw,Colangelo:2017fiz,Hoferichter:2018kwz,Gerardin:2019vio,Bijnens:2019ghy,Colangelo:2019uex,Blum:2019ugy,Colangelo:2014qya}\\
    \hline
    $y_\mu$ & muon Yukawa coupling & $1.12 \pm 0.2$ & \multirow{2}{*}{\cite{Workman:2022ynf,ATLAS:2022vkf,CMS:2022dwd}}\\
    $y_\tau$ & $\tau$ Yukawa coupling & $0.94 \pm 0.07$ & \\
    \hline
    $BR(\mu \rightarrow e \gamma)$ & $\mu \rightarrow e \gamma$ branching fraction & $< 4.2 \times 10^{-13}$ & \cite{MEG:2016leq}\\
    $BR(\tau \rightarrow e \gamma)$ & $\tau \rightarrow e \gamma$ branching fraction & $< 3.3 \times 10^{-8}$ & \cite{BaBar:2009hkt}\\
    $BR(\tau \rightarrow \mu \gamma)$ & $\tau \rightarrow \mu \gamma$ branching fraction & $< 4.2 \times 10^{-8}$ & \cite{Belle:2021ysv}\\
    $\Gamma^{\textrm{conv}}_{\textrm{Au}} / \Gamma^{\textrm{capt}}_{\textrm{Au}}$  & $\mu-e$ conversion in gold nuclei & $< 7 \times 10^{-13}$ & \cite{SINDRUMII:2006dvw}\\
    \hline
    \end{tabular}
    \caption{The physical observables used to estimate the log-likelihood for models produced by our agent. Upper limits are quoted as 90\% CL bounds.}
    \label{tab:observables}
\end{table}

The observables that we have considered here will experience BSM contributions sensitive to parameters related to the masses and couplings of the vector-like leptons. Contributions to the electroweak precision observables will emerge from tree-level corrections to SM leptons' electroweak couplings, which will in turn be sensitive solely to the mass matrix of the complete lepton sector, rather than processes involving the sub-GeV dark Higgs or dark photon.\footnote{In theory, kinetic mixing between the dark photon and the SM hypercharge boson will yield small corrections to the electroweak precision observables, however in practice for the mass range we consider any parameter space compatible with dedicated dark photon searches will render these contributions insignificant \cite{Filippi:2020kii}.} Meanwhile, those observables which do feature significant dark photon and dark Higgs contributions, namely the anomalous lepton magnetic moments and the lepton flavor violating decays, will have their contributions appear in loops with much heavier $O(\text{TeV})$ vector-like leptons. As a result, the dark Higgs and dark photon masses do not enter here either-- they are effectively zero. Furthermore, because at the loop energies the longitudinal mode of the dark gauge boson dominates the couplings appearing in the loop and is set (by Goldstone boson equivalence) by the dark Higgs coupling, even the dark gauge coupling will not enter any of our observables. To summarize, then, the observables that we consider here are essentially independent of the details of the sub-GeV BSM field content. In fact, this characteristic is enormously helpful: Observations constraining the parameter space of the dark photon gauge coupling or sub-GeV dark sector masses, such as dedicated dark photon searches \cite{Filippi:2020kii} or dark matter relic abundance calculations \cite{Izaguirre:2015yja}, will in general be dependent on aspects of the model that we have not specified, such as the nature of the sub-GeV dark matter, its mass, and its charge under the dark $U(1)$. Given that the dark sector in this class of models possesses a rich parameter space sensitive to a number of parameters that do not affect the observables related to the vector-like lepton sector, for the purposes of our model building task we shall assume that identifying a viable parameter space for the sub-GeV dark sector parameters is an orthogonal problem to ours-- in practice it is well-known that viable regions of parameter space persist in these constructions \cite{Izaguirre:2015yja,Wojcik:2021xki,Rizzo:2021lob}. More concretely, for our numerical analysis we shall simply select plausible values for the dark Higgs mass $m_{h_D} \lesssim O(\textrm{GeV})$, the dark gauge coupling $g_D \sim O(1)$, and the dark Higgs vev $v_D \lesssim O(\textrm{GeV})$ and treat these as fixed parameters during our reinforcement learning scan.\footnote{Technically speaking, the electroweak precision observables will be somewhat sensitive to $O(1)$ rescalings of the dark Higgs vev $v_D$, since these will directly alter the charged lepton mass matrix. However, by setting $v_D$ to the upper end of the region we consider (that is, $v_D = 1 \; \textrm{GeV}$), we shall approximately maximize these effects and render our analysis comfortably conservative. The more dominant experimental constraints from lepton flavor violation and anomalous magnetic moments are insensitive to $v_D$ at leading order.}

Finally, we note that by virtue of using precision observables with known analytical expressions for our likelihood computations, the likelihood that we compute here will in principle be differentiable with respect to the underlying continuous model parameters, suggesting that one might identify optimal continuous parameters simply by gradient descent, as discussed in, \eg, \cite{Matchev:2024ash}. However, this characteristic is clearly \emph{not} generic -- other experimental data, such as likelihoods extracted from astrophysical or collider counting experiments, may not be as well-behaved. As our purpose in this work is not strictly to maximize the efficacy of the particular model class that we are studying, but rather to infer the characteristics and behavior of a reinforcement learning scan of our type in general, we do not leverage the differentiability of our likelihood function here.

\subsection{Graph Representation}\label{sec:vll-graph}

We conclude our discussion of the class of models that we are employing in our study by specifying the concrete graph representations that our reinforcement learning agent shall use to express these models. Following the procedure outlined in Section \ref{sec:graph}, we note that our models' symmetry group is simply the SM gauge group extended by a single dark $U(1)_D$ (with the usual 4-dimensional Poincare group as the spacetime symmetry). In theory, there are 8 representations into which the various BSM fields of the model fall: 3 different $U(1)_D$ charge states for vector-like $SU(2)_L$ doublets, 3 for vector-like $SU(2)_L$ singlets, as well as the dark scalar and dark photon fields. However, as discussed in Section \ref{sec:observables}, the BSM contributions to our observables are to excellent approximation solely sensitive to the masses and dark Higgs couplings of the vector-like leptons. This eliminates the need for us to produce specialized nodes for the dark photon and dark Higgs, allowing us to keep their internal parameters fixed and so leave them implicit in the graph. We can now take stock of the parameters that we shall need to specify in order to uniquely define a BSM action of the form in Eq.(\ref{eq:model-action}) with these 6 field representations.
\begin{itemize}
    \item For each vector-like electroweak doublet (singlet) fermion with a dark charge of $\pm 1$, we must specify its vector-like mass $M_k^{L(E), \pm}$, as well as a vector of the three Yukawa couplings $\lambda_{L(E)}^{\pm}$ which mix these fermions with the three generations of SM electroweak doublets (singlets) through coupling with the dark Higgs.
    \item For each vector-like electroweak doublet (singlet) fermion with a dark charge of $0$, we must specify the mass $M_k^{L(E), 0}$ and a vector of three Yukawa couplings $y_{L(E)}^0$ which mix these fermions with  the three generations of SM electroweak doublets (singlets) through coupling with the SM Higgs.
    \item The following interaction couplings between different vector-like leptons must be specified: $y_{LE, EL}^{0, \pm}$ couple vector-like electroweak doublets to electroweak singlets through the SM Higgs, and $\lambda^{0 \pm}_{L(E)}$ and $\lambda^{\pm 0}_{L(E)}$ couple vector-like doublets (singlets) of dark charge $\pm 1$ to those with dark charge of $0$.
    \item The Yukawa coupling terms between two vector-like fields must treat the two incoming fields differently, otherwise terms such as the $y_{EL}$ couplings would not be distinguished from the $y_{LE}$ couplings. Therefore, for each different type of interaction node we consider, we must have two different edge types.
\end{itemize}
To get a sense of the effect of different graph representation choices on the performance of our reinforcement learning agent, we develop two graph grammars, which we label A and B, which are summarized in Tables \ref{tab:rep-A} and \ref{tab:rep-B}, respectively. Both are heterogeneous graphs, featuring different node types for coupling nodes and particle nodes, and therefore don't require explicit edge features (these are instead described as different edge types). Particles are represented with four continuous parameters, which denote the mass and the three Yukawa couplings with the SM leptons, while dark charge is described with a single discrete feature that can take one of three values: 0, +1, or -1. Notably, in both graph grammars particles with different dark charges will use the \emph{same} continuous parameters in their graph representation, allowing for the same weights that are trained on particles of one dark charge to be applied to message-passing featuring a different dark charge. The two graph grammars differ in their handling of particles with different representations under the electroweak group. In graph representation A, the electroweak representation of a vector-like lepton is encoded as a discrete feature within the feature vector, while in graph representation B, electroweak doublets and electroweak singlets are represented as separate node types. As discussed in Section \ref{sec:graph}, it is precisely this type of discretionary choice which is the principal ambiguity in creating graph representations for a class of models. By sharing continuous parameters between electroweak doublet and singlet nodes, graph representation A therefore permits significant knowledge to be shared between the agent's treatment of the two types of vector-like leptons, while graph representation B learns entirely separate weights for each node type. We shall compare the results of training reinforcement learning agents with the two different graph representations in Section \ref{sec:results}.

\begin{table}[]
    \centering
    \begin{tabular}{| c | c | c | c | c |}
    \hline
    \multicolumn{3}{| c |}{Nodes} & \multicolumn{2}{| c |}{Edges}\\
    \hline
    Node Type & Field/Coupling & Feature Vector & Edge Type & Particle $F$ to Coupling $\{ g_1, g_2 \}$\\
    \hline
    \multirow{4}{*}{Particle} & $L_0$ & $\{ M^{L,0}, \vec{y}^0_L, 0, 0 \}$ & \multirow{4}{*}{$e_1$} & \multirow{4}{*}{$g_1 \overline{F} P_R \square + g_2 \overline{\square} P_R F$}\\
    & $L_\pm$ & $\{ M^{L, \pm}, \vec{\lambda}^\pm_L, \pm 1, 0 \}$ & &\\
    & $E_0$ & $\{ M^{E, 0}, \vec{y}^0_E, 0, 1 \}$ & &\\
    & $E_\pm$ & $\{ M^{E, \pm}, \vec{\lambda}^\pm_E, \pm 1, 1 \}$ & &\\
    \hline
    \multirow{4}{*}{Coupling} & $y^{0}_{LE}, \, y^{0}_{EL}$ & $\{ y^{0}_{LE}, y^{0}_{EL}, 0 \}$ & \multirow{4}{*}{$e_2$} & \multirow{4}{*}{$g_1 \overline{\square} P_R F + g_2 \overline{F} P_R \square$}\\
    & $y^{\pm}_{LE}, \, y^{\pm}_{EL}$ & $\{ y^{\pm}_{LE}, y^{\pm}_{EL}, 0 \}$ & &\\
    & $\lambda^{0 \pm}_L, \, \lambda^{\pm 0}_L$ & $\{ \lambda^{0 \pm}_L, \lambda^{\pm 0}_L, 1 \}$ & &\\
    & $\lambda^{0 \pm}_E, \, \lambda^{\pm 0}_E$ & $\{ \lambda^{0 \pm}_E, \lambda^{\pm 0}_E, 1 \}$ & &\\
    \hline
    \end{tabular}
    \caption{The graph grammar used to represent models of the class given in Eq.(\ref{eq:model-action}) using graph representation A. Symbols match the notation of Eq.(\ref{eq:model-action}), with $\vec{y}$ and $\vec{\lambda}$ labeling three-component vectors of the Yukawa couplings between vector-like fermions and SM fields. There are two heterogeneous graph node types-- particles and couplings. Discrete node features are used to identify the electroweak representation and dark $U(1)$ charge of particle nodes, and a discrete node feature is used to distinguish between dark Higgs and SM Higgs Yukawa couplings among the coupling nodes. Note that coupling nodes contain two edge types, which both connect particle nodes with coupling nodes, are identified in order to distinguish between the two ways in which a particle node may appear in the contraction of an interaction term, with the form of the coupling denoted by an edge depicted in the table.}\label{tab:rep-A}
\end{table}

\begin{table}[]
    \centering
    \bgroup
    \def\arraystretch{1.5}
    \begin{tabular}{| c | c | c |}
    \hline
    \multicolumn{3}{| c |}{Nodes}\\
    \hline
    Node Type & Field/Coupling & Feature Vector \\
    \hline
    \multirow{2}{*}{Doublet} & $L_0$ & $\{ M^{L,0}, \vec{y}^0_L, 0 \}$ \\
    & $L_\pm$ & $\{ M^{L, \pm}, \vec{\lambda}^\pm_L, \pm 1 \}$\\
    \hline
    \multirow{2}{*}{Singlet} & $E_0$ & $\{ M^{E, 0}, \vec{y}^0_E, 0 \}$\\
    & $E_\pm$ & $\{ M^{E, \pm}, \vec{\lambda}^\pm_E, \pm 1 \}$\\
    \hline
    \multirow{2}{*}{$h$ Yukawa} & $y^{0}_{LE}, \, y^{0}_{EL}$ & $\{ y^{0}_{LE}, y^{0}_{EL} \}$\\
    & $y^{\pm}_{LE}, \, y^{\pm}_{EL}$ & $\{ y^{\pm}_{LE}, y^{\pm}_{EL} \}$\\
    \hline
    $h_D$ Doublet Yukawa & $\lambda^{0 \pm}_L, \, \lambda^{\pm 0}_L$ & $\{ \lambda^{0 \pm}_L, \lambda^{\pm 0}_L \}$\\
    \hline
    $h_D$ Singlet Yukawa & $\lambda^{0 \pm}_E, \, \lambda^{\pm 0}_E$ & $\{ \lambda^{0 \pm}_E, \lambda^{\pm 0}_E \}$\\
    \hline
    \multicolumn{3}{| c |}{Edges}\\
    \hline
    Edge Type & Connects & $F$-$\{ g_1, g_2 \}$ Coupling\\
    \hline
    $e_D$ & Doublet-$h$ Yukawa & $y^{0, \pm}_{LE} \overline{F} P_R \square + y^{0, \pm}_{EL} \overline{\square} P_R F$\\
    \hline
    $e_S$ & Singlet-$h$ Yukawa & $y^{0, \pm}_{LE} \overline{\square} P_R F + y^{0, \pm}_{EL} \overline{F} P_R \square$\\
    \hline
    \multirow{2}{*}{$e_{DD}$} & \multirow{2}{*}{Doublet-$h_D$ Doublet Yukawa} & $\lambda_L^{0 \pm} \overline{F} P_R \square + \lambda_L^{\pm 0} \overline{\square} P_R F, \;\; Q_D = 0$\\
     & & $\lambda_L^{0 \pm} \overline{\square} P_R F + \lambda_L^{\pm 0} \overline{F} P_R \square, \;\; Q_D = \pm 1$\\
     \hline
     \multirow{2}{*}{$e_{SS}$} & \multirow{2}{*}{Singlet-$h_D$ Singlet Yukawa} & $\lambda_E^{0 \pm} \overline{F} P_R \square + \lambda_E^{\pm 0} \overline{\square} P_R F, \;\; Q_D = 0$\\
     & & $\lambda_E^{0 \pm} \overline{\square} P_R F + \lambda_E^{\pm 0} \overline{F} P_R \square, \;\; Q_D = \pm 1$\\
     \hline
    \end{tabular}
    \egroup
    \caption{As Table \ref{tab:rep-A}, except depicting the graph grammar for graph representation B. There are 5 different node types, denoting vector-like doublets, vector-like singlets, SM Higgs Yukawa interactions, and dark Higgs Yukawa interactions that couple either vector-like singlets or vector-like doublets. Edges follow the pattern outlined in the Table, with 4 edge types. Note that even though two edge types, $e_{DD}$ and $e_{SS}$, denote two different possible places for the fermion $F$ to be inserted in the contraction, the graph representation is unambiguous because the particles connected to the $h_D$ Yukawa nodes are distinguished by their dark charge $Q_D$.}\label{tab:rep-B}
\end{table}

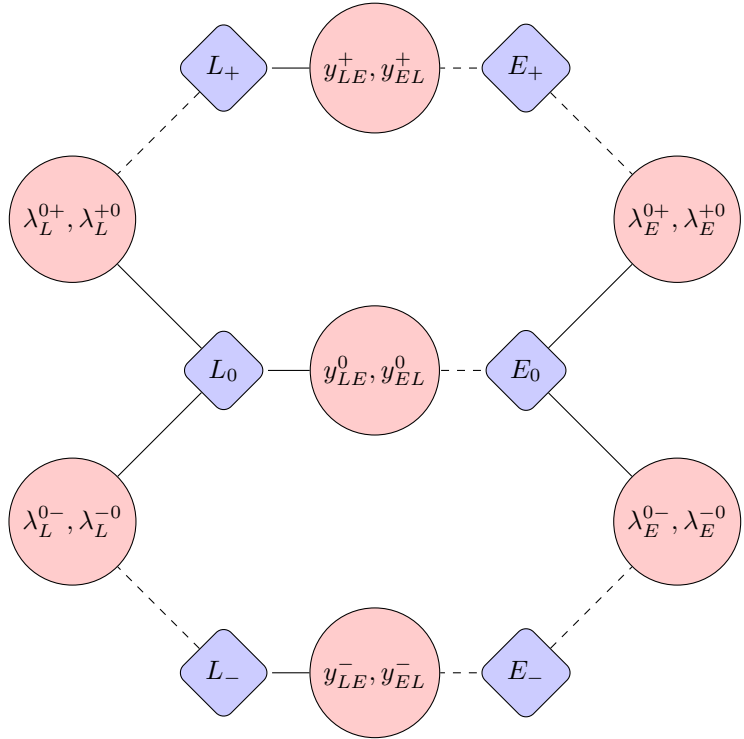
\begin{figure}
    \centering
    \begin{tikzpicture}
        \node at (0,4) [particle, fill=blue!20] (lp-id) {$L_+$};
        \node at (0,0) [particle, fill=blue!20] (l0-id) {$L_0$};
        \node at (0,-4) [particle, fill=blue!20] (lm-id) {$L_-$};
        \node at (4,4) [particle, fill=blue!20] (ep-id) {$E_+$};
        \node at (4,0) [particle, fill=blue!20] (e0-id) {$E_0$};
        \node at (4,-4) [particle, fill=blue!20] (em-id) {$E_-$};
        \node at (-2,2) [coupling, fill=red!20] (lamLP-id) {$\lambda^{0+}_L, \lambda^{+0}_L$};
        \node at (-2,-2) [coupling, fill=red!20] (lamLM-id) {$\lambda^{0-}_L, \lambda^{-0}_L$};
        \node at (2,4) [coupling, fill=red!20] (yLEP-id) {$y^+_{LE},y^+_{EL}$};
        \node at (2,0) [coupling, fill=red!20] (yLE0-id) {$y^0_{LE},y^0_{EL}$};
        \node at (2,-4) [coupling, fill=red!20] (yLEM-id) {$y^-_{LE},y^-_{EL}$};
        \node at (6,2) [coupling, fill=red!20] (lamEP-id) {$\lambda^{0+}_E, \lambda^{+0}_E$};
        \node at (6,-2) [coupling, fill=red!20] (lamEM-id) {$\lambda^{0-}_E, \lambda^{-0}_E$};
        \draw[-] (l0-id) -- (lamLP-id);
        \draw[dashed] (lp-id) -- (lamLP-id);
        \draw[-] (l0-id) -- (lamLM-id);
        \draw[dashed] (lm-id) -- (lamLM-id);
        \draw[-] (lp-id) -- (yLEP-id);
        \draw[dashed] (ep-id) -- (yLEP-id);
        \draw[-] (l0-id) -- (yLE0-id);
        \draw[dashed] (e0-id) -- (yLE0-id);
        \draw[-] (lm-id) -- (yLEM-id);
        \draw[dashed] (em-id) -- (yLEM-id);
        \draw[-] (e0-id) -- (lamEP-id);
        \draw[dashed] (ep-id) -- (lamEP-id);
        \draw[-] (e0-id) -- (lamEM-id);
        \draw[dashed] (em-id) -- (lamEM-id);
    \end{tikzpicture}
    \caption{A visual depiction of a graph in the class of theories given by the action in Eq.(\ref{eq:model-action}) with six total BSM fermions (one of each possible group representation that our model builder will consider), following the graph grammar A summarized in Table \ref{tab:rep-A}. Diamonds denote particle nodes, which contain the particle mass and Yukawa couplings to SM particles (through either the SM Higgs or dark Higgs) as features. Circles denote Yukawa couplings, with their coupling constants following the notation conventions of Eq.(\ref{eq:model-action}). Different edge types are denoted with different line styles: A solid line denotes an edge of type $e_1$, while a dashed line denotes an edge type of $e_2$.}
    \label{fig:vll-graph-A}
\end{figure}

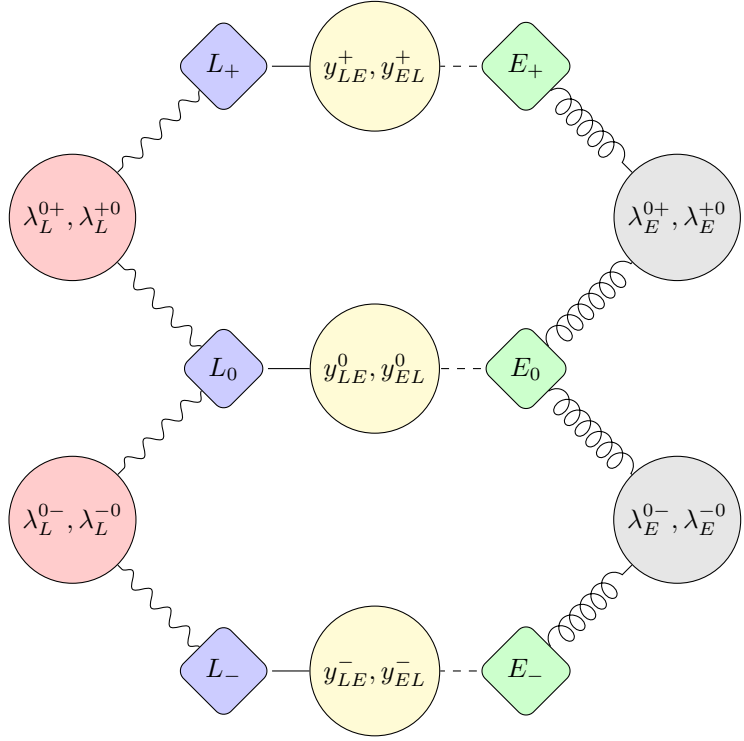
\begin{figure}
    \centering
    \begin{tikzpicture}
        \node at (0,4) [particle, fill=blue!20] (lp-id) {$L_+$};
        \node at (0,0) [particle, fill=blue!20] (l0-id) {$L_0$};
        \node at (0,-4) [particle, fill=blue!20] (lm-id) {$L_-$};
        \node at (4,4) [particle, fill=green!20] (ep-id) {$E_+$};
        \node at (4,0) [particle, fill=green!20] (e0-id) {$E_0$};
        \node at (4,-4) [particle, fill=green!20] (em-id) {$E_-$};
        \node at (-2,2) [coupling, fill=red!20] (lamLP-id) {$\lambda^{0+}_L, \lambda^{+0}_L$};
        \node at (-2,-2) [coupling, fill=red!20] (lamLM-id) {$\lambda^{0-}_L, \lambda^{-0}_L$};
        \node at (2,4) [coupling, fill=yellow!20] (yLEP-id) {$y^+_{LE},y^+_{EL}$};
        \node at (2,0) [coupling, fill=yellow!20] (yLE0-id) {$y^0_{LE},y^0_{EL}$};
        \node at (2,-4) [coupling, fill=yellow!20] (yLEM-id) {$y^-_{LE},y^-_{EL}$};
        \node at (6,2) [coupling, fill=gray!20] (lamEP-id) {$\lambda^{0+}_E, \lambda^{+0}_E$};
        \node at (6,-2) [coupling, fill=gray!20] (lamEM-id) {$\lambda^{0-}_E, \lambda^{-0}_E$};
        \draw[-, decorate,decoration=snake] (l0-id) -- (lamLP-id);
        \draw[-, decorate,decoration=snake] (lp-id) -- (lamLP-id);
        \draw[-, decorate,decoration=snake] (l0-id) -- (lamLM-id);
        \draw[-, decorate,decoration=snake] (lm-id) -- (lamLM-id);
        \draw[-] (lp-id) -- (yLEP-id);
        \draw[dashed] (ep-id) -- (yLEP-id);
        \draw[-] (l0-id) -- (yLE0-id);
        \draw[dashed] (e0-id) -- (yLE0-id);
        \draw[-] (lm-id) -- (yLEM-id);
        \draw[dashed] (em-id) -- (yLEM-id);
        \draw[-, decorate,decoration={coil, aspect=0.8, segment length=2mm, amplitude=1.5mm}] (e0-id) -- (lamEP-id);
        \draw[-, decorate,decoration={coil, aspect=0.8, segment length=2mm, amplitude=1.5mm}] (ep-id) -- (lamEP-id);
        \draw[-, decorate,decoration={coil, aspect=0.8, segment length=2mm, amplitude=1.5mm}] (e0-id) -- (lamEM-id);
        \draw[-, decorate,decoration={coil, aspect=0.8, segment length=2mm, amplitude=1.5mm}] (em-id) -- (lamEM-id);
    \end{tikzpicture}
    \caption{A visual depiction of a graph in the class of theories given by the action in Eq.(\ref{eq:model-action}) with six total BSM fermions (one of each possible group representation that our model builder will consider), following the graph grammar B summarized in Table \ref{tab:rep-B}. Diamonds denote particle nodes, which contain the particle mass and Yukawa couplings to SM particles (through either the SM Higgs or dark Higgs) as features. Circles denote Yukawa couplings, with their coupling constants following the notation conventions of Eq.(\ref{eq:model-action}). Different heterogeneous node types are denoted by different colors. For couplings: Yellow denotes an $h$ Yukawa, red denotes $h_D$ doublet Yukawa, gray denotes $h_D$ singlet Yukawa. For particles: Blue denotes an electroweak doublet vector-like lepton, while green denotes an electroweak singlet. Different edge types are denoted with different line styles: A solid line denotes an edge of type $e_D$, while a dashed line $e_S$, a wavy line $e_{DD}$, and a coiled line $e_{SS}$.}
    \label{fig:vll-graph-B}
\end{figure}

To highlight the usage of our graph grammars, in Figures \ref{fig:vll-graph-A} and \ref{fig:vll-graph-B}, we present example graphs following graph representations A and B, respectively, of a model with exactly one instance of each BSM vector-like fermion representation that can be included in the model-- that is, three electroweak doublets with dark charges $Q_D=0, +1, -1$ and three electroweak singlets with the same dark charges.

\section{Putting It Together: Exploring Portal Matter Model Space with Reinforcement Learning}\label{sec:experiment-setup}

With our model established and appropriate graph grammars constructed, we can now outline our strategy for leveraging reinforcement learning to explore this BSM model space. Broadly, we follow a similar procedure to \cite{Halverson:2019tkf, Nishimura:2020nre} in that each episode starts as a randomly initialized model graph, and at each time step the reinforcement learning agent makes modifications to the graph suggested by its policy. After $N_{\textrm{steps}}$ number of steps, the network is trained according to the PPO algorithm. An episode ends when it has either achieved a phenomenologically promising state (the metric for which we will define later) or a maximum number of steps $N_{\textrm{max}}$ has been taken. If an episode terminates due to arriving at a promising model, we record the model parameters for later analysis.
% To mitigate variance in the training efficacy, a total of $N_{\textrm{envs}}$ episodes will be run in parallel during training.

To implement this strategy, we require two components: First, we must define a metric by which we judge models as promising and devise reward functions that incentivize arriving at promising states, and second, we must construct our reinforcement learning environment, consisting of a state space $\mathcal{S}$, an action space $\mathcal{A}$, and a neural network which will parameterize our policy $\pi_{\theta}$.

\subsection{Evaluating Models and Designing Rewards}\label{sec:model-evaluation}

Our first task is to specifically define what constitutes a promising model in our framework, or in other words, specify what our agent is tasked to search for. Using such a metric, we can then devise reward functions for our agent to construct high-value states, and the form of the reward function will in turn, because all rewards must follow the Markov property, influence the construction of the state space for our reinforcement learning environment. As discussed in Section \ref{sec:observables}, a natural metric for evaluating models will be the difference in log-likelihood (or equivalently, the $\Delta \chi^2$) between our BSM models and the SM. In addition to log-likelihood as a metric, we wish to incentivize the agent to find simpler models. To that end, we penalize models of greater complexity in a similar manner to the Akaike information criterion (AIC): Specifically, we subtracting the number of BSM particles in the model (scaled by a constant) from the difference in log-likelihoods to create a score function $K$ of a state $s$. Mathematically, $K$ is defined as
\begin{align}\label{eq:model-eval-metric}
    K(s) \equiv \log \bigg( \frac{L(\textrm{Data} | s)}{L(\textrm{Data} | \textrm{SM}) } \bigg) - k n_{\textrm{particles}},
\end{align}
where $n_{\textrm{particles}}$ is the total number of BSM particles in a model and $k$ is a (positive) constant, while $L(\textrm{Data} | s, \textrm{SM} )$ is the likelihood of the data given the BSM theory $s$, or the SM.  This AIC-like criterion introduces an additional hyperparameter $k$ into our analysis, however we have found that for $k \lesssim 1$, the specific choice of $k$ has little impact on the results, so for our experiments we specialize to $k = 0.5$. With an evaluation metric $K(s)$ defined, we can also define the conditions under which we might identify a model as a terminal state. Naively, we note that a model which addresses the muon $g-2$ anomaly, the largest discrepancy from the SM prediction among our observables (while not accounting for the effect on any other observable), will give a log-likelihood difference with the SM of approximately $+9.05$,\footnote{As discussed in Section \ref{sec:observables}, we are using the older $\sim 4.25 \sigma$ significance measurement for the muon anomalous magnetic dipole moment; this difference is immaterial to our analysis.}. Meanwhile, we know from previous work \cite{Wojcik:2022woa,CarcamoHernandez:2019ydc} that this anomaly can be addressed in models with just two vector-like leptons. With these points in mind we estimate that a terminal state can be defined by
\begin{align}\label{eq:terminal-score}
    K(s_{\textrm{terminal}}) \geq 9 - 2 k, \;\; k = 0.5 \rightarrow K(s_{\textrm{terminal}}) \geq 8,
\end{align}
suggesting that we desire our agent to find models which explain the muon anomalous magnetic moment with at most two additional particles. Notably, by defining different criteria for terminal states, we would anticipate that the reinforcement learning agent will return different models. Furthermore, even though some prior knowledge of particle content which might address the anomaly has been incorporated in defining our terminal state, it is by no means a strong specification of the particle content that the model will suggest: If possible, the agent is incentivized to produce models with fewer than two particles that can describe the anomaly, and if a better fit can be achieved with a larger particle content, the agent will produce such a model as long as the improvement is greater than the penalty associated with adding complexity.

Our model evaluation metric provides us with a framework on which we can construct reward functions for our agent. Given that our environment is episodic, the simplest conceivable reward function we might devise would simply provide a positive reward for terminal model states, and no reward otherwise. However, such a minimalistic reward function leaves a great deal of information about intermediate states unused: For example, the agent would not experience any reward for producing a model which offers a significant improvement over the SM by our evaluation metric $K(s)$, but is not quite terminal. Instead, in addition to providing a flat positive reward for terminal states, it behooves us to devise a system of intermediate rewards, such as employed in \cite{Halverson:2019tkf,Dersy:2022bym,Nishimura:2020nre,Harvey:2021oue}. As in \cite{Dersy:2022bym}, which employed a reinforcement agent to simplify polylogarithm expressions, we find significant success with reward functions that provide a positive reward based on improving on the maximum score that the agent has achieved in the episode so far. To better explore the impact of choosing different reward functions for our agent, we shall consider three different reward functions in our work, all of which reward the agent for improving on its maximum score for a given episode. The first of these reward functions is
\begin{align}\label{eq:reward-I}
    R_{\textrm{I}}(K_t, K_{\textrm{max}}) = \begin{cases}
        0 & K_t \leq K_{\textrm{max}})\\
        \log (K_{\textrm{max}}/K(s_t) )/\textrm{max}(\log|K_t|, \log|K_{\textrm{max}}) & K_{\textrm{max}} < K_t < 0\\
        K_t + (K_t - \theta(K_{\textrm{max}})) & K_t > K_{\textrm{max}}, K_t > 0
    \end{cases},\\
    K_{\textrm{max}} \equiv \max_{\tau < t} K_\tau, \nonumber
\end{align}
where $K_t$ is the score $K$ of the model, as defined in Eq.(\ref{eq:model-eval-metric}) at timestep $t$. For positive $K$, this reward function is straightforward: For each time step where it attains a new maximum score, the function gives a base reward of the new score $K_t$, plus an additional reward equal to the difference between the current score and the maximum score attained so far during the episode (to avoid unreasonably huge numerical rewards, if the old maximum score had a lower $K$ than the SM, which by construction has $K=0$, the old maximum score is taken to be $0$). However, as the exceptionally strong constraints on lepton flavor violation will result in potentially enormous negative $K$ for models which don't respect lepton flavor conservation, using the same reward metric for negative $K$ values can result in unworkably large rewards (given that our neural network must include a critic which estimates the expected cumulative rewards for an episode, a variation of these rewards over 20 orders of magnitude will certainly degrade critic performance): Eliminating a single vector-like lepton from the model which has significant mixing with both muons and electrons, for example, might result in a log-likelihood difference of more than $10^{20}$! Instead, for negative $K$ values a proportional difference in the logs of these values are used, in order to reduce the rewards received while still in the negative environment to between 0 and 1, in addition to allowing for the reward function to significantly differentiate between improvements of radically different magnitudes in this regime.

Our second and third reward functions adopt a different philosophy for rewarding improvement. Specifically, rather than rewarding each improvement, regardless of the increment, these functions provide set rewards for passing certain ``milestone'' values-- exceeding certain $K$ values for the first time in the episode will result in rewards. Mathematically, we have
\begin{align}\label{eq:reward-II}
    R_{\textrm{IIa}} (K_t, K_{\textrm{max}}) = \begin{cases}
        0 & K_t \leq K_{\textrm{max}}\\
        \sum_{j} [j \theta(K_t - m_i) - j \theta(K_{\textrm{max}}-m_i)] & K_t > K_{\textrm{max}}
    \end{cases}, \nonumber\\
    R_{\textrm{IIb}} (K_t, K_{\textrm{max}}) = \begin{cases}
        0 & K_t \leq K_{\textrm{max}}\\
        \sum_{j} [j \theta(K_t - m'_i) - j \theta(K_{\textrm{max}}-m'_i)] & K_t > K_{\textrm{max}}
    \end{cases},\\
    \vec{m} \equiv [-10, -5, -2, 1, 2, 3, 4, 5, 6, 7, 8], \;\; \vec{m}' \equiv [-10^{15}, -10^{10}, -10^{5}, -100, \vec{m}]. \nonumber
\end{align}
In both reward functions $R_{IIa}$ and $R_{IIb}$, passing the $j^{\textrm{th}}$ milestone will give a positive reward of $j$ the first time in the episode it is accomplished, giving modest positive rewards for early milestones, and much larger rewards for achieving scores close to our cutoff value for a terminal state. Given that a reward function devised in this manner has a significant number of hyperparameters (namely, the specification of the milestones themselves), we have devised the two reward functions $R_{IIa}$ and $R_{IIb}$ in order to explore different specifications for them-- in $R_{IIa}$, rewards will only accrue once the model has reached a model for which tension with observables is at best modest, while $R_{IIb}$ includes milestones which reward improvements even to models with spectacularly poor scores.

In theory, reward functions I and II offer potentially different advantages and disadvantages. While reward function I provides positive feedback for even incremental improvements to the model (weighted by the degree of the improvement), differing paths to the same terminal state can yield radically different rewards. It is feasible that such a construction might lead to nontrivial ``reward hacking''-- for example, a model might learn to maximize its reward using incremental steps around models with high likelihood. Given that our episodes are of finite truncated length, we might expect such a behavior to negatively impact the number of terminal models that the agent produces. Meanwhile, the reward functions $R_{IIa}$ and $R_{IIb}$ provide a constant cumulative reward for all trajectories that lead to a terminal state, guaranteeing that the optimal reward is always to achieve a promising model within a single episode, but this comes at the price of requiring a significantly greater number of hyperparameters and diminished sensitivity to incremental model improvements. In our experiments, we shall aim to determine the level of importance that these particularities of reward function design have in our experiment, and by extension gain a sense of their importance in more sophisticated scans of BSM parameter spaces with reinforcement learning.

\subsection{Environment}\label{sec:environment}
With our reward functions established, the next step is the construction of the reinforcement learning environment. Following Section \ref{sec:reinforcement}, this means that we must specify a state space $\mathcal{S}$, an action space $\mathcal{A}$, and an agent with a policy $\pi_\theta$. We begin with $\mathcal{S}$, the state space. Obviously, each state must include the graph of a given BSM model, and as the number of BSM particles in turn plays a central role in our model evaulation metric Eq.~(\ref{eq:model-eval-metric}), we also directly pass the agent metadata about the number of particles of each possible vector-like lepton species in the new model -- even though this information is already contained in the model graph itself, we simplify the agent's learning task by including it explicitly. Beyond what is contained in the model itself, we can also supplement the state with additional information. To determine what additional information must be included in the state, we recall that as a Markov decision process, we must require that given an input state $s_t$, the expectation value of the sum of rewards accumulated over the rest of the episode will depend \emph{only} on information available in the state $s_t$ and the policy $\pi_{\theta}$ which governs the agent's actions. In addition to a flat positive reward for reaching a terminal state, we note that our reward function is supplemented by intermediate rewards which are given by one of the functional forms given in Eqs.(\ref{eq:reward-I}) and (\ref{eq:reward-II}). These intermediate rewards depend on both the model under consideration and the highest score as defined in Eq.(\ref{eq:model-eval-metric}) that has been achieved so far in the episode. To satisfy the Markov property, then, our state must include the maximum episode score $K_{\textrm{max}}$, in addition to the model graph itself. Finally, because our agent will always be working in a regime in which episode length is truncated at a finite number of steps, we note that estimation of the long-term accumulated rewards for the remainder of an episode will depend on the number of steps remaining in the episode. To satisfy the Markov property, then, we also supplement the state space with the number of steps that have already been taken in the episode.\footnote{It should be noted that this practice differs from the more familiar implementation of actor-critic architectures, in which a network trained using episodes of truncated length is then deployed to perform the same task without a step limit. In these cases, even when an episode is truncated the critic will estimate the long-term rewards of the state had it continued, and the step count need not be included in the state in order to satisfy the Markov property. In our case, however, because training with a truncated state is the only environment in which the agent is deployed, a more sensible approach is to include the step count within the state and treat truncated states as having no expectation of longer-term rewards.} We can then write an element of the state space as
\begin{align}
    s_t = \{ \mathcal{G}, \vec{n}_\textrm{particles}, K_{\textrm{max}}, n_{\textrm{steps}} \} \in \mathcal{S},
\end{align}
where $\mathcal{G}$ represents a BSM model graph, $\vec{n}_{\textrm{particles}}$ represents a 6-dimensional vector of integers representing the number of BSM particles of each possible species that appear in the model, $K_{\textrm{max}}$ represents the maximum score achieved in the episode so far, and $n_{\textrm{steps}}$ represents the number of steps that have already been taken in the episode. 

While what we have thus far presented constitutes a complete definition of a state space, in our practical construction we shall find it useful to impose several boundaries on the space of model parameters, both to mimic collider constraints that we have not included in our likelihood computations and because a neural network may struggle when confronted with a non-compact space of training data (or more accurately, with the possibility that inputs may be outside of a finite range). To simulate current collider limits on vector-like lepton production, we limit our model space by requiring all vector-like leptons to have a mass of $\geq 1.5 \; \textrm{TeV}$-- we note that this is extremely conservative and well in excess of current LHC constraints on even the most constrained possible BSM lepton species (in our case, predominantly $\mu$ or $e$-coupled leptons with  nonzero dark charge) \cite{OsmanAcar:2021plv,ATLAS:2015qoy,ATLAS:2023sbu,Guedes:2021oqx,Kawamura:2023zuo,Bhattiprolu:2019vdu}, and should therefore allow us to easily ignore collider constraints when computing our likelihoods. To avoid positing particles that are too heavy to be realistically observed at a collider experiment in the foreseeable future, we also place a (somewhat arbitrary) upper limit on vector-like lepton mass of $7 \; \textrm{TeV}$, a mass which may be in reach via pair production, for example, in future multi-TeV muon colliders. We also limit the various Yukawa coupling constants in the model. For simplicity, we assume that all Yukawa coupling constants are real, but may have either positive or negative sign. Furthermore, we generally restrict them to have a magnitude $\leq 10$ and $\geq 10^{-10}$, with the exception of those SM Higgs  Yukawa coupling constants which mix the SM leptons with zero dark-charge vector-like leptons, where to avoid a numerically troublesome regime in which SM leptons' mixing with the BSM states can become large (due to the SM Higgs vev being significantly larger than the mass scale of the SM leptons), we limit its range to $\leq 10^{-2}$ and $\geq 10^{-12}$. Finally, we note that although our graph architecture permits us, in theory, to consider models with arbitrarily large particle content, the complexity of the numerical likelihood calculation will increase polynomially with larger numbers of BSM particles. In the interest of accelerating our computations, we then place a modest limit on the number of additional BSM particles in the model, requiring that each proposed model have at most 6 vector-like leptons of a given electroweak representation (that is, no more than 6 doublets and no more than 6 singlets). In practice, as this is significantly in excess of the number of particles that terminal states will have, we do not expect this truncation to have any serious effect on our results, other than to prevent the agent from proposing computationally taxing large models early in its training, since the agent is rapidly incentivized toward smaller models by our reward functions. We summarize all of these parameter space constraints in Table \ref{tab:limits}, using the notation of our original action defining this model class given in Eq.(\ref{eq:model-action}).

\begin{table}[]
    \centering
    \renewcommand{\arraystretch}{1.2}
    \begin{tabular}{| c | c | c |}
    \hline
    Parameter & Lower Limit & Upper Limit\\
    \hline
    $M^{E, 0}, M^{E, \pm}, M^{L, 0}, M^{L, \pm}$ & $1.5 \; \textrm{TeV}$ & $7 \; \textrm{TeV}$\\
    \hline
    $|y^0_{L}, y^0_{E}|$ & $10^{-12}$ & $10^{-2}$\\
    \hline
    $|\lambda^{\pm}_E, \lambda^{\pm}_L |$ & $10^{-10}$ & $10$\\
    \hline
    $|y^0_{LE}, y^{\pm}_{LE}, y^{0}_{EL}, y^{\pm}_{EL}|$ & $10^{-10}$ & $10$\\
    \hline
    $|\lambda^{0 \pm}_{L}, \lambda^{\pm 0}_{L}, \lambda^{0 \pm}_{E}, \lambda^{\pm 0}_{E} |$ & $10^{-10}$ & $10$\\
    \hline
    $n_{\textrm{particles}}$ & 1 & N/A\\
    \hline
    $n_{\textrm{doublets}}$ & 0 & 6\\
    \hline
    $n_{\textrm{singlets}}$ & 0 & 6\\
    \hline
    \end{tabular}
    \renewcommand{\arraystretch}{1}

    \caption{The bounds on the model space for our reinforcement learning probe, using the notation in Eq.~(\ref{eq:model-action}). $n_{\textrm{particles}}$ denotes the total number of vector-like leptons in the model, while $n_{\textrm{doublets}}$ and $n_{\textrm{singlets}}$ denote the total number of vector-like electroweak doublets and singlets, respectively.}
    \label{tab:limits}
\end{table}

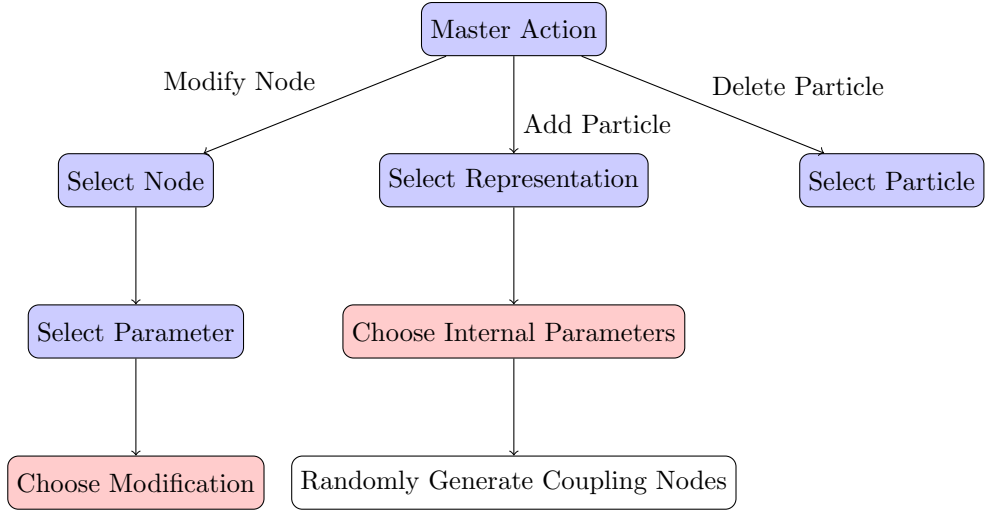
\begin{figure}
    \centering
    \begin{tikzpicture}
        \node at (5,4) [action, fill=blue!20] (m-id) {Master Action};
        \node at (0,2) [action, fill=blue!20] (mc-id) {Select Node};
        \node at (0,0) [action, fill=blue!20] (c-id) {Select Parameter};
        \node at (0,-2) [action, fill=red!20] (p-id) {Choose Modification};
        \node at (5,2) [action, fill=blue!20] (ad-id) {Select Representation};
        \node at (5,0) [action, fill=red!20] (ac-id) {Choose Internal Parameters};
        \node at (5,-2) [action] (y-id) {Randomly Generate Coupling Nodes};
        \node at (10,2) [action, fill=blue!20] (dc-id) {Select Particle};
        \draw[->] (m-id) -- (mc-id) node[midway,above left] {Modify Node};
        \draw[->] (mc-id) -- (c-id);
        \draw[->] (c-id) -- (p-id);
        \draw[->] (m-id) -- (ad-id) node[midway, below right] {Add Particle};
        \draw[->] (ad-id) -- (ac-id);
        \draw[->] (ac-id) -- (y-id);
        \draw[->] (m-id) -- (dc-id) node[midway,above right] {Delete Particle};
    \end{tikzpicture}
    \caption{The structure of the action space in our reinforcement learning environment, with blue boxes representing discrete actions and red particles representing continuously-parameterized actions. The single white box denotes random sampling with a uniform prior, and so not informed by a learned policy.}
    \label{fig:action}
\end{figure}

With our state space defined, we now turn to the action space $\mathcal{A}$ of the learning environment. In our context, the action space must describe changes that the agent can make to the BSM theory, which, for a well-trained agent, will produce a more phenomenologically promising theory than the previous state. As noted at the end of Section \ref{sec:reinforcement}, the action space for our problem is hierarchical, which in turn motivates our use of the H-PPO modification of the more conventional PPO reinforcement learning algorithm. The specific action space that our agent takes is highly complicated, and somewhat dependent on the graph grammar (representation A or B discussed in Section \ref{sec:vll-graph}). However, we can broadly summarize our action space by noting that an agent must select between one of three classes of actions, which each have associated subactions:
% The actual action space is extremely complex and hierarchical, involving elaborate trees of discrete and continuous decisions, and varies depending on our choice of graph representation, discussed in Section \ref{sec:vll-graph}. In the interest of clarity, we shall present a simplified description of the action space in the main text here, which summarizes how the agent can make decisions, and direct the reader to Appendix \ref{appendix:Action} for a more detailed account of our implementation of this space, including the treatment of hierarchically different numerical parameters and how the action space differs with differing graph representations. Our essential action space consists of three broad categories of actions:
%
\begin{enumerate}
    \item Delete a particle: This action has one associated discrete choice: The agent must specify which particle in the model to delete. Once the particle is deleted, all coupling nodes that are associated with it are automatically deleted as well.
    \item Add a particle: If this action is selected, the agent must then make discrete decisions about the electroweak representation (singlet or doublet) of the new particle, as well as its charge under the dark $U(1)_D$ group ($-1$, $0$, or $+1$). Then, the agent must select the four continuous features describing a particle node (its mass and its Yukawa couplings to the three charged SM leptons). The addition of a particle node also requires specifying the Yukawa couplings between the new particle and any existing BSM particles. Because the model likelihoods themselves aren't limited to extremely small-volume parameter spaces in these variable (in the way that they are for couplings to SM particles, due to lepton flavor violation constraints), we elect to sample $O(1)$ values for these coupling parameters randomly with a uniform prior 
    for the purposes of this study, for simplicity. We stress, however, that if necessary in other model-building contexts specifying coupling parameters between newly-created particles and existing ones is advisable, there is no {\it a priori } reason that this can't be accommodated using our procedure based on the H-PPO algorithm.
    \item Modify a continuous feature on an existing node: In this case, the agent must first make a discrete decision about which node to modify. Then, it must select the parameter of the node to modify. Finally, it must select a number (from a continuous interval), which will be added to the feature which it has chosen to modify.
\end{enumerate}
In Figure \ref{fig:action}, we summarize this action space in a diagram. It is clear that the space is highly hierarchical, in that different actions require radically different sets of sub-actions to uniquely specify a modification to be made to a model. Furthermore, in contrast to previous work in reinforcement learning to explore BSM parameter spaces, many of the discrete actions, such as selecting nodes for modification or deletion, have different numbers of options depending on the underlying model. For the price of this added complexity, however, we have described an action space which can be trivially generalized to any BSM theory that is expressible as a graph in the manner described in Section \ref{sec:graph}.

To complete our action space, we modify the framework we have outlined here in two modest ways: First, because the graph grammars we have employed to represent the model use heterogeneous graphs, with different node types, we allow the agent to treat the top-level modification actions for different node types as separate top-level actions-- that is, ``modify a Yukawa coupling'' is considered a different action from ``modify a particle''. This splits our top-level node modification actions into two different actions in graph grammar A (in which there are two node types, particles and couplings), and five different actions in graph grammar B (in which the node types are electroweak doublet particles, electroweak singlet particles, SM Higgs Yukawa couplings, dark Higgs Yukawa couplings among doublets, and dark Higgs Yukawa couplings among singlets). In the case of graph grammar B, because there are multiple particle types, we also split the action to add a particle into two different top-level actions: Adding an electroweak doublet and adding an electroweak singlet. Our second modification of the action space we have described here relates to the way in which the agent handles hierarchically different model parameters. The coupling constants in our model can vary immensely in magnitude: A calculation would suggest, for example, that in order to satisfy lepton flavor violation constraints a vector-like dark-charged lepton which has an $O(1)$ dark Higgs Yukawa coupling to the SM muon would require a Yukawa coupling to the electron of smaller than $O(10^{-6})$. To understand these hierarchies meaningfully, Yukawa couplings in the model (both those which mix the vector-like leptons among themselves and those which mix the vector-like leptons with the SM leptons) are passed to the agent in scientific notation, independently storing both an $O(1)$ magnitude variable and an exponent of 10 by which the value is multiplied, so that $3 \times 10^{-4}$ could be represented as $(3, -4)$, for example. Actions on these continuous parameters, then, can instead be either continuous modifications of the $O(1)$ parameter in this representation or a discrete modification of the exponential one. In practice, this means that when a coupling parameter is modified, the agent must choose whether to modify its continuous or exponential component, suggesting a continuous modification in the case of the former and a discrete modification in the case of the latter.

% The agent can delete a particle from the model, it can add a particle to the model, or it can modify a continuous parameter of the model using one of the nodes. In turn, each of these tasks feature sub-actions that the agent must specify. If a particle is to be deleted, the agent must select which particle should be removed from the model. If a model parameter is to be modified, the agent must select the graph node (either particle or coupling) to modify, then select the feature within that node to modify, and then specify a modification to make. Finally, if a new particle is to be added, the agent must specify its parameters-- its gauge group representation, its mass, and its Yukawa couplings to SM particles. 

Our final task is describing the implementation of the policy, $\pi_\theta$, which takes the form of a neural network that takes the environment state as an input, and outputs both a prediction for the long-term rewards that the agent can expect from the state over the course of the episode, and probability distributions from which we can sample actions. As described in greater detail in Appendix \ref{appendix:PPO}, proper implementation of the H-PPO algorithm  requires a neural network to output probability distributions for \emph{all} parameters that appear in the hierarchical action space, and these parameters are then trained using separate policy gradient loss functions, each of which is computed using only those steps for which a given parameter is applicable (so, the probability distribution of possible modifications to, \eg, BSM particle mass will not enter the PPO loss for a step in which a particle was deleted instead of modified). This means that our neural network must output a large number of probability distributions over both continuous and discrete action parameters. For each continuous action parameter, the agent must output both a mean and a variance, which together define a normal distribution from which a continuous action parameter will be sampled. For each discrete action parameter, the agent must output a log-probability associated with each choice, which defines a probability over the discrete options from which the action will be sampled. 

To devise a neural network which can capture our policy, we begin with a graph neural network which takes the graph representing the model as an input. To begin, we note that because our BSM models are represented as graphs, we must do much of our processing via message-passing layers, as described in Section \ref{sec:graph}. Due to the fact that no comparative analysis is done on different message-passing layers, we shall avoid detailed comment on the form of our message-passing layers here and refer the curious reader to Appendix \ref{appendix:experiment-details} for details.
% In the interest of simplicity, we limit ourselves to message-passing layers using the graph convolution proposed in \cite{2018arXiv181002244M}, which mimics the form of our simple linear example message-passing layer given in Eq.(\ref{eq:message-passing}) in Section \ref{sec:graph}, with no edge features. To accelerate training, we include a GraphNorm layer after each message-passing layer, with the GraphNorm operation defined in \cite{2020arXiv200903294C}, ReLU activation is then applied to the nodes after the GraphNorm layer to introduce nonlinearities. To accommodate our graphs' heterogeneous structure, we aggregate the results of convolutions with different node types by summing each output before applying the activation function.

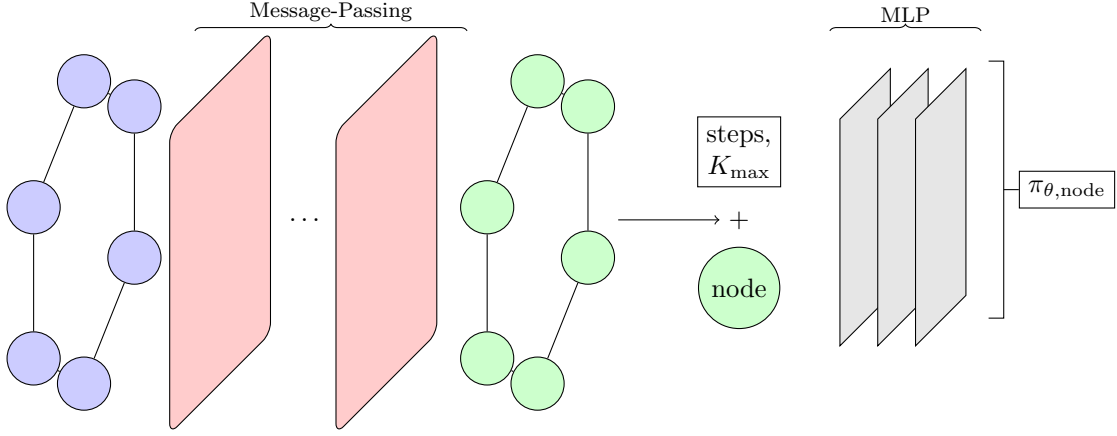
\begin{figure}
    \centering
    \begin{tikzpicture}
        \begin{scope}[canvas is zy plane at x=0, every node/.style={draw, circle, fill=blue!20, minimum height = 2em}]
          \node at (0, 0) (n1-id) {};
          \node at (1.73205, 1.) (n2-id) {};
          \node at (3.4641, 0) (n3-id) {};
          \node at (3.4641, -2.) (n4-id) {};
          \node at (1.73205, -3.) (n5-id) {};
          \node at (0, -2.) (n6-id) {};
          \draw[-] (n1-id) -- (n2-id);
          \draw[-] (n2-id) -- (n3-id);
          \draw[-] (n3-id) -- (n4-id);
          \draw[-] (n4-id) -- (n5-id);
          \draw[-] (n5-id) -- (n6-id);
          \draw[-] (n6-id) -- (n1-id);
        \end{scope}
        \begin{scope}[canvas is zy plane at x=1.8]
          \draw[fill=red!20, rounded corners] (0,1.) rectangle (3.4641, -3.);
        \end{scope}
        \node at (2.28,-1.5, 0) {\ldots};
        \draw[decorate, decoration = brace] (0.8, 1.) -- (4.4, 1.) node [midway, above] {\footnotesize{Message-Passing}};
        \begin{scope}[canvas is zy plane at x=4.]
          \draw[fill=red!20, rounded corners] (0,1.) rectangle (3.4641, -3.);
        \end{scope}
        \begin{scope}[canvas is zy plane at x=6., every node/.style={draw, circle, fill=green!20, minimum height = 2em}]
          \node at (0, 0) (n1-id) {};
          \node at (1.73205, 1.) (n2-id) {};
          \node at (3.4641, 0) (n3-id) {};
          \node at (3.4641, -2.) (n4-id) {};
          \node at (1.73205, -3.) (n5-id) {};
          \node at (0, -2.) (n6-id) {};
          \draw[-] (n1-id) -- (n2-id);
          \draw[-] (n2-id) -- (n3-id);
          \draw[-] (n3-id) -- (n4-id);
          \draw[-] (n4-id) -- (n5-id);
          \draw[-] (n5-id) -- (n6-id);
          \draw[-] (n6-id) -- (n1-id);
        \end{scope}
        \node at (8, -1.5, 0) (out-node) {+};
        \node at (8, -2.4, 0) [draw, circle, fill=green!20] {node};
        \node at (8, -0.6, 0) [draw, rectangle, align=center] {steps, \\ $K_{\textrm{max}}$};
        \draw[->] (6.4, -1.5, 0) -- (out-node);
        \begin{scope}[canvas is zy plane at x = 10.]
          \draw[fill=gray!20] (0,0.5) rectangle (1.73205, -2.5);
        \end{scope}
        \begin{scope}[canvas is zy plane at x = 10.5]
          \draw[fill=gray!20] (0,0.5) rectangle (1.73205, -2.5);
        \end{scope}
        \begin{scope}[canvas is zy plane at x = 11.]
          \draw[fill=gray!20] (0,0.5) rectangle (1.73205, -2.5);
        \end{scope}
        \draw[decorate, decoration=brace] (9.2,  1., 0.) -- (11.2, 1., 0.) node [midway, above] {\footnotesize{MLP}};
        \draw[-] (11.3, 0.6, 0) -- (11.5, 0.6, 0) -- (11.5, -2.8, 0.) -- (11.3, -2.8, 0.);
        \draw[-] (11.5, -1.1, 0) -- (11.7, -1.1, 0) node[right, draw, rectangle] {$\pi_{\theta, \textrm{node}}$};
    \end{tikzpicture}
    \caption{A diagram depicting how the graph neural network is used to output probability distributions for action parameters for node-level tasks, such as parameter modification or selecting particles to delete or modify. An input state's model graph is transformed by message-passing layers, and then each transformed node's feature vector is supplemented with the supplementary state information and used as the input to a multilayer perceptron. This produces outputs for each node which are used to generate the policy probabilities for node-level actions.}
    \label{fig:node-action}
\end{figure}

The message-passing layers produce a new, transformed graph from which we will produce probability distributions for our actions. In order to give outputs corresponding to all policy parameters, we require some outputs which differ for each graph node (namely, parameters controlling selection of individual nodes for modification or deletion, and the parameters controlling modification actions made to each node), and some which must be given only once for each graph (the selection of whether to add a particle, modify a particle, or delete a particle, as well as the parameters of any particle added to the model). In the language of graph neural networks, we can say that these two categories of action parameters are node-level and graph-level tasks, respectively. Our strategies for these different tasks must differ accordingly. Fortunately graph neural networks are well-equipped to address learning tasks at both levels. In the case of node-level outputs, after transforming the model graph using stacked message-passing layers, the transformed node feature vectors are then supplemented with the state information that is not contained in the graph itself (the highest score and the number of steps taken in the episode), and each is used as an input to a 3-layer multilayer perceptron (MLP) with a rectified linear unit (ReLU) activation function,\footnote{Because 3-level MLP's of arbitrary width are universal function approximators \cite{LESHNO1993861}, this depth allows the final processing of a transformed node's outputs to capture nonlinear relationships between the state's supplementary information and the transformed node feature vectors.} the outputs of which will be used to generate the probability distributions for discrete and continuous actions. This procedure is used to generate log-probabilities for selecting individual nodes for removal, log-probabilities for selecting nodes for modification, and means and variances for proposed parameter modifications to each node. For clarity, we depict the basic strategy for generating node-level probabilities diagramatically in Figure \ref{fig:node-action}.

\begin{figure}
    \centering
    \begin{tikzpicture}
        \begin{scope}[canvas is zy plane at x=0, every node/.style={draw, circle, fill=blue!20, minimum height = 2em}]
          \node at (0, 0) (n1-id) {};
          \node at (1.73205, 1.) (n2-id) {};
          \node at (3.4641, 0) (n3-id) {};
          \node at (3.4641, -2.) (n4-id) {};
          \node at (1.73205, -3.) (n5-id) {};
          \node at (0, -2.) (n6-id) {};
          \draw[-] (n1-id) -- (n2-id);
          \draw[-] (n2-id) -- (n3-id);
          \draw[-] (n3-id) -- (n4-id);
          \draw[-] (n4-id) -- (n5-id);
          \draw[-] (n5-id) -- (n6-id);
          \draw[-] (n6-id) -- (n1-id);
        \end{scope}
        \begin{scope}[canvas is zy plane at x=1.8]
          \draw[fill=red!20, rounded corners] (0,1.) rectangle (3.4641, -3.);
        \end{scope}
        \node at (2.28,-1.5, 0) {\ldots};
        \draw[decorate, decoration = brace] (0.8, 1.) -- (4.4, 1.) node [midway, above] {\footnotesize{Message-Passing}};
        \begin{scope}[canvas is zy plane at x=4.]
          \draw[fill=red!20, rounded corners] (0,1.) rectangle (3.4641, -3.);
        \end{scope}
        \begin{scope}[canvas is zy plane at x=6., every node/.style={draw, circle, fill=green!20, minimum height = 2em}]
          \node at (0, 0) (n1-id) {};
          \node at (1.73205, 1.) (n2-id) {};
          \node at (3.4641, 0) (n3-id) {};
          \node at (3.4641, -2.) (n4-id) {};
          \node at (1.73205, -3.) (n5-id) {};
          \node at (0, -2.) (n6-id) {};
          \draw[-] (n1-id) -- (n2-id);
          \draw[-] (n2-id) -- (n3-id);
          \draw[-] (n3-id) -- (n4-id);
          \draw[-] (n4-id) -- (n5-id);
          \draw[-] (n5-id) -- (n6-id);
          \draw[-] (n6-id) -- (n1-id);
        \end{scope}
        \draw[->] (6.4, -1.5, 0) -- (8.2, -1.5, 0) node [midway, above] {\footnotesize{Pooling}};
        \node at (8.65, 0.5, 0) [draw, rectangle, align=center] {steps, \\ $K_{\textrm{max}}$};
        \node at (8.65, -0.25, 0) {+};
        \draw[fill=green!20] (8.4, -0.5, 0) rectangle (8.9, -4, 0);
        \draw[-] (8.4, -0.9375, 0) -- (8.9, -0.9375, 0);
        \draw[-] (8.4, -1.375, 0) -- (8.9, -1.375, 0);
        \draw[-] (8.4, -1.8125, 0) -- (8.9, -1.8125, 0);
        \draw[-] (8.4, -2.25, 0) -- (8.9, -2.25, 0);
        \draw[-] (8.4, -2.6875, 0) -- (8.9, -2.6875, 0);
        \draw[-] (8.4, -3.125, 0) -- (8.9, -3.125, 0);
        \draw[-] (8.4, -3.5625, 0) -- (8.9, -3.5625, 0);
        \begin{scope}[canvas is zy plane at x = 10.3]
          \draw[fill=gray!20] (0,0.5) rectangle (1.73205, -2.5);
        \end{scope}
        \begin{scope}[canvas is zy plane at x = 10.8]
          \draw[fill=gray!20] (0,0.5) rectangle (1.73205, -2.5);
        \end{scope}
        \begin{scope}[canvas is zy plane at x = 11.3]
          \draw[fill=gray!20] (0,0.5) rectangle (1.73205, -2.5);
        \end{scope}
        \draw[decorate, decoration=brace] (9.5,  1., 0.) -- (11.5, 1., 0.) node [midway, above] {\footnotesize{MLP}};
        \draw[-] (11.8, 1., 0) -- (12., 1., 0) -- (12., -4.1, 0) -- (11.8, -4.1, 0);
        \draw[-] (12., -1.55,0) -- (12.2, -1.55, 0) node[right, draw, rectangle] {$\pi_{\theta, \textrm{graph}}$};
    \end{tikzpicture}
    \caption{A diagram depicting how the graph neural network is used to output probability distributions for action parameters graph-level tasks, such as selecting the top-level action or determining the feature vector for a particle to add to the model. An input state's model graph is transformed by message-passing layers, and then the transformed nodes' feature vectors are pooled by summing each node of the same heterogeneous graph type (see Section \ref{sec:vll-graph} for details on our graph grammars), and concatenating the results for each node type into a single vector. The output is supplemented with additional state information and used as the input to a multilayer perceptron. This produces outputs for each node which are used to generate the policy probabilities for graph-level actions. This strategy is also used for producing the critic output, a necessary component of the PPO algorithm.}
    \label{fig:graph-action}
\end{figure}
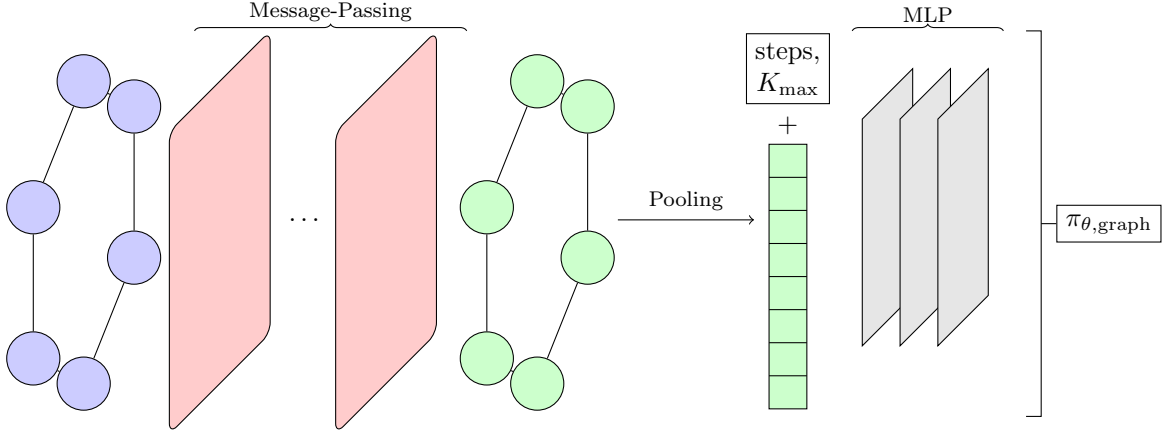

For graph-level tasks, including all graph-level action parameters as well as the critic output (which, as discussed in Section \ref{sec:reinforcement}, estimates the long-term rewards that the agent will expect over the course of the episode), we must adopt a slightly different procedure. In this case, after the nodes are transformed via a stack of message-passing layers, we aggregate their values through a pooling operation, in which all nodes are aggregated into a single finite-dimensional vector. Several different pooling operations are common in graph neural networks, and for our purposes we simply take the sum of all node features. Because the graph grammars presented in Section \ref{sec:vll-graph} both use heterogeneous graph architecture, we must also devise a strategy to aggregate the pooled vectors from different node types-- in this case we simply concatenate the pooled vectors for each different node type to produce one long pooled vector, which is finally supplemented with the additional state information not contained in the graph. To generate outputs for graph-level action parameters, the final vector is used as the input for a 3-layer multi-layer perceptron, the output of which will give the critic's evaluation of the state and the probability distributions for graph-level action parameters. We have schematically depicted this strategy in Figure \ref{fig:graph-action}.

To finalize the specification of our agent, we note that some discrete actions, such as selecting the action to remove a particle when only one total particle remains in the model, are clearly ``illegal'', meaning that they can't be implemented without leaving the model space, or at least leaving the area defined by the parameter limits given in Table \ref{tab:limits}. To marginally improve the performance of the model, after generating the action probabilities via the neural network the agent will automatically modify its outputs to assign probabilities of 0 for discrete actions that are illegal given the current model state. Because illegal continuously parameterized actions, such as decreasing a particle mass parameter that is already at its minimum value, are not so trivially removed from the action parameter distributions, we shall disincentivize these actions by assigning them a small negative reward in our experiments, supplementing the large positive reward that the agent receives for achieving a terminal state and the intermediate rewards discussed in Section \ref{sec:model-evaluation}.

\section{Experiments and Results}\label{sec:experiments}

To test the efficacy of our reinforcement learning scan, we perform a number of trials with the reinforcement learning environment and agent outlined in the previous Section. For the environment, we implement a customized learning environment built with the Python library Gymnasium \cite{towers_gymnasium_2023}, and implement our graph neural network using the library Pytorch Geometric \cite{2019arXiv190302428F}. Our agent's neural network follows a modular structure consisting of units of the forms discussed in \ref{sec:environment}. In the interest of brevity we defer a detailed summary of its construction to Appendix \ref{appendix:experiment-details}, however here we summarize that the graph neural network takes an initial graph representing a BSM model and generates a shared latent representation through four message-passing layers, after which this shared graph representation is used as input for the modules for the various node-level and graph-level parameters in the action space. When computing model likelihoods, we shall use the computational techniques discussed in detail in Appendix \ref{appendix:Observables}, as well as the numerical parameters identified in Table \ref{tab:inputs} in that Appendix.

To evaluate our agents, we randomly sample 32 parallel environments from the parameter space described in Table \ref{tab:limits} with a uniform prior over all variables. The agent then evolves the model according to its policy for 50 time steps before being trained according to the H-PPO algorithm, after which point it will continue. Episodes that have not attained a terminal state will end after 250 total steps. We allow each agent to train simultaneously over the 32 parallel environments, and continue until 1000 rounds of 50 steps each have been completed, so that in total, the agent samples 1.6 million models. We record all terminal states that the agent produces over the course of its training. To mitigate the significant variance in performance due to initial trajectories (which in turn will heavily depend on the initial training model states, the agent's initial weight parameters, and even the random sampling from the agent's policy probabilities early in training), we also perform each experiment 10 independent times with 10 independent initializations.

In our experiments we compare the agent's performance with differing choices across six factors, from which we aim to get a qualitative picture of the sensitivity of the agent to a variety of training factors:\footnote{In addition to these settings, there are a number of other training hyperparameters which we hold constant, either because they are minimally influential or typical settings in comparable tasks result in good performance. We summarize these additional parameters in Appendix \ref{appendix:experiment-details}.}

\begin{itemize}
    \item \textbf{Graph Representation:} We consider the two different graph representations of our vector-like lepton models, discussed in Section \ref{sec:vll-graph} and described as representations A (in which different electroweak lepton representations are distinguished only by a single discrete parameter) and B (in which isospin singlet and doublet vector-like leptons are assigned different node types in a heterogeneous graph).
    \item \textbf{Skip Layers:} A potential concern in multi-layer graph neural networks is the loss of any difference between graph nodes after several layers of message-passing. To determine if this effect is a significant factor within our framework, we compare the performance of agents with ``skip layers'' implemented to those without. When implementing these skip layers, we supplement our existing neural network architecture by appending each node's original feature vectors to their transformed feature vectors at certain layers. When skip layers are implemented in our model, we perform these skips at two points in the neural network: First, immediately after the neural network's shared message-passing layers (those which generate the graph representation shared as input to all the agent's modules) complete their transformations, and second, onto the final node feature vectors which are used as inputs for a multilayer perceptron in node-level tasks, such as selecting particles for deletion or modifying particle parameters.
    \item \textbf{Training Epochs:} As discussed in Appendix \ref{appendix:PPO}, this hyperparameter controls the number of training epochs performed at each step of the H-PPO algorithm. In theory, a larger number of training epochs will allow the agent to extract a greater amount of information out of each set of sampled trajectories that it collects, at the cost of potentially biasing the agent toward overtraining on its latest training sample, which may suffer from statistical errors in the algorithm's estimation of expected reward.
    \item \textbf{Intermediate Rewards:} The agent always receives a reward of $+100$ for finding a terminal state and a reward of $-1$ for taking a forbidden action (that is, an action which, due to boundaries on the model space, cannot be executed, such as reducing a parameter when it is already at its minimum value or removing a particle when only one BSM particle remains in the model). However, the structure of intermediate rewards varies. We consider intermediate rewards of each of the forms given in Eqs.(\ref{eq:reward-I}) and (\ref{eq:reward-II}).
    \item \textbf{Entropy Regularization:} The coefficient $\beta$, discussed in Section \ref{sec:reinforcement} and Appendix \ref{appendix:PPO}, is a training hyperparameter that controls the degree to which the agent is incentivized to explore new parameter spaces or exploit regions that it has already found to be fruitful. A higher value of $\beta$ will provide a larger reward for agent outputs with higher entropy, and therefore higher $\beta$ incentivizes greater exploration over exploitation. We shall scan over the values $\beta=0.001, 0.01, 0.05, 0.1, 0.2, 0.5$.
    \item \textbf{Reward Normalization:} Because rewards can attain large magnitudes, in other tasks it has proven beneficial to adjust rewards to limit the magnitude of the rewards that the agent receives \cite{2016arXiv160207714V}. We shall compare the results of running a trials without a reward normalization scheme to trials in which we perform a normalization on both the immediate rewards that the agent receives, as well as the critic's estimates of the cumulative total reward that the agent expects over the course of a single episode.\footnote{More specifically, when normalizing the rewards, we shall implement normalization at two points in training: First, leveraging gymnasium's built-in reward normalization capabilities, we shall normalize the exponential moving average of the agent's rewards in order to have a fixed variance, following the default parameters of gymnasium's NormalizeRewards wrapper. Second, we normalize the \emph{advantage} function (defined in Eq.(\ref{eq:advantage}) at each training step to have a mean of 0 and variance of 1. Both of these changes will ultimately work to reduce the magnitude of the state values that the critic network must approximate. Because we observed little independent effect of either normalization scheme, we simply implement both normalization schemes at once or implement neither, depending on our experiment parameters.}
\end{itemize}

While a scan over these options is hardly a comprehensive probe of the space of all discretionary choices that can be made over the course of constructing a learning environment, we believe that these choices allow us to probe a number of potential factors in our experiments. We see that the first three choices we have listed above focus on the potential effects of altering architecture of our neural network and how it learns: Our graph representation B is significantly more complex than A, and allows us to probe the utility of grouping similar parameters as the same features when representing a BSM model graph. Implementing skip layers, meanwhile, can inform us about the effect of possible information loss over the course of our message-passing layers. Finally, the number of training epochs will give us a sense of the benefit (or detriment) of extracting more information from each training example in our rather exotic action space. The latter three choices in our above list do not affect the neural network directly, but instead alter the loss function that the agent attempts to optimize. Altering our intermediate reward structure and entropy regularization constant will give us a picture of the sensitivity of the agent's performance to differing reward structures. Meanwhile, examining the effects of normalizing our environment's rewards will allow us to gauge whether our task can benefit from calibrating reward magnitudes to quantities more numerically palatable to a neural network.

In short, by probing potentially influential learning hyperparameters, we aim to get a reasonable picture of the sensitivity of our agent to these types of choices, at least in the simple model building task which we pursue here. By observing these sensitivities, we can in turn comment on the advisability of certain choices in more general, and potentially more complex, model building tasks to which a version of this procedure might be applied-- including which of these different factors might be significant and which can be safely ignored without seriously degrading an agent's performance.

\subsection{Results}\label{sec:results}

We now present the results of our various reinforcement learning scans. Broadly, our technique proves effective in generating promising models: Depending on hyperparameter selections, we find that agents may achieve as few as several terminal states or as many as several thousand over the course of their training. We divide our detailed analysis of our results into three sections. First, as a simple elucidation of our techniques' utility, we discuss the space of viable models that the agents collectively identify, and discuss what physical insights we can glean from the results of our scans, as also highlighted in our short companion paper\cite{Wojcik:2024azu}. Then, to further explore the efficacy of our frameworks, we shall discuss the performance of two of our hyperparameter sets that have proven highly successful at generating a large number or diversity of viable models, and explore what further understanding of the parameter space can be extracted from the data of these scans. Finally, we shall explore the relative importance to the agent's performance of the various training hyperparameters that we have varied. 

\subsubsection{Models Identified by the Agents}

A necessary precondition to establishing the utility of our framework is demonstrating that some new knowledge of physics can be extracted from the results of our reinforcement learning scans. Before inspecting individual scans' performance, however, we can gain significant insight simply by identifying the particle content of different terminal states that are found across all of our various scans. As discussed also in \cite{Wojcik:2024azu}, our reinforcement learning agents have identified six distinct sets of new physics particle content, each with just two BSM particles, which achieve a high enough evaluation score by the standard given in Eq.(\ref{eq:terminal-score}) to be deemed terminal states. We summarize the particle contents of the six models in Table \ref{tab:distinct_chiral}, using the same notation for the vector-like leptons as given in Eq.(\ref{eq:model-action}), and assigning each distinct model a corresponding letter identifier from `a' to `f'. 

\begin{table}[]
\centering
\begin{tabular}{|c|c|c|c|c|c|c|}
\hline
      & \multicolumn{6}{c|}{Number of particles}       \\ \hline
Label & $L_0$ & $L_+$ & $L_-$ & $E_0$ & $E_+$ & $E_-$ \\ \hline
a     & 0     & 1     & 0     & 0     & 1     & 0     \\ \hline
b     & 0     & 0     & 1     & 0     & 0     & 1     \\ \hline
c     & 1     & 1     & 0     & 0     & 0     & 0     \\ \hline
d     & 1     & 0     & 1     & 0     & 0     & 0     \\ \hline
e     & 0     & 0     & 0     & 1     & 1     & 0     \\ \hline
f     & 0     & 0     & 0     & 1     & 0     & 1     \\ \hline
\end{tabular}
\caption{The BSM particle contents for distinct chiral models, following the notation of Eq.(\ref{eq:model-action}) to denote different vector-like lepton representations under the SM and dark gauge groups.}\label{tab:distinct_chiral}
\end{table}

As discussed in \cite{Wojcik:2024azu}, of the models listed in Table \ref{tab:distinct_chiral}, we see that models `a' and `b' arise from precisely the chiral enhancement mechanism previously identified in \cite{Wojcik:2022woa}: A model with two vector-like leptons sharing the same (non-zero) dark charge, one being an electroweak singlet and the other being an electroweak doublet, is capable of providing chirally-enhanced one-loop dark photon and dark Higgs contributions to the muon anomalous magnetic moment from the SM Higgs coupling between the two vector-like lepton representations. The sole discrepancy between `a' and `b' is whether the BSM fermions have dark charges of $+1$ or $-1$, the only two nontrivial dark charge values that we consider in this analysis. Models `c'-`f', however, identify a new chiral enhancement mechanism. In these models, the vector-like leptons are invariably of a single electroweak representation (either singlet or doublet), and instead differ by one unit of dark charge (specifically, one lepton has $Q_D = 0$, while the other has $Q_D = \pm 1$. Inspired by the appearance of these classes of models in our agents' scans, we can construct a toy model of classes `e' and `f' to develop an analytical intuition for the physics. Following the notation of Eq.(\ref{eq:model-action}), the correction to the SM prediction for the muon magnetic moment should be given by
\begin{align}
    a_{\mu}^{e,f} \approx - \frac{m_\mu^2}{16 \sqrt{2} \pi^2} \frac{M^{E,\pm}}{m_\mu} \frac{\lambda_{E, \mu}^{\pm} \lambda_E^{\pm 0} }{(M^{E, \pm})^2} \frac{y_{E, \mu}^0 v}{M^{E, 0}}.
\end{align}
Intuitively, we can understand that this chiral enhancement of the one-loop magnetic moment contribution occurs because, in models `e' and `f', left-handed SM leptons mix primarily with the $Q_D = 0$ vector-like lepton, through couplings with the SM Higgs, while right-handed SM leptons mix primarily with the $Q_D = \pm 1$ vector-like lepton. A chirality-flipping operator that generates a magnetic moment then appears at leading order from the mixing between the two vector-like lepton species, from the dark Higgs Yukawa coupling $\lambda^{\pm 0}_E$. The generalization to the same construction with electroweak doublet vector-like leptons (that is, models `c' and `d') function in perfect analogy to models `e' and `f'.

The differences between these chiral enhancement mechanisms have significant implications to the model builder. Different electroweak representations of vector-like leptons will have differing production cross-sections, with vector-like electroweak doublets somewhat more tightly constrained by collider searches \cite{Rizzo:2022lpm,OsmanAcar:2021plv}. Furthermore, a variety of sources have noted that the decay channels of vector-like leptons with $Q_D = \pm 1$ will be overwhelmingly dominated by channels featuring a dark photon emission, rather than the emission of an SM gauge boson, and as a result are subject to dramatically different (and generally stronger) experimental constraints than those states with $Q_D = 0$. In addition, these models suggest differing UV completions: Models `a' and `b', for example, might suggest an extension to a left-right symmetric model of sorts, where the BSM particle content possesses states in both the electroweak singlet and doublet representations. Meanwhile, models `c'-`f' suggest that the structure of vector-like leptons might be limited to only one of the two electroweak representations that we observe among the SM leptons, reminiscent of, for example, an $E_6$ grand unified theory.

In this (admittedly simple) framework, our reinforcement learning scan has succeeded both in recreating known models (namely those of \cite{Wojcik:2022woa}) and identifying models previously not enumerated in the literature. While for a model-building task such as this one, with reasonably finite distinct possible BSM particle content and significant simplifying assumptions that we did not specify to the agent, first among them flavor conservation, the reader may consider the use of machine learning techniques (and the associated demand for computing power) here to be excessive. However, we emphasize again that there is \emph{no} theoretical barrier to applying these same techniques in dramatically more diverse classes of theories, and correspondingly more complex model building tasks. In this light, we can consider the success of the agent in our simple framework as a proof of the underlying concept of our techniques, and leave more ambitious implementations of our procedure to future work.

\subsubsection{Scan Performance: Optimal Configurations}\label{sec:optimal-configurations}

To further understand the utility of the reinforcement learning scan, we must now inspect the results of specific scans. To begin, we defer a detailed discussion of our various hyperparameter selections somewhat and instead consider two specific configurations that are, by differing metrics, ``optimal'' results: One which, averaged over ten independent experiments, produces the largest total number of terminal states, and another which achieves the largest number of models with distinct particle content over the course of a single scan (averaged, again, over ten independent experiments). These two samples will be referred to as the ``optimal terminal states'' and ``optimal distinct states'' samples in our subsequent discussion.\footnote{We note that while both sets of hyperparameters are, by their respective metrics, the optimal performers among our scans, there exist multiple other configurations with comparable performance to these settings. For practical purposes, a reader need not be overly concerned with the possibility that every use case of this methodology requires a grid scan of hyperparameters of comparable complexity to the one we have performed in this work.} By considering the results of these scans in some detail, we can get a sense for the potential capabilities of this method, both in generating distinct sets of particle content that satisfy a theorist's desired parameters and in scanning the parameter space of these models. In Table \ref{tab:optimals}, we summarize the selections for the various training hyperparameters for both scans. The results of all ten independent scans for each configuration are then summarized in Table \ref{tab:optimal-results}. 

\begin{table}[]
    \centering
    \begin{tabular}{| c | c | c | c | c | c | c |}
    \hline
    Configuration & Graph & Skip Layers & Epochs & Reward & $\beta$ & Reward Normalization\\
    \hline
    Optimal Terminal States & A & No & 5 & $R_{IIa}$ & 0.05 & No\\
    \hline
    Optimal Distinct States & A & Yes & 5 & $R_{IIa}$ & 0.2 & Yes\\
    \hline
    \end{tabular}
    \caption{The various hyperparameter selections for the two optimal configurations that we find in our analysis. The meaning of each selection category in the various columns is outlined in Section \ref{sec:experiments}.}
    \label{tab:optimals}
\end{table}

\begin{table}[]
    \centering
    \begin{tabular}{| c | c | c | c | c |}
    \hline
    \multirow{2}{*}{} & \multicolumn{2}{| c |}{Optimal Terminal States} & \multicolumn{2}{| c |}{Optimal Distinct States}\\
    \cline{2-5}
     & Terminal States & Distinct States & Terminal States & Distinct States\\
     \hline
     Trial 1 & 9559 & 2 & 131 & 6\\
     \hline
     Trial 2 & 10246 & 4 & 69 & 5\\
     \hline
     Trial 3 & 5578 & 3 & 33 & 6\\
     \hline
     Trial 4 & 8182 & 2 & 150 & 5\\
     \hline
     Trial 5 & 4377 & 3 & 57 & 6\\
     \hline
     Trial 6 & 5750 & 3 & 49 & 6\\
     \hline
     Trial 7 & 8917 & 2 & 29 & 6\\
     \hline
     Trial 8 & 4473 & 4 & 89 & 6\\
     \hline
     Trial 9 & 10171 & 2 & 82 & 5\\
     \hline
     Trial 10 & 3591 & 2 & 17 & 6\\
     \hline
     Mean & $7100 \pm 2600$ & $2.7 \pm 0.8$ & $71 \pm 44$ & $5.7 \pm 0.5$\\
     \hline
    \end{tabular}
    \caption{The performance, including the number of terminal states identified and the number of models with distinct particle content, achieved by the two optimal agents across ten independent trials.}
    \label{tab:optimal-results}
\end{table}

Given the fact that a total of 1.6 million total states were sampled by the agent during each scan, a reader may be concerned that the total number of terminal states generated seems quite small. However, we emphasize that the parameter space that the agent must learn is quite complex: Starting from allowing each BSM lepton to have 3 independent couplings to each SM generation, during each scan the agent must independently learn to avoid extremely stringent constraints on lepton flavor violation \emph{in addition} to identifying parameters which are appropriate to generate the observed correction to the anomalous magnetic moment of the muon. 

By inspecting our model space, we might anticipate a liberal upper bound on the number of terminal states that a random sampler might generate given the same number of attempts. If we restrict ourselves to only models with at least one BSM vector-like lepton, and a maximum of 6 vector-like leptons of a given electroweak representation (which will benefit a random sampler far more than our reinforcement agent, which due to its intermediate reward structure heavily favors sampling among models with many fewer BSM particles than the maximum number it is allowed), there are 7055 distinct sets of particle content within the model space, of which 6 are viable. As our reinforcement learning agent stores Yukawa coupling parameters in scientific notation, with the exponent of 10 ranging across 12 discrete values, we would require a random sampler to make appropriate selections for the exponents of the two particles in a viable model. Inspecting the parameters of terminal states in our model, we find that generously, randomly sampling among the possible orders of magnitude grants a given BSM lepton a probability of 1/6 to have a muon Yukawa coupling large enough to significantly affect the muon anomalous magnetic moment, a probability of 1/2 of having an electron Yukawa coupling small enough to avoid lepton flavor-violating constraints (roughly $\lesssim O(10^{-5})$ for $Q_D = \pm 1$ and $O(10^{-7})$ for $Q_D = 0$ leptons), and a probability of at most 5/6 of having a small enough $\tau$ Yukawa coupling to avoid similar LFV constraints for the third generation. Since each viable model has only two BSM particles, this suggests that a random sampler would have a probability of $\sim 4 \times 10^{-6}$ of simply producing a model with a viable particle content that avoids lepton flavor constraints. 

Finally, we note that in all viable models, the magnitude of $\Delta a_\mu$ is linearly dependent on the magnitude of a Yukawa coupling between BSM leptons, which generates a chirality-flipping dimension-8 parameter. In order to generate the correct order of magnitude of this coupling, given the orders of magnitude for all prior parameters we can estimate that only one of the 12 possible magnitudes will be viable, reducing our probability to $\sim 3 \times 10^{-7}$. Later in this section, we shall see that terminal states generally populate a narrow band of muon anomalous magnetic moments in the region $2.3 \times 10^{-9} \lesssim \Delta a _{\mu} \lesssim 2.7 \times 10^{-9}$. This suggests that around a central value, the magnitude of the chirality-flipping Yukawa coupling may be modified by around $\pm 7 \%$ in either direction without ruining the model, allowing us to generously estimate that, assuming a permitted value of this Yukawa coupling exists given all other continuously sampled parameters in the model, there might be a $\sim (1/2) \times (1/6) \sim 1/12$ probability of a uniform sampler sampling a viable $O(1)$ value and sign for this critical Yukawa coupling. Taken together, this suggests that, very generously, the probability of a given random sample of the parameter space being a terminal state will be $\sim 3 \times 10^{-8}$, suggesting that a random sampler would require $\sim 4 \times 10^7$ samples  to produce a single viable model. Under this metric, then, our optimal terminal states configuration is on average, very conservatively, $O(10^5)$ times more efficient at producing terminal states than a random scan, while even our optimal distinct states configuration is $O(10^3)$ times more efficient. As these approximate arguments are heavily favorable to the random scan, positing, for example, that there is no correlation between the order of magnitude of a BSM lepton's muon Yukawa coupling and the corresponding limit on its electron or $\tau$ Yukawa couplings and limiting the number of BSM particles in a manner that, as discussed in the previous subsection, is anticipated to have little effect on the performance of the reinforcement learning agent, we anticipate that the true discrepancy between random sampling and the reinforcement learning agent's exploration will be considerably greater.

Having established that our reinforcement learning agent is capable of an intelligent exploration of the parameter space, we now inspect these results further. Table \ref{tab:optimal-results} clearly shows that there is a significant trade-off in the scan hyperparameters between  configurations that maximize the number of terminal states that are generated and those that maximize the number of distinct particle contents that are found. The optimal terminal states configuration produces on average two orders of magnitude more terminal states than the optimal distinct states configuration, however the optimal distinct states configuration either finds all 6 viable particle contents that are represented collectively in the scan, or finds 5 of them-- in stark contrast to the optimal terminal states configuration which never identifies more than 4. 

\begin{figure}
    \centerline{\includegraphics[width=3.2in]{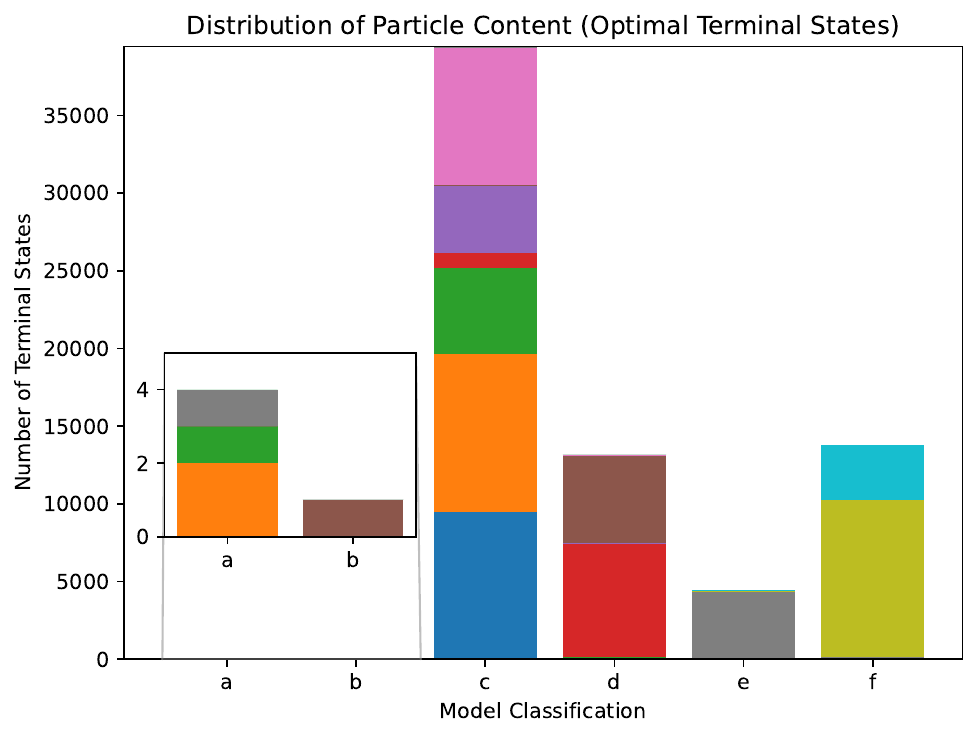}
    \hspace{-0.25cm}
    \includegraphics[width=3.2in]{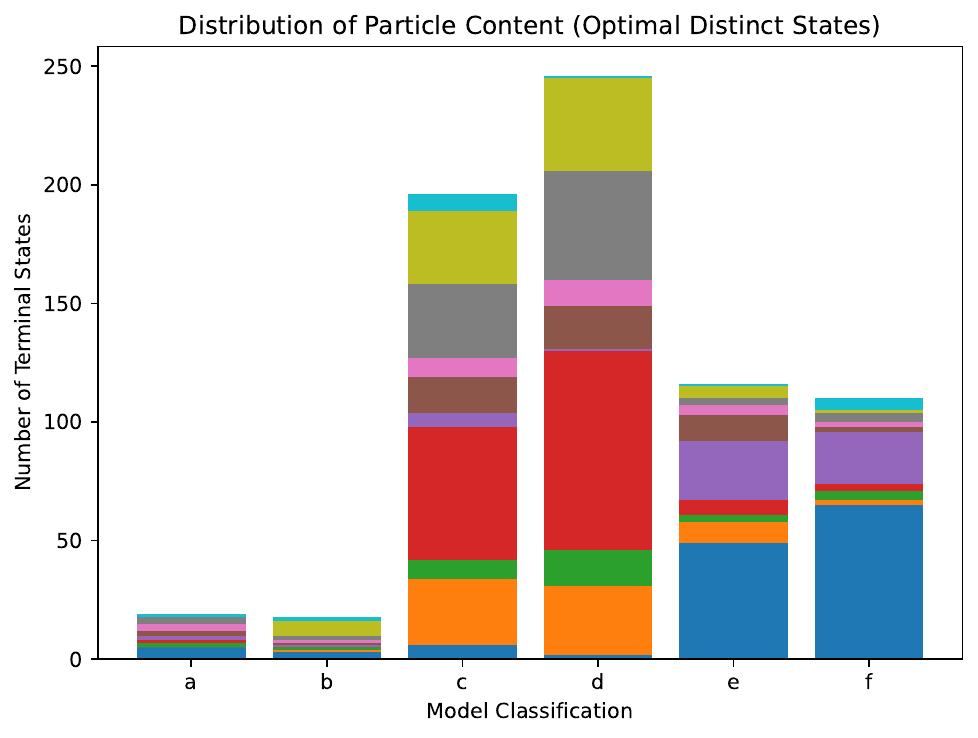}}
    \caption{Stacked histograms of the different models, following the naming convention in Table \ref{tab:distinct_chiral}, achieved by the agent over 10 independent trials for the optimal terminal states configuration (left) and the optimal distinct states configuration (right). Different colors denote the contributions of individual trials.}
    \label{fig:optimal-chiral-models}
\end{figure}

Apart from identifying the number of distinct models that both configurations produce, we can also extract how the different viable particle contents are distributed among the terminal states that each model produces. In Figure \ref{fig:optimal-chiral-models}, we depict stacked histograms over our 10 independent trials which show the number of viable models of each of the types outlined in Table \ref{tab:distinct_chiral} that the agents produce. Notably, we see that although individual trials in the optimal terminal states configuration invariably fail to produce all possible models, all are ultimately represented among the aggregate of all 10 independent trials. This suggests that some of the disadvantage in model diversity that the optimal terminal states configuration exhibits in comparison to the optimal distinct states configuration can be mitigated simply by repeated scans. However, this technique is not perfect. In particular, we see that across both configurations, the agents preferentially sample models `c'-`f' (namely, those which have one $Q_D = 0$ and one $Q_D = \pm 1$ vector-like lepton of the same electroweak representation) over models `a' and `b' (those models which rely on one electroweak doublet lepton and one electroweak singlet, which share a dark charge of $Q_D = \pm 1$). The hierarchy is far more pronounced in the case of the optimal terminal states configuration, however, where for example only a single terminal state of configuration `b' is produced across all 10 scans. Without a more detailed analysis of the models of Table \ref{tab:distinct_chiral}, beyond the scope of this work, it is difficult to determine why both of these sets of scans so strongly prefer models `c'-`f' to models `a' and `b', however it is clear that this bias is considerably more pronounced in the optimal terminal states configuration than in the optimal distinct states one, and the difference between the two cannot be completely erased by aggregating multiple trials of the former configuration.

\begin{figure}
    \centering
    \includegraphics[width=6in]{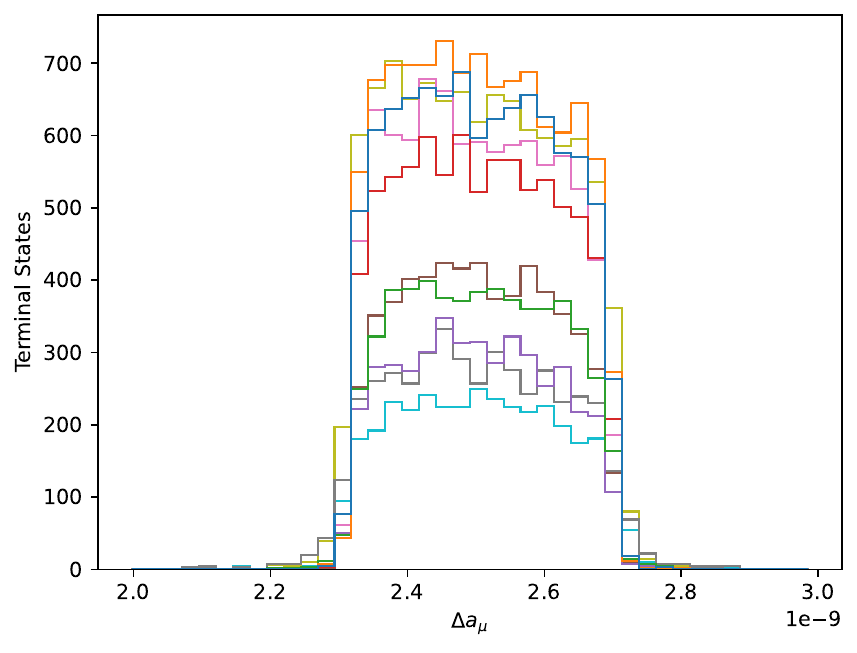}
    \caption{A histogram of the correction to the muon anomalous magnetic moment, $\Delta a_{\mu}$, in different models generated as terminal states from the ``optimal terminal states'' configuration of our reinforcement learning scan, described in Table \ref{tab:optimals}. Different colors denote the results of different independent trials using the same hyperparameters. Recall that our numerical likelihood calculations use the experimental data up through \cite{Muong-2:2021ojo}, which suggests $\Delta a_\mu = (2.51 \pm 0.59) \pm 10^{-9}$.}
    \label{fig:optimal-g-2}
\end{figure}

Beyond simply counting model states, it is instructive to explore the parameter space of terminal states, both in order to reassure us that the agents' outputs are sensible and to get a sense for the behavior of the agent's sampling of the model's continuous parameters. In particular, a user might be interested in whether any insights, either into the agent's limitations or the preferred parameter space of the class of models, can be extracted from the distributions of the continuous model parameters. Since the number of total terminal states in the optimal distinct states configuration is limited, for this component of our analysis we shall restrict ourselves to the optimal terminal states configuration, where thousands of models have been sampled. To begin, in Figure \ref{fig:optimal-g-2} we depict a histogram of the corrections to the muon anomalous magnetic moment, $\Delta a_{\mu}$, which appear in the terminal states for our ``optimal terminal states'' configuration. As is clear in the Figure, the terminal states of the model roughly uniformly sample an area that is around 1/3 of a standard deviation from the central value of the anomaly. This range coincides quite precisely with the model score cutoff of Eq.(\ref{eq:terminal-score}) that we use to define terminal states-- computing the change in the model score from addressing the muon magnetic moment anomaly exactly, and ignoring the effects of all other observables we find that a two-particle theory will produce a model score above the cutoff value for a model which recreates the observed muon anomalous magnetic moment to within $\sim 0.31 \sigma$. The fact that the number of terminal states drops off so sharply outside of this range suggests that the agent is not capable of precisely recreating the observed anomalous magnetic moment correction to within this precision, since it does not appear to have a strong preference for producing states near the central value, as would be anticipated given the fact that various model parameters are selected stochastically, and the agent is incentivized to maximize the probability of sampling within the desired range for $\Delta a_\mu$. Instead, a much broader range of values for $\Delta a_\mu$ are likely achieved by the agent's actions, which are then subject to a sharp cutoff when we collect only the terminal states.

\begin{figure}
    \centerline{\includegraphics[width=3.2in]{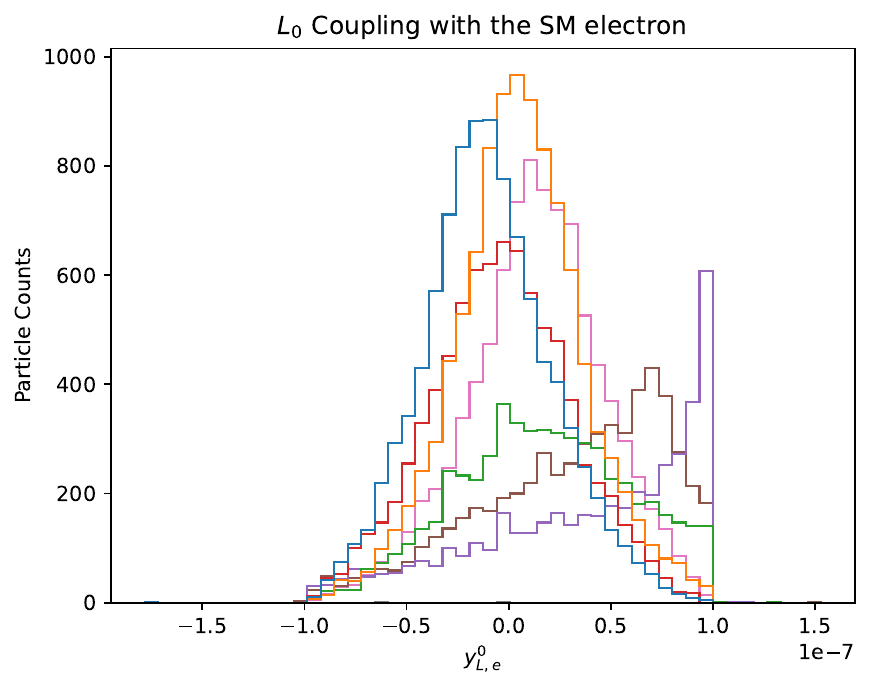}
    \hspace{-0.25cm}
    \includegraphics[width=3.2in]{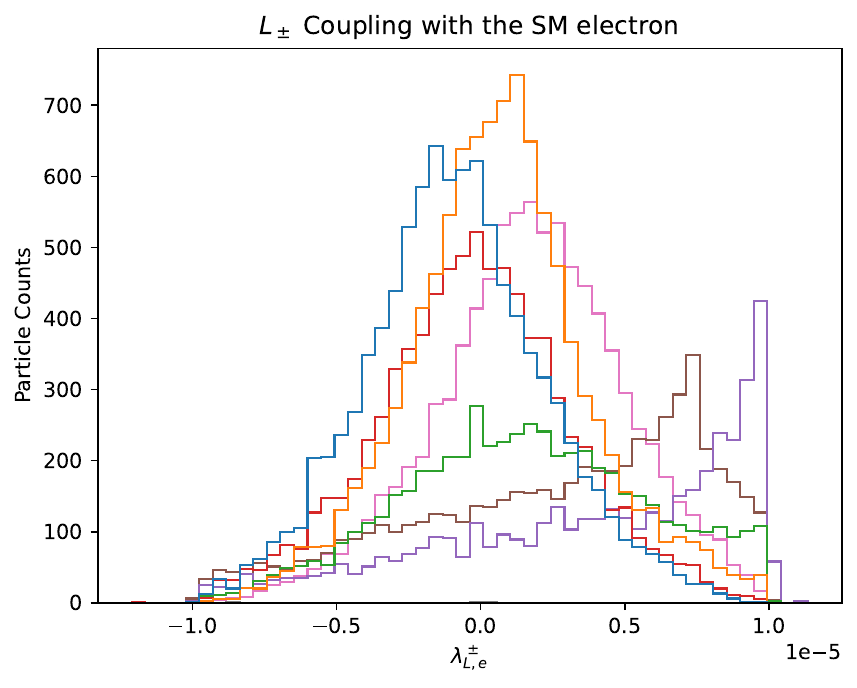}}
    \centerline{\includegraphics[width=3.2in]{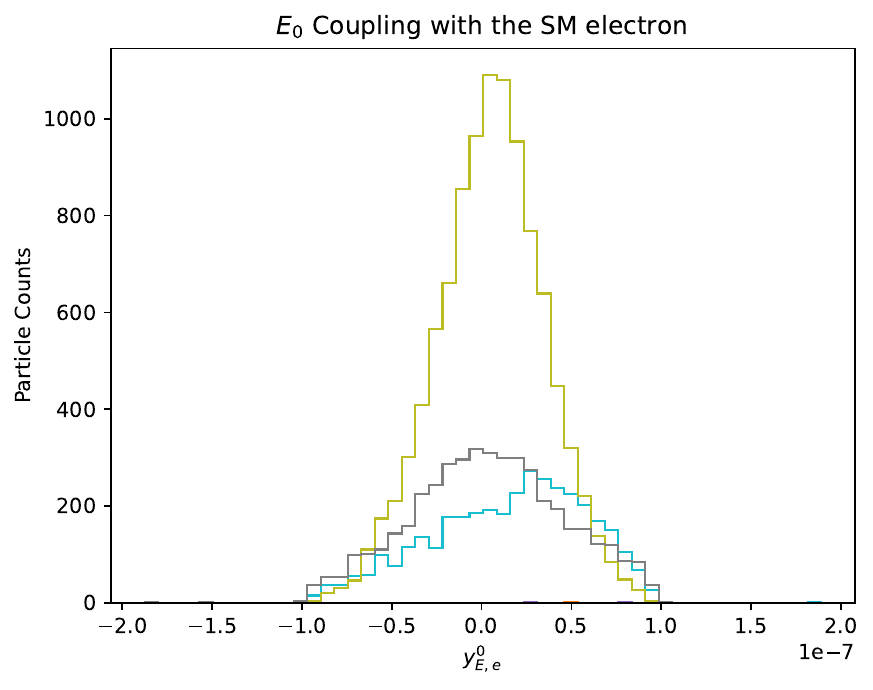}
    \hspace{-0.25cm}
    \includegraphics[width=3.2in]{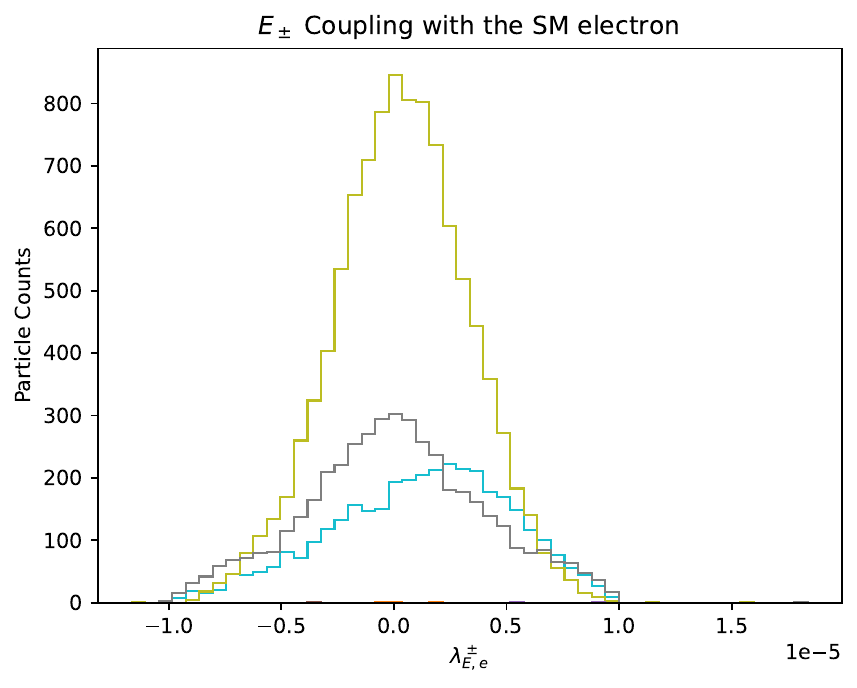}}
    \caption{Histograms of vector-like leptons' Yukawa couplings with electrons for different trials in the optimal terminal states hyperparameter configuration. Following the notation of Eq.(\ref{eq:model-action}, the histograms depict $y^0_{L, e}$ (top left), $\lambda^{\pm}_{L, e}$ (top right), $y^0_{E, e}$ (bottom left), and $\lambda^{\pm}_{E, e}$ (bottom right). Different colors denote different independent trials.}
    \label{fig:electron-yukawa}
\end{figure}

Inspecting individual parameter values, we note that we can extract some limited physical insight into the constraints on these models from the distributions of continuous parameters. As an example, we consider the BSM leptons' Yukawa couplings with the electron, which in theory need only be below a particular order of magnitude to avoid lepton flavor violation constraints. In Figure \ref{fig:electron-yukawa}, we depict histograms of the various BSM leptons' Yukawa couplings with the electron in the terminal states. To fully understand this Figure, we remind the reader that the agent specifies Yukawa couplings in scientific notation, providing both a discrete exponent of 10 and a continuous $O(1)$ value. The results of Figure \ref{fig:electron-yukawa} are approximately visually consistent with normally distributed sampling of the $O(1)$ value around a value with a \emph{fixed} order of magnitude. Notably, the sharp cutoffs in the distributions for two of the trials suggest that a higher order-of-magnitude parameter, near the boundary of the experimental constraints, is preferred in these scans. The fact that such a sharp cutoff exists for these trials in turn suggests that the agent still has difficulty estimating parameters beyond the $O(1)$ level, but it also suggests that we might get a numerical sense of the constraint on these couplings, which we might refine with further analysis outside of the reinforcement learning scan. 

In addition to this information on the model, we can take a far more general lesson away from Figure \ref{fig:electron-yukawa}: The form of the distribution of model parameters depends strongly on the design of the action space. For example, we note that consistently across independent trials, the agent selects (for its terminal states) BSM-to-electron Yukawa couplings on the order of $10^{-7}$ for those to the SM Higgs, and $10^{-5}$ for those to the dark Higgs, in spite of the fact that significantly lower (indeed, down to the model space's magnitude cutoffs of $10^{-12}$ and $10^{-10}$, respectively) orders of magnitude are phenomenologically viable. The reason for the agent selecting these magnitudes, as opposed to lower ones, is unclear, and is likely an artifact of specifics of our implementation. The broader implication is clear: The population of terminal states from a reinforcement learning scan represents a nontrivially distributed sample of the viable model parameter space, and this distribution may not readily match a theorist's priors on a given parameter (for these Yukawa couplings, a model builder might more readily anticipate that they would be zero, or suppressed by known loop factors that give a natural magnitude quite distinct from the agent's guess). As a corollary to this observation, we can comment on the specific use case of this technology: The reinforcement learning agent can readily produce examples of viable models that a theorist might use as inspiration, but the distribution of terminal states in the parameter space that it finds can be subject to a number of factors unrelated to the underlying physics of the problem. It seems probable, then, that the more valuable metric for the model builder may be the number of distinct models (that is, with different particle content) that a scan produces, after which a more formally rigorous numerical scan of a fixed parameter space featuring these models can be performed. In the language of this Section, a model builder is more likely to value the optimal \emph{distinct} states hyperparameter configuration, which produces a larger diversity of models that may be used as inspiration for the model builder, over the optimal \emph{terminal} states configuration, which produces a nontrivially biased sample of different models' parameter spaces.

\subsubsection{Scan Performance: Comparing Hyperparameter Choices}\label{sec:hyperparameter-choices}

We shall now discuss the impact of different training factors on the agent's performance. Again, we will evaluate the agent's performance using the two introduced performance metrics: ``optimal terminal states" and ``optimal distinct states." As discussed in Section \ref{sec:optimal-configurations}, for practical use cases maximizing the total number of terminal states is less useful as a performance target than maximizing the number of distinct states, but the number of terminal states is still important, since reaching a terminal state in our context is equivalent to finding a viable BSM model. Although this metric cannot measure the diversity of the terminal states, it measures the frequency at which these states are attained and can thus serve as a measure of the agent's ability to explore the continuous part of the parameter space and therefore learn effectively. As shown in Table~\ref{tab:distinct_chiral}, the distinct states here are defined as distinct combinations of 2 isospin singlet or doublet particles that carry different dark $U(1)$ charges and have significant chiral enhancements contributing to the muon anomalous magnetic moment. There are only 6 such combinations for this simple vector-like lepton model, given that we want our terminal states to be models with at most two BSM fermions. Qualitatively this metric serves as a measure of the agent's ability to explore the discrete action space and identify distinct promising models which might be further explored through other means. 

\begin{figure}
    \centering
    \includegraphics[scale=0.37]{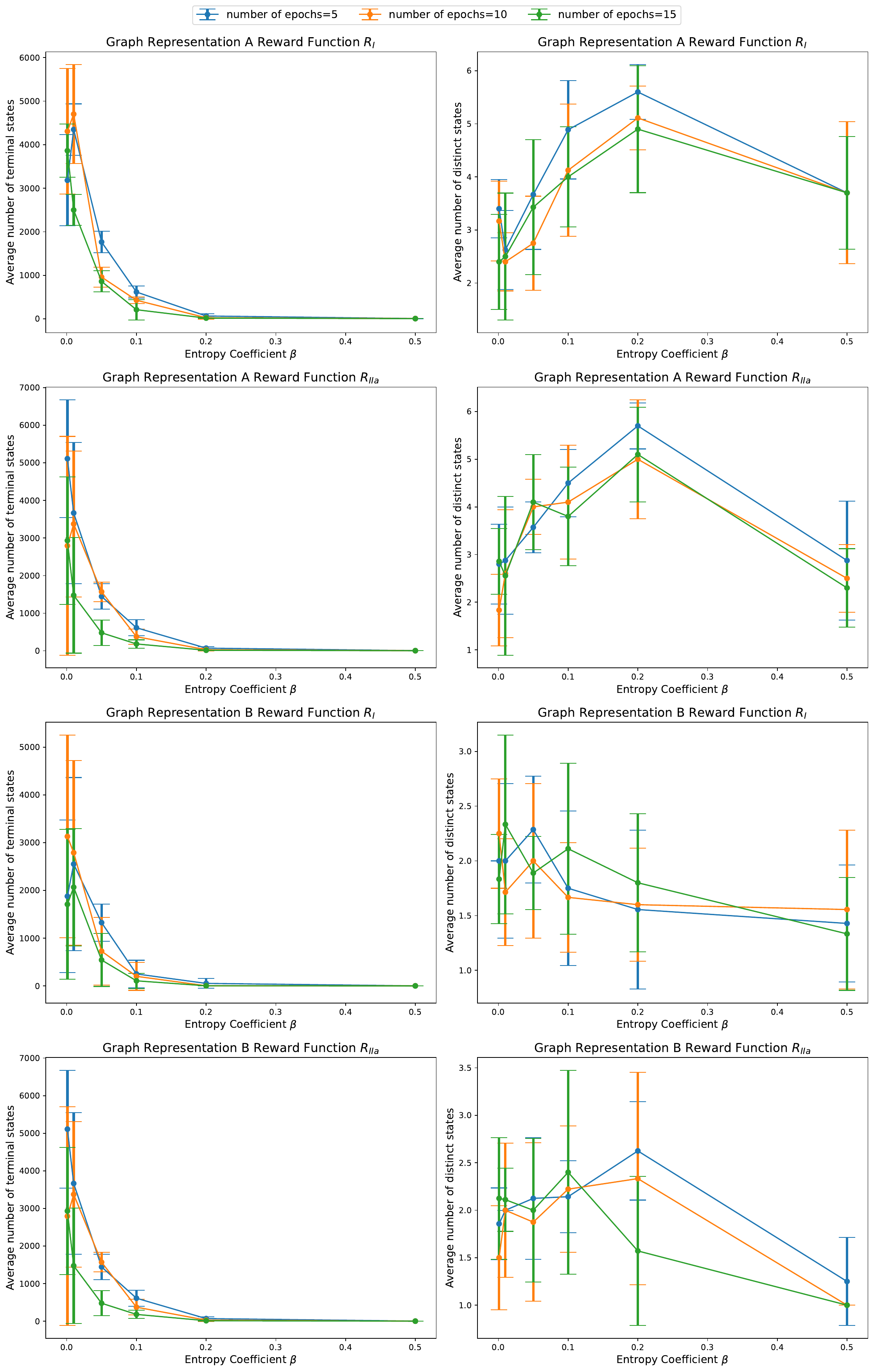}
    \caption{The number of terminal states (left) and distinct states (right), obtained as the mean of 10 independent trials for each data point, with standard deviation as a function of the entropy coefficient $\beta$ for both graph representations A and B and reward functions $R_I$ and $R_{IIa}$.}
    \label{fig:entropy_terminal}
\end{figure}

In Figure~\ref{fig:entropy_terminal}, the total number of terminal states and distinct states, averaged over 10 independent identical trials for each hyperparameter configuration, are plotted for both graph representation A and B and reward function $R_I$ and $R_{IIa}$, as a function of a scan of entropy coefficient $\beta=0.001$, $0.01$, $0.05$, $0.1$, $0.2$, $0.5$. These values are taken for experiments with different numbers of training epochs, and with skip layers and reward normalization implemented. Out of the 10 different independent experiments for each number of epochs, the experiments that return null results are excluded -- this principally affects the results of trials with graph representation B. It is clear that both performance metrics depend on the choice of entropy coefficient $\beta$. This dependency is particularly strong for the average total number of terminal states. For both graph representation with both reward functions, we can see that the average total number of terminal states peaks at around $O(0.001 - 0.01)$. This is unsurprising: If it has a low entropy coefficient, an agent will have a strong incentive to output probability distributions that are heavily weighted toward states which it has previously seen yield high rewards, and therefore will be likely to produce a large number of terminal states, as long as it finds at least one before its training is completed. Despite being the optimal range for finding maximum average total number of terminal states, this range is not optimal for finding the average total number of distinct states, a metric which clearly favors exploration of the model space over exploitation of known regions. For graph representation A, $\beta \sim 0.2$ seems to be optimal for both reward functions. In contrast, for graph representation B, the performance is comparable for $\beta=0.001, 0.01, 0.05, 0.1, 0.2$-- we might attribute this to the fact that the total magnitude of the number of discrete states is somewhat smaller for trials in representation B and may therefore be more sensitive to statistical fluctuations. An important message is that depending on the task and performance metric to be optimized, we should always tune on the entropy coefficient $\beta$. For the remainder of this section, we always optimize the entropy coefficient while presenting results and discussing the roles of other factors. 

In Figure~\ref{fig:entropy_terminal}, we can also see that both the average total number of terminal states and distinct states possess a large standard deviation over the trials, as depicted in the error bars around each point. This indicates that the results are very sensitive to the initial starting conditions, and hence each experiment with the same set of hyperparameters selection can return very different results.\footnote{For practical implementations of this graph-based reinforcement learning strategy, this sensitivity means that ensuring reproducibility between runs can be difficult. In particular, the PyTorch machine learning library's random number generation utilities will produce different random sequences even when passed the same seed when operating on different machines, even when computations are not performed on a CPU. For best practices, we recommend maximizing the number of independent trials when exploring a model space in order to defray this sensitivity.} For the total number of distinct states, graph representation B generally has a larger standard deviation than graph representation A, which indicates that the agent's performance is even more sensitive to the starting condition. This is probably due to the fact that in representation B, different particles (electroweak isospin doublets and singlets) are represented as different node types, resulting in more separate weights to be learned. From the figure, it is also clear that graph representation A generally finds more terminal states and distinct states than graph representation B, regardless of the type of reward function used. 

\begin{table}[t!]
\centering
\scalebox{0.8}{
\begin{tabular}{|c|c|c|c|cc|cc|}
\hline
 &  &  &  & \multicolumn{2}{c|}{Number of terminal states} & \multicolumn{2}{c|}{Number of distinct states} \\ \hline
\begin{tabular}[c]{@{}c@{}} Epochs\end{tabular} & \begin{tabular}[c]{@{}c@{}}Reward\end{tabular} & \begin{tabular}[c]{@{}c@{}}Skip\\  Layers\end{tabular} & \begin{tabular}[c]{@{}c@{}}Reward\\ Normalization\end{tabular} & \multicolumn{1}{c|}{\begin{tabular}[c]{@{}c@{}}Representation \\ A\end{tabular}} & \begin{tabular}[c]{@{}c@{}}Representation\\  B\end{tabular} & \multicolumn{1}{c|}{\begin{tabular}[c]{@{}c@{}}Representation\\  A\end{tabular}} & \begin{tabular}[c]{@{}c@{}}Representation\\  B\end{tabular} \\ \hline
\multirow{12}{*}{5} & \multirow{4}{*}{$R_I$} & No & No & \multicolumn{1}{c|}{7756.0} & 6368.3 & \multicolumn{1}{c|}{3.5} & 2.3 \\ \cline{3-8} 
 &  & Yes & No & \multicolumn{1}{c|}{6101.8} & 4827.2 & \multicolumn{1}{c|}{3.7} & 2.0 \\ \cline{3-8} 
 &  & No & Yes & \multicolumn{1}{c|}{5690.4} & 3847.0 & \multicolumn{1}{c|}{4.7} & 2.0 \\ \cline{3-8} 
 &  & Yes & Yes & \multicolumn{1}{c|}{4348.1} & 2551.4 & \multicolumn{1}{c|}{5.6} & 2.3 \\ \cline{2-8} 
 & \multirow{4}{*}{$R_{IIa}$} & No & No & \multicolumn{1}{c|}{8229.3} & 6453.3 & \multicolumn{1}{c|}{2.9} & 2.3 \\ \cline{3-8} 
 &  & Yes & No & \multicolumn{1}{c|}{6345.0} & 6314.4 & \multicolumn{1}{c|}{3.7} & 2.1 \\ \cline{3-8} 
 &  & No & Yes & \multicolumn{1}{c|}{5715.0} & 2201.3 & \multicolumn{1}{c|}{3.8} & 1.9 \\ \cline{3-8} 
 &  & Yes & Yes & \multicolumn{1}{c|}{5109.6} & 3540.9 & \multicolumn{1}{c|}{5.7} & 2.6 \\ \cline{2-8} 
 & \multirow{4}{*}{$R_{IIb}$} & No & No & \multicolumn{1}{c|}{4932.0} & 4477.0 & \multicolumn{1}{c|}{3.3} & 2.3 \\ \cline{3-8} 
 &  & Yes & No & \multicolumn{1}{c|}{6322.1} & 6011.0 & \multicolumn{1}{c|}{3.9} & 2.2 \\ \cline{3-8} 
 &  & No & Yes & \multicolumn{1}{c|}{4376.0} & 1294.4 & \multicolumn{1}{c|}{3.6} & 1.7 \\ \cline{3-8} 
 &  & Yes & Yes & \multicolumn{1}{c|}{4230.2} & 3361.5 & \multicolumn{1}{c|}{5.2} & 2.0 \\ \hline
\multirow{12}{*}{10} & \multirow{4}{*}{$R_{I}$} & No & No & \multicolumn{1}{c|}{4691.8} & 5866.0 & \multicolumn{1}{c|}{4.5} & 2.0 \\ \cline{3-8} 
 &  & Yes & No & \multicolumn{1}{c|}{6975.7} & 3929.3 & \multicolumn{1}{c|}{3.0} & 1.9 \\ \cline{3-8} 
 &  & No & Yes & \multicolumn{1}{c|}{3734.2} & 3189.0 & \multicolumn{1}{c|}{4.3} & 2.2 \\ \cline{3-8} 
 &  & Yes & Yes & \multicolumn{1}{c|}{4702.8} & 3129.5 & \multicolumn{1}{c|}{5.1} & 2.3 \\ \cline{2-8} 
 & \multirow{4}{*}{$R_{IIa}$} & No & No & \multicolumn{1}{c|}{5423.9} & 3086.7 & \multicolumn{1}{c|}{3.1} & 2.3 \\ \cline{3-8} 
 &  & Yes & No & \multicolumn{1}{c|}{4922.4} & 5271.8 & \multicolumn{1}{c|}{4.3} & 2.1 \\ \cline{3-8} 
 &  & No & Yes & \multicolumn{1}{c|}{3478.2} & 3344.0 & \multicolumn{1}{c|}{4.0} & 2.2 \\ \cline{3-8} 
 &  & Yes & Yes & \multicolumn{1}{c|}{3373.6} & 2230.4 & \multicolumn{1}{c|}{5.0} & 2.3 \\ \cline{2-8} 
 & \multirow{4}{*}{$R_{IIb}$} & No & No & \multicolumn{1}{c|}{3960.1} & 4240.2 & \multicolumn{1}{c|}{4.1} & 2.0 \\ \cline{3-8} 
 &  & Yes & No & \multicolumn{1}{c|}{4842.0} & 4216.4 & \multicolumn{1}{c|}{4.3} & 2.4 \\ \cline{3-8} 
 &  & No & Yes & \multicolumn{1}{c|}{1018.2} & 794.0 & \multicolumn{1}{c|}{4.0} & 2.0 \\ \cline{3-8} 
 &  & Yes & Yes & \multicolumn{1}{c|}{4647.0} & 2698.7 & \multicolumn{1}{c|}{4.4} & 2.0 \\ \hline
\multirow{12}{*}{15} & \multirow{4}{*}{$R_{I}$} & No & No & \multicolumn{1}{c|}{4154.3} & 3718.0 & \multicolumn{1}{c|}{3.9} & 2.0 \\ \cline{3-8} 
 &  & Yes & No & \multicolumn{1}{c|}{4359.2} & 3310.9 & \multicolumn{1}{c|}{3.7} & 2.0 \\ \cline{3-8} 
 &  & No & Yes & \multicolumn{1}{c|}{3440.9} & 1642.0 & \multicolumn{1}{c|}{4.1} & 2.0 \\ \cline{3-8} 
 &  & Yes & Yes & \multicolumn{1}{c|}{3861.8} & 2067.2 & \multicolumn{1}{c|}{4.9} & 2.3 \\ \cline{2-8} 
 & \multirow{4}{*}{$R_{IIa}$} & No & No & \multicolumn{1}{c|}{4510.3} & 2315.9 & \multicolumn{1}{c|}{3.5} & 2.2 \\ \cline{3-8} 
 &  & Yes & No & \multicolumn{1}{c|}{3746.3} & 3160.5 & \multicolumn{1}{c|}{4.0} & 2.2 \\ \cline{3-8} 
 &  & No & Yes & \multicolumn{1}{c|}{1557.2} & 1970.2 & \multicolumn{1}{c|}{3.9} & 2.0 \\ \cline{3-8} 
 &  & Yes & Yes & \multicolumn{1}{c|}{2931.7} & 1813.0 & \multicolumn{1}{c|}{5.1} & 2.4 \\ \cline{2-8} 
 & \multirow{4}{*}{$R_{IIb}$} & No & No & \multicolumn{1}{c|}{3034.1} & 2262.0 & \multicolumn{1}{c|}{3.3} & 2.3 \\ \cline{3-8} 
 &  & Yes & No & \multicolumn{1}{c|}{2304.9} & 2576.3 & \multicolumn{1}{c|}{4.3} & 2.2 \\ \cline{3-8} 
 &  & No & Yes & \multicolumn{1}{c|}{2432.8} & 821.3 & \multicolumn{1}{c|}{3.5} & 2.0 \\ \cline{3-8} 
 &  & Yes & Yes & \multicolumn{1}{c|}{1740.1} & 2015.3 & \multicolumn{1}{c|}{3.8} & 2.0 \\ \hline
\end{tabular}}
\caption{The number of terminal states and distinct models with/without the implementation of skip layers and/or reward normalization, for different number of epochs and reward functions, when the optimal value of entropy coefficient $\beta$ is picked across each metric, in graph representations A and B.}\label{tab:skipandnorm}
\end{table}

Table~\ref{tab:skipandnorm} shows the total number of terminal states and distinct states with or without the implementation of skip layers and/or reward normalization, averaged across 10 independent trials with non-null results, when the optimal value of entropy coefficient $\beta$ is chosen across each metric.  
Here we neglect the fact that results for different $\beta$ are very different, as our aim is just to examine the overall effects of the implementation of skip layers and reward normalization. An immediate observation is that these parameters (skip layers, reward normalization) can have an impact on the results, but the amount of impact depends non-trivially on the global selection of other discrete hyperparameters. For example, looking at the total number of terminal states for the optimal case, the agents always found more terminal states when there was no reward normalization. However, depending on the choice of reward function and the number of epochs, the implementation of skip layers may be favorable or not, so again different performance metrics are sensitive to different choices of hyperparameters. If we instead consider the optimal total number of distinct states, graph representation A always found the most number of distinct models when skip layers and reward normalization were both implemented, while for graph representation B, the optimal result varies non-trivially with different choices of the number of epochs and the reward function. 

In addition to showing the results of selecting different optimal $\beta$ for each set of experiments with different hyperparameter settings, we also present Table~\ref{tab:smallskipandnorm}, in which the number of training epochs is 10, and the entropy coefficient $\beta=0.2$ is selected for representation A and  $\beta=0.001$ is selected for representation B, averaged over 10 independent identical trials with null results excluded. For this set of data with a fixed $\beta$ value chosen across the four architecture choices (with/without skip layers and/or reward normalization), it clearly shows no consistent behavior in the dependency on skip layers and reward normalization for different choices of graph representations and reward functions, except that for reward function $R_I$ and $R_{IIa}$, representation A performs the best when both are implemented. This further justifies our claim that these two architecture choices are very model-dependent. Therefore, any other model building and analysis similar to this project should experiment with these techniques to see if they truly improve the results before deploying them. 

From Tables~\ref{tab:skipandnorm}--\ref{tab:smallskipandnorm}, if we compare the performance of the agent with the same hyperparameter settings but different choices of reward functions, we see that the results are insensitive to the selection of the reward function. In fact, comparing all the results, we find that each reward function might have its own set of optimal hyperparameters that maximize both or one of the performance metrics. However, for most of the hyperparameter ranges that we have covered, both graph representations has reasonable performance under all three reward functions, so fine-tuning is not necessary. 

\begin{table}[h]
\centering
\scalebox{0.8}{
\begin{tabular}{|c|c|c|cc|cc|}
\hline
 &  &  & \multicolumn{2}{c|}{Number of terminal states} & \multicolumn{2}{c|}{Number of distinct states} \\ \hline
\begin{tabular}[c]{@{}c@{}}Reward\end{tabular} & \begin{tabular}[c]{@{}c@{}}Skip\\  Layers\end{tabular} & \begin{tabular}[c]{@{}c@{}}Reward\\ Normalization\end{tabular} & \multicolumn{1}{c|}{\begin{tabular}[c]{@{}c@{}}Representation \\ A\\ ($\beta = 0.2$)\end{tabular}} & \begin{tabular}[c]{@{}c@{}}Representation\\  B\\ ($\beta = 0.001$)\end{tabular} & \multicolumn{1}{c|}{\begin{tabular}[c]{@{}c@{}}Representation\\  A\\ ($\beta = 0.2$)\end{tabular}} & \begin{tabular}[c]{@{}c@{}}Representation\\  B\\ ($\beta=0.001$)\end{tabular} \\ \hline
\multirow{4}{*}{$R_I$} & No & No & \multicolumn{1}{c|}{11.7} & 5866.0 & \multicolumn{1}{c|}{4.5} & 2.0 \\ \cline{2-7} 
 & Yes & No & \multicolumn{1}{c|}{1828.4} & 3929.3 & \multicolumn{1}{c|}{3.0} & 1.8 \\ \cline{2-7} 
 & No & Yes & \multicolumn{1}{c|}{8.6} & 2623.7 & \multicolumn{1}{c|}{4.3} & 2.2 \\ \cline{2-7} 
 & Yes & Yes & \multicolumn{1}{c|}{24.8} & 3129.5 & \multicolumn{1}{c|}{5.1} & 2.3 \\ \hline
\multirow{4}{*}{$R_{IIa}$} & No & No & \multicolumn{1}{c|}{1429.9} & 2588.9 & \multicolumn{1}{c|}{3.0} & 2.3 \\ \cline{2-7} 
 & Yes & No & \multicolumn{1}{c|}{3608.8} & 5271.8 & \multicolumn{1}{c|}{3.4} & 2.0 \\ \cline{2-7} 
 & No & Yes & \multicolumn{1}{c|}{7.9} & 2559.6 & \multicolumn{1}{c|}{4.0} & 2.2 \\ \cline{2-7} 
 & Yes & Yes & \multicolumn{1}{c|}{29.0} & 2093.2 & \multicolumn{1}{c|}{5.0} & 1.5 \\ \hline
\multirow{4}{*}{$R_{IIb}$} & No & No & \multicolumn{1}{c|}{1634.7} & 3105.2 & \multicolumn{1}{c|}{4.1} & 1.8 \\ \cline{2-7} 
 & Yes & No & \multicolumn{1}{c|}{1914.7} & 4216.4 & \multicolumn{1}{c|}{2.7} & 2.4 \\ \cline{2-7} 
 & No & Yes & \multicolumn{1}{c|}{3.5} & 794.0 & \multicolumn{1}{c|}{2.3} & 1.4 \\ \cline{2-7} 
 & Yes & Yes & \multicolumn{1}{c|}{8.8} & 2698.7 & \multicolumn{1}{c|}{3.9} & 2.0 \\ \hline
\end{tabular}}
\caption{The average total number of terminal states and distinct models with/without the implementation of skip layers and/or reward normalization, for 10 epochs, and $\beta=0.2$ for graph representation A and $\beta=0.001$ for graph representation B.}\label{tab:smallskipandnorm}
\end{table}

\begin{table}[]
\resizebox{\columnwidth}{!}{%
\begin{tabular}{|c|cccc|cccc|}
\hline
 & \multicolumn{4}{c|}{Optimal Terminal States (A)} & \multicolumn{4}{c|}{Optimal Distinct States (A)} \\ \hline
 & \multicolumn{2}{c|}{Terminal States} & \multicolumn{2}{c|}{Distinct States} & \multicolumn{2}{c|}{Terminal States} & \multicolumn{2}{c|}{Distinct States} \\ \hline
 & \multicolumn{1}{c|}{A} & \multicolumn{1}{c|}{B} & \multicolumn{1}{c|}{A} & B & \multicolumn{1}{c|}{A} & \multicolumn{1}{c|}{B} & \multicolumn{1}{c|}{A} & B \\ \hline
Trial 1 & \multicolumn{1}{c|}{9559} & \multicolumn{1}{c|}{7797} & \multicolumn{1}{c|}{2} & 2 & \multicolumn{1}{c|}{131} & \multicolumn{1}{c|}{29} & \multicolumn{1}{c|}{6} & 2 \\ \hline
Trial 2 & \multicolumn{1}{c|}{10246} & \multicolumn{1}{c|}{1} & \multicolumn{1}{c|}{4} & 1 & \multicolumn{1}{c|}{69} & \multicolumn{1}{c|}{12} & \multicolumn{1}{c|}{5} & 3 \\ \hline
Trial 3 & \multicolumn{1}{c|}{5578} & \multicolumn{1}{c|}{0} & \multicolumn{1}{c|}{3} & 0 & \multicolumn{1}{c|}{33} & \multicolumn{1}{c|}{7} & \multicolumn{1}{c|}{6} & 2 \\ \hline
Trial 4 & \multicolumn{1}{c|}{8182} & \multicolumn{1}{c|}{6620} & \multicolumn{1}{c|}{2} & 2 & \multicolumn{1}{c|}{150} & \multicolumn{1}{c|}{0} & \multicolumn{1}{c|}{5} & 0 \\ \hline
Trial 5 & \multicolumn{1}{c|}{4377} & \multicolumn{1}{c|}{5069} & \multicolumn{1}{c|}{3} & 2 & \multicolumn{1}{c|}{57} & \multicolumn{1}{c|}{6} & \multicolumn{1}{c|}{6} & 3 \\ \hline
Trial 6 & \multicolumn{1}{c|}{5750} & \multicolumn{1}{c|}{7600} & \multicolumn{1}{c|}{3} & 2 & \multicolumn{1}{c|}{49} & \multicolumn{1}{c|}{0} & \multicolumn{1}{c|}{6} & 0 \\ \hline
Trial 7 & \multicolumn{1}{c|}{8917} & \multicolumn{1}{c|}{0} & \multicolumn{1}{c|}{2} & 0 & \multicolumn{1}{c|}{29} & \multicolumn{1}{c|}{28} & \multicolumn{1}{c|}{6} & 3 \\ \hline
Trial 8 & \multicolumn{1}{c|}{4473} & \multicolumn{1}{c|}{2525} & \multicolumn{1}{c|}{4} & 2 & \multicolumn{1}{c|}{89} & \multicolumn{1}{c|}{2} & \multicolumn{1}{c|}{6} & 2 \\ \hline
Trial 9 & \multicolumn{1}{c|}{10171} & \multicolumn{1}{c|}{6816} & \multicolumn{1}{c|}{2} & 2 & \multicolumn{1}{c|}{82} & \multicolumn{1}{c|}{3} & \multicolumn{1}{c|}{5} & 3 \\ \hline
Trial 10 & \multicolumn{1}{c|}{3591} & \multicolumn{1}{c|}{6540} & \multicolumn{1}{c|}{2} & 2 & \multicolumn{1}{c|}{17} & \multicolumn{1}{c|}{128} & \multicolumn{1}{c|}{6} & 3 \\ \hline
Mean & \multicolumn{1}{c|}{7100 $\pm$ 2600} & \multicolumn{1}{c|}{4297 $\pm$ 3147} & \multicolumn{1}{c|}{2.7 $\pm$ 0.8} & 1.5 $\pm$ 0.9 & \multicolumn{1}{c|}{71 $\pm$ 44} & \multicolumn{1}{c|}{22 $\pm$ 37} & \multicolumn{1}{c|}{5.7 $\pm$ 0.5} & 2.1 $\pm$ 1.2 \\ \hline
\end{tabular}%
}
\caption{The optimal performance of the agent in graph representation A, including the number of terminal states identified and the number of models with distinct particle content and the comparison between results for graph representations A and B, across 10 independent trials. } \label{tab:optimalA}
\end{table}

\begin{table}[]
\resizebox{\columnwidth}{!}{%
\begin{tabular}{|c|cccc|cccc|}
\hline
 & \multicolumn{4}{c|}{Optimal Terminal States (B)} & \multicolumn{4}{c|}{Optimal Distinct States (B)} \\ \hline
 & \multicolumn{2}{c|}{Terminal States} & \multicolumn{2}{c|}{Distinct States} & \multicolumn{2}{c|}{Terminal States} & \multicolumn{2}{c|}{Distinct States} \\ \hline
 & \multicolumn{1}{c|}{A} & \multicolumn{1}{c|}{B} & \multicolumn{1}{c|}{A} & B & \multicolumn{1}{c|}{A} & \multicolumn{1}{c|}{B} & \multicolumn{1}{c|}{A} & B \\ \hline
Trial 1 & \multicolumn{1}{c|}{6268} & \multicolumn{1}{c|}{5045} & \multicolumn{1}{c|}{2} & 3 & \multicolumn{1}{c|}{14} & \multicolumn{1}{c|}{8} & \multicolumn{1}{c|}{4} & 1 \\ \hline
Trial 2 & \multicolumn{1}{c|}{5973} & \multicolumn{1}{c|}{6449} & \multicolumn{1}{c|}{3} & 2 & \multicolumn{1}{c|}{41} & \multicolumn{1}{c|}{3} & \multicolumn{1}{c|}{4} & 2 \\ \hline
Trial 3 & \multicolumn{1}{c|}{8669} & \multicolumn{1}{c|}{3468} & \multicolumn{1}{c|}{2} & 2 & \multicolumn{1}{c|}{217} & \multicolumn{1}{c|}{136} & \multicolumn{1}{c|}{2} & 2 \\ \hline
Trial 4 & \multicolumn{1}{c|}{4872} & \multicolumn{1}{c|}{5733} & \multicolumn{1}{c|}{3} & 2 & \multicolumn{1}{c|}{115} & \multicolumn{1}{c|}{28} & \multicolumn{1}{c|}{5} & 2 \\ \hline
Trial 5 & \multicolumn{1}{c|}{5825} & \multicolumn{1}{c|}{4260} & \multicolumn{1}{c|}{2} & 2 & \multicolumn{1}{c|}{100} & \multicolumn{1}{c|}{10} & \multicolumn{1}{c|}{4} & 5 \\ \hline
Trial 6 & \multicolumn{1}{c|}{8179} & \multicolumn{1}{c|}{6791} & \multicolumn{1}{c|}{5} & 2 & \multicolumn{1}{c|}{203} & \multicolumn{1}{c|}{18} & \multicolumn{1}{c|}{5} & 3 \\ \hline
Trial 7 & \multicolumn{1}{c|}{0} & \multicolumn{1}{c|}{4858} & \multicolumn{1}{c|}{0} & 2 & \multicolumn{1}{c|}{322} & \multicolumn{1}{c|}{4} & \multicolumn{1}{c|}{4} & 2 \\ \hline
Trial 8 & \multicolumn{1}{c|}{0} & \multicolumn{1}{c|}{4684} & \multicolumn{1}{c|}{0} & 2 & \multicolumn{1}{c|}{321} & \multicolumn{1}{c|}{6} & \multicolumn{1}{c|}{4} & 2 \\ \hline
Trial 9 & \multicolumn{1}{c|}{5407} & \multicolumn{1}{c|}{4814} & \multicolumn{1}{c|}{2} & 2 & \multicolumn{1}{c|}{230} & \multicolumn{1}{c|}{7} & \multicolumn{1}{c|}{4} & 3 \\ \hline
Trial 10 & \multicolumn{1}{c|}{5567} & \multicolumn{1}{c|}{6467} & \multicolumn{1}{c|}{2} & 2 & \multicolumn{1}{c|}{260} & \multicolumn{1}{c|}{4} & \multicolumn{1}{c|}{2} & 2 \\ \hline
Mean & \multicolumn{1}{c|}{5076 $\pm$ 2930} & \multicolumn{1}{c|}{5257 $\pm$ 1019} & \multicolumn{1}{c|}{2.1 $\pm$ 1.4} & 2.1 $\pm$ 0.3 & \multicolumn{1}{c|}{182 $\pm$ 110} & \multicolumn{1}{c|}{22 $\pm$ 39} & \multicolumn{1}{c|}{3.8 $\pm$ 1.0} & 2.4 $\pm$ 1.1 \\ \hline
\end{tabular}%
}
\caption{The optimal performance of agent with representation B, including the number of terminal states identified and the number of models with distinct particle content and the comparison between results for representation A and B, across ten independent trials. } \label{tab:optimalB}
\end{table}

The last factor to be discussed, and potentially the most important one, is the choice of the graph representation and how it affects the agent's performance. In Table~\ref{tab:optimalA}, we select the ``optimal terminal states'' and ``optimal distinct states'' hyperparameter configurations as discussed in Section \ref{sec:optimal-configurations}, and compare the results of these trials to trials with the same hyperparameter configurations using graph representation B.
Comparing these results, it is apparent that graph representation A always outperforms graph representation B; however, this could simply be a product of the fact that our other hyperparameters have optimized the results for graph representation A. To examine the robustness of this finding, we present Table~\ref{tab:optimalB}, in which we select the optimal set of hyperparameter configurations for graph representation B. We can see that for the optimal total number of terminal states, the best set of results for graph representation B slightly outperforms graph representation A, but the values final averages are easily statistically identical (albeit with graph representation B's result having a substantially lower variance). However, in terms of finding distinct states, graph representation A is always better. The best performance of the agent with graph representation B finds 5 distinct states in one out of the ten trials, while A easily finds 5-6 distinct states for all trials under its optimal hyperparameter setting. Even when working with B's optimal hyperparameters, the average number of distinct states that the graph representation A agent finds significantly outpaces that of the graph representation B.

\begin{figure}[t]
    \centering
    \includegraphics[scale=0.45]{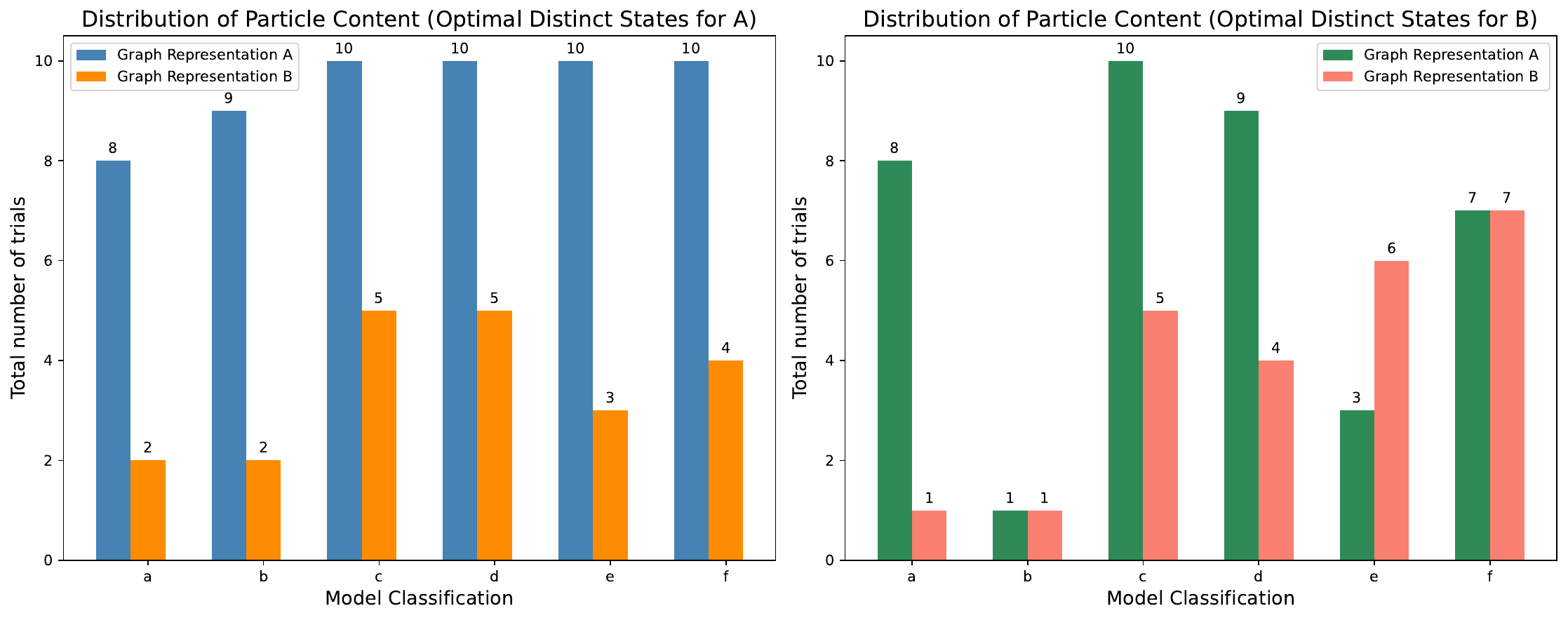}
    \caption{The number of times each model was achieved by the agent over 10 independent trials for the optimal distinct states configuration of graph representation A and its comparison to the results of graph representation B with the same hyperparameter settings (left). The number of times each model was achieved by the agent over 10 independent trials for the optimal distinct states configuration of graph representation B and its comparison to the results of graph representation A with the same hyperparameter settings (right).}
    \label{fig:histo_chiral}
\end{figure}

\begin{figure}
    \centering
    \includegraphics[scale=0.45]{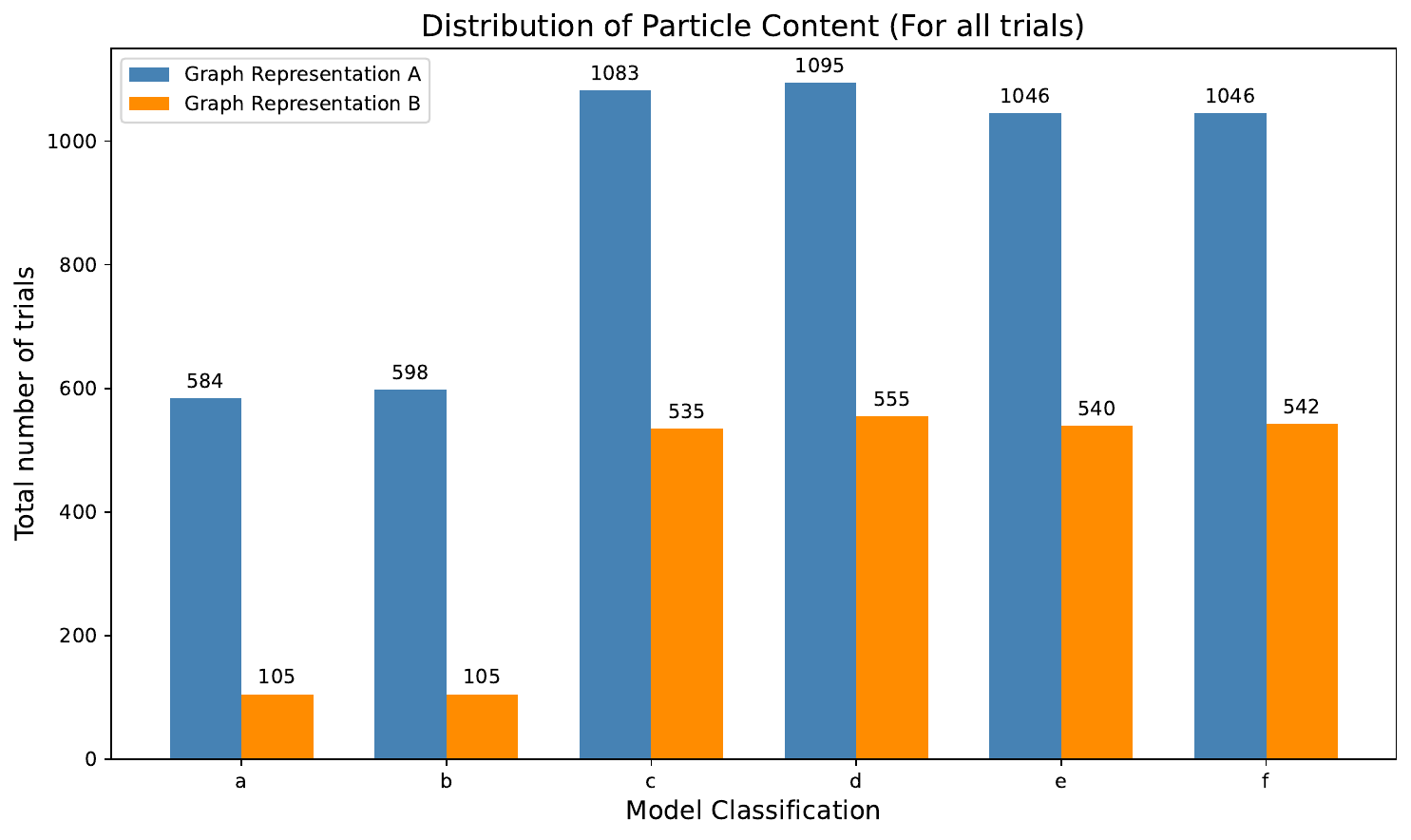}
    \caption{The number of times each model was achieved by the agent over all independent trials with different hyperparameter configurations we probed, for graph representations A and B.}
    \label{fig:histo_chiral_total}
\end{figure}

The fact that graph representation A outperforms B is not surprising, as A shares weights between related parameters of isospin singlet and isospin doublet leptons, such that the agent has more information and fewer separate weights to learn. To see this, we plot the total number of trials (out of 10) in which each model appears in both representations for the hyperparameter setting that returns the optimal number of distinct states for A (left subplot) and B (right subplot) in Figure~\ref{fig:histo_chiral}. Looking at the optimal configuration for graph representation B, the agent of B has a clear preference for models `c'-`f', while graph representation A's agent does not. Meanwhile the difference is somewhat slighter in the optimal configuration of graph representation A, but the agent of B still somewhat favors `c'-`f'. To further justify this finding, we show the total number of times our agents produced `a' and `b' terminal states compared to `c'-`f' ones over all trials with all different hyperaparameter configurations probed with graph representation A and graph representation B in Figure~\ref{fig:histo_chiral_total}. Clearly, both the A and B agents favor `c'-`f', but the difference is much significant in graph representation B. 
Recalling the particle content for these models in Table~\ref{tab:distinct_chiral}, models `c'--`f' consist of one charged and one uncharged particle under the dark gauge group, but with the same electroweak representation, while models `a' and `b' consists of one doublet and one singlet that are both charged under the dark gauge group. This discrepancy between the two graph representations in the favoritism of specific models can likely be explained by the sharing of weights between isospin doublets and singlets in graph representation A. Unlike in graph representation B, the agent does not need to learn the weight for these two types of particles independently, which helps to better identify the possible connections between them.

From all the discussions and results presented with varying factors and hyperparameter settings, we conclude that the performance of the agent is strongly influenced by the graph representation and the entropy coefficient $\beta$. If a suitable graph representation that contains appropriate amount of information about the underlying physics is chosen, the agent can achieve good performance with little dependence on the other hyperparameter settings that we have probed. In this case, we can see that agent with graph representation A nearly always outperforms agent with graph representation B, regardless of the hyperparameters chosen, including regarding the choices of environment or agent architecture, such as the implementation of reward normalization or skip layers. These additional settings might help improve performance, but only with subleading effects in our study. In the case of the entropy parameter $\beta$, it is reasonably straightforward to tune this parameter via a simple scan over various values. However, choosing an appropriate graph representation for a class of BSM models is, as we have discussed in Section \ref{sec:graph}, somewhat nontrivial and involves discretionary choices informed by the physics of the system. That said, in spite of the graph representation's apparent primacy, it is worth mentioning that our suboptimal graph representation (B) still successfully identified a multitude of terminal states and was capable of identifying all 6 distinct models over the course of 10 trials for multiple hyperparameter configurations. This suggests that while an effective choice of graph representation can provide a significant benefit to the agent's performance, the procedure we have outlined can achieve success without necessarily finding an optimal representation, at least for the simple model building task we have analyzed in this paper.

\section{Discussion and Conclusions}\label{sec:conclusion}

In this work, we have aimed to communicate several key conclusions. First, we argued that for machine learning tasks across BSM actions, graphs represent a natural and well-suited format for expressing almost arbitrary BSM theories-- to that end, we presented a generic recipe for developing a graph grammar for these BSM theories. Then, as a case study for this technology, we then employed such a graph grammar in developing a reinforcement learning scan of a BSM parameter space. In contrast to previous studies of reinforcement learning scans of BSM parameter spaces, such as those of \cite{Harvey:2021oue,Nishimura:2020nre}, we found that we could leverage the graph representation of BSM models in order to perform a learning task across a space with an indefinite (and in principle arbitrarily large) particle content and therefore an indefinite parameter space dimensionality. In our study, we restricted ourselves to a particular comparatively simple class of BSM models featuring vector-like leptons which may or may not be charged under a dark $U(1)$  symmetry, inspired by models of sub-GeV vector portal/kinetic mixing dark matter and associated portal matter phenomenology. By evaluating models based on their log-likelihood difference with the SM, the agent was capable of generating both known and novel constructions which addressed the most significant experimental tension with the SM in our input data, which is the anomalous magnetic moment of the muon. The nature of our results suggests some significant inferences with implications to both reinforcement learning scans of BSM model spaces, and the use of graphs to represent BSM models in general machine learning tasks. To start, we shall restrict our discussion to the case of reinforcement learning scans, and then conclude this Section with a discussion of the lessons we might transfer to other learning tasks.

% While our use case of the graph technology represents an application in a somewhat simple model-building paradigm, we note that there are a variety of conclusions that we might infer from our results, both for the task of using reinforcement learning as a tool for exploring BSM parameter spaces, and for the use of the graph technology we have outlined here for learning tasks over BSM theories more broadly.
By considering a variety of training hyperparameters in our analysis, we can comment on various configurations that appear to have significant influence on the agent's performance, and conjecture some suggested ``best practices'' when applied to more complicated learning tasks. We note that consistent with similar tasks in \cite{Dersy:2022bym,2021arXiv210703961Z}, applying intermediate rewards based on \emph{improving} a score over the maximum that has been achieved over an episode is highly effective, but the sensitivity of the agent's performance to the particular form of such a reward function is limited. Furthermore, we found that among the hyperparameters that we have tuned, the entropy regularization constant $\beta$ and the graph grammar used to represent a BSM theory have the most unambiguous effect on the quality of a scan's results, while the other hyperparameters that we consider have considerably more limited effects. In our case, we found that performance was generally superior for our graph representation A, which combined features (mass and SM coupling parameters of the electroweak doublet and singlet vector-like leptons) which were kept separate in our graph representation B, permitting greater transfer learning between features. Meanwhile, the hyperparameter $\beta$ had an enormous effect on the number and diversity of viable models that an agent would produce, simply by parameterizing a scan's trade-off between exploration and exploitation. Given limited computing power, then, we can conjecture that best practices would be to scan over various values of $\beta$, and design a graph grammar that uses the same node features for as many parameters subject to similar rules as possible, while leaving other training hyperparameters fixed. Among models that the reinforcement learning agent generates, we find that the continuous parameters that it samples will generally be subject to nontrivial biases embedded in the design of the agent and the reinforcement learning environment, suggesting that this technique is perhaps ill-suited to probing specific models' parameter spaces. However, it excels at identifying viable particle content, suggesting that its outputs can be readily used as inspiration for a model-builder to consider a wider diversity of constructions.

We emphasize that although our particular reinforcement learning scan explores a simple model, the generalization of the techniques to scan virtually \emph{any} BSM theory space is straightforward, and requires solely the creation of a suitable graph grammar and learning environment. A reader may be concerned that the precise scaling of the efficacy of these techniques to more complicated models remains unclear, and as the model space that is considered in this study is quite simple, this technique may fail for more complex constructions. However, we note that while indeed the model space we have considered is simple, we introduced a significant degree of additional complexity by allowing the BSM vector-like leptons to take on generic couplings to all SM flavors, producing a parameter space which is difficult to efficiently explore and evinces a level of complexity perhaps more in keeping with more complex realistic use cases. Furthermore, considerable efficiency gains in this technique could likely be achieved by combining reinforcement learning with more conventional Monte Carlo scans, in the manner already employed in the studies of \cite{Harvey:2021oue,Nishimura:2020nre}. In particular, we have found that the agent is effective at identifying an appropriate order of magnitude of various parameters, but can struggle with precisely recreating certain predictions. By using the reinforcement learning agent to identify solely the particle content and the orders of magnitude of different parameters, and then doing a conventional Monte Carlo maximum-likelihood fit based on these values, we might dramatically improve the performance of this scanning technique.

We note that even as is, the agent that we have created here represents a critical step in the development of a complete artificial model builder using reinforcement learning. In particular, the agent is capable of proposing models with an arbitrary particle content, rather than performing a scan over a pre-specified space. If the construction of an appropriate graph grammar can be automated, there is no underlying reason that a tool which, given a theorist-specified gauge group and a set of possible representations of particles under it, can automatically explore the theory space defined by those specifications and identify promising models. With refinements, such as incorporating Monte Carlo scans to tune the proposed model's continuous parameters, such a tool has the potential to be highly effective in exploring highly complicated model spaces and generate viable models at a far greater rate than a human theorist.

More broadly, we wish to highlight the utility of mathematical graphs in representing BSM models for a variety of machine learning tasks beyond a reinforcement learning scan. The procedure for developing a graph grammar that we have outlined can be applied for developing an input format for a graph neural network trained for \emph{any} learning task over BSM models. Possible applications beyond reinforcement learning could include, for example, supervised training to learn computationally intensive likelihood computations (creating a model which would be applicable over theories with different particle content) or even developing an alternate form of artificial model builder using generative techniques-- specifically a variational graph autoencoder (VGAE) \cite{2016arXiv161107308K}. The generality of the graph grammar can also be extended by leaving more symmetry groups in the model unspecified. By devising a graph grammar based solely on the Poincare group, one could even develop a universal (at least over the space of 4D quantum field theories) graph grammar through which arbitrary models might be represented for learning tasks. In fact, the key difficulty with many of these applications is not the viability of the graph structure, but rather the quantity of training data available. It remains unclear whether, for example, even aggregating all BSM models produced in the literature over the last half century would represent a large enough training set to train a generic generative model builder. From our reinforcement learning study, we can however infer that the efficiency of a learner might be dramatically improved by leveraging an appropriately designed graph grammar, as evidenced by the greater efficiency of our graph representation A, which represented analogous parameters for doublet and singlet vector-like fermions with the same feature. Designing a suitable graph grammar for more data-intensive machine learning tasks is likely to be highly non-trivial, and its feasibility will likely depend on the task in question, but there is no obvious theoretical barrier to doing so.

\section*{Acknowledgements}
We thank R.~Ximenes for helpful and informative discussions. This work was supported by the U.S. Department of Energy under the contract number DE-SC0017647.

\appendix
\section{Details on PPO and H-PPO}\label{appendix:PPO}

This Appendix discusses the mathematical details of the PPO reinforcement learning algorithm, and those of its generalization H-PPO. Leveraging a policy gradient strategy, PPO trains an agent's policy network with a constrained step size (policy updates) to find a balance between efficiency and stability. The goal of policy-gradient strategy is to find parameters $\theta$ that maximize the expected return. In the context of neural network, this parameter $\theta$ is the weights and biases of the neural network that maps actions to probability distributions. The policy will then be tuned so that ``good" actions that maximize the return are sampled more frequently in the future. This goal is achieved by maximizing an objective function:
\begin{align}
    L^{PG}(\theta) =\hat{\mathbb{E}}_\tau \big[ R(\tau)] = \sum_\tau P(\tau;\theta)R(\tau) = \sum_\tau \left[ \prod _{t=0} P(s_{t+1} | s_t,a_t)\pi_\theta(a_t|s_t) \right]R(\tau)  .
\end{align}
where the expected return $\hat{\mathbb{E}}_\tau \big[ R(\tau)]$ is the weighted average of all possible values that the return $R(\tau)$ can get. $P(\tau;\theta)$ is the probability of each possible trajectory $\tau$ as a function of parameter $\theta$. It is determined by the state distribution, namely the possibility of state $s_{t+1}$ getting selected from state $s_t$ if action $a_t$ is selected, and the policy $\pi_\theta(a_t|s_t)$, which is just the probability of the agent selecting action $a_t$ from state $s_t$ given our policy. Instead of calculating the actual gradient of this objective function, which requires calculating the probability of each possible trajectory, a gradient estimation is performed instead with a sample-based estimate. This objective function is also not directly differentiable since it involves differentiation of state distribution that might be unknown. Using the Policy Gradient Theorem, it can be shown that the gradient of the objective function can be re-expressed into this form:
%
% \begin{align}
%     \grad_{\theta} L^{PG}(\theta) =\hat{\mathbb{E}}_t \big[ \grad_{\theta}\log \pi_\theta (a_t | s_t) \hat{A}_t \big].
% \end{align}
\begin{align}
    \grad_{\theta} L^{PG}(\theta) =\hat{\mathbb{E}}_t \big[ \grad_{\theta}\log \pi_\theta (a_t | s_t) R(\tau) \big].
\end{align}
where the expectation $\hat{\mathbb{E}}_t$ is the empirical average over a finite batch of samples.  Now the weight of the policy is updated following the direction of $\grad_{\theta}\log \pi_\theta (a_t | s_t) $; this is the direction of steepest increase of the log probability of selecting action $a_t$ at state $s_t$ scored by the return.

% The quantity $\hat{A}_t$ (which we use to replace the return $R(\tau)$) is the \emph{advantage function}, which is an estimate of the expected accumulated reward for taking action $a_t$ from state $s_t$. Conceptually it measures the extra reward we get if we choose certain action at that state compared to the mean reward we get at that state.

In an actor-only method that depends solely on policy gradient, the gradient estimator may have a large variance. This is due to the fact that the return (long-term cumulative reward) used to score the log-probability of state-action pair can have very different values in a stochastic environment with a stochastic policy. In this case, an actor-critic approach could be employed to mitigate this large variance problem, and could thus lead to faster convergence of the algorithm. To demonstrate this, let's start by introducing a function $V(s_t)$, which is the state value that estimate the expected return of being in state $s_t$:
\begin{align}
 V(s_t) = \hat{\mathbb{E}}_\tau \big[R_t |s_t \big]    
\end{align}
A critic network can, if passed a state $s_t$, be trained to estimate this value function. Then to compute the policy gradient, we can define a state action value $Q(s_t,a_t)$, also known as $Q$-function, which is the expected return of taking action $a_t$ in state $s_t$ under the current policy. If the trajectory of agent interacting with its environment is recorded over $T$ time steps, $Q(s_t,a_t)$ takes the following form:
%
% \begin{align}
%     V_t^{\text{target}} = r_t + \gamma r_{t+1} + \cdots + \gamma^{T-t-1}r_{T-1} + \gamma^{T-t} V(s_T)
% \end{align}
\begin{align}
Q(s_t,a_t) = \hat{\mathbb{E}}_\tau \big[R_t |s_t,a_t \big]  =   r_t + \gamma r_{t+1} + \cdots + \gamma^{T-t-1}r_{T-1} + \gamma^{T-t} V(s_T)
\label{eqn:Q-function}
\end{align}
where $\gamma \leq 1$ is the discount factor that gauge the importance between immediate and future rewards -- if $\gamma$ is small, the agent will prioritize immediate rewards, since far-future rewards will contribute less to $Q$, while if $\gamma$ is large, the agent will be more responsive to longer-term rewards. $V(s_T)$ is the state value estimated at the terminal state $s_T$ of the trajectory. Introducing $Q$-function of this form with the discount factor enables us to look a finite number of steps ahead and still anticipate long-term rewards, yet reduce variance by down-weighting future rewards that corresponds to delayed effects (at the expense of introducing bias into our estimator of cumulative rewards over an episode, which increases with smaller $\gamma$) \cite{2015arXiv150602438S}. We can reduce the variance even further by introducing the advantage function, which is a measure of the relative advantage for taking action $a_t$ at state $s_t$ compared to taking other possible actions at that state. This relative advantage is quantified by measuring the extra reward we get if we choose certain action at that state compared to the mean reward we get at that state: 
\begin{align}\label{eq:advantage}
    A(s_t,a_t) =  Q(s_t,a_t) - V(s_t),
\end{align}
Since this is the difference between the estimated target value for a state-action pair and the value function of the state itself, $A(s_t,a_t)$ measure the benefit of taking a particular action $a_t$ at state $s_t$ over the expected value of following the current policy from state $s_t$. Subtracting out $V(s_t)$ does not alter the maximum of the expected return since it is a constant as a function of the action $a_t$.

In the actor-critic approach, both the action network and critic network are trained together. The critic learns to approximate the advantage function (or more precisely, the value function) which is then used to update the actor's policy parameters. Conceptually, the actor network learns the policy that controls how our agent acts and the critic network learns the value function to assist the policy update by evaluating how good the action taken is. Because smaller sample trajectories rather than an entire episode, are used to evaluate the expected rewards, the variance of these expectation values is dramatically reduced. This approach also allows the agent to learn tasks with longer, or even infinite-length, episodes.

PPO is essentially an architecture with an actor-critic mechanism that also improves the agent's training stability by controlling the step size of the policy update. With appropriately-sized small policy updates during the training, we can avoid having step size that is too small that slows down the training significantly, while also avoid having a step size that is so large that the policy might be updated to a state so far from optimality that convergence to a good policy again might be severely delayed or prevented entirely. PPO possesses several advantages which we can leverage in this task: It is capable of learning policies with a long time horizon, its stability allows agents to be trained for multiple epochs on a single episode's training data, and it can readily be adapted to train simultaneously on multiple episodes in parallel, which is essential to maximize the exploration of our space of theories.

Using gradient ascent, PPO maximizes the overall objective functions for policy and value functions, which have the same local extremum as the parameter $\theta$, supplemented by an entropy bonus term to ensure sufficient exploration: 
\begin{align}
    L = L_{PPO} - L_{\textrm{critic}} + \beta S[\pi_\theta]
\end{align}
where $L_{PPO}$ denotes the clipped surrogate objective function used to train the policy network. $L_{\textrm{critic}}$ is the loss function for training the critic network and $S[\pi_\theta]$ is the mean entropy of the policy's output distributions over the sampled trajectory, scaled by a hyperparameter $\beta$. $\beta$ plays the role as a regularizer during policy training and also tuning it allows us to adjust the degree of exploration of the state space versus exploitation of a single optimal strategy. The Adam optimizer is used to perform this stochastic gradient ascent. 

The clipped surrogate objective function takes the form
\begin{align}\label{eq:PPO-obj}
    L_{PPO}(\theta) = \hat{\mathbb{E}}_t \bigg[ \textrm{min}( r_t(\theta) \hat{A}_t,  \textrm{clip}(r_t(\theta), 1-\epsilon, 1 + \epsilon) \hat{A}_t) \bigg],\\
    r_t(\theta) = \frac{\pi_\theta (a_t | s_t)}{ \pi_{\theta,\textrm{old}} (a_t | s_t)} \nonumber.
\end{align}
Here, $\pi_\theta (a_t | s_t)$ represents the policy's (given by the set of parameters $\theta$) probability of selecting action $a_t$ given an input state of $s_t$, while $\pi_{\theta, \textrm{old}}(a_t | s_t)$ represents the probability of the same action given the model \emph{before} its current round of training. The ratio function $r_t(\theta)$ denotes the probability ratio between the current and old policy, that tells us whether an action $a_t$ is more or less likely in the current policy than the old one. $\epsilon$ is a hyperparameter that defines the clip range of the probability ratio, so that we can avoid a policy update that is too large: if the current network's probability for a given action differs too much from that of the policy when the action is sampled (which can occur if the agent is trained on a given trajectory for multiple epochs), this ratio is clipped and becomes independent of the policy parameters, and therefore that action becomes irrelevant to training the network further. $\hat{A}_t$ is the advantage estimator.

The objective function of the critic network takes the form
\begin{align}
    L_{\text{critic}}(\theta) = \hat{\mathbb{E}}_t \bigg[ (V_\theta(s_t) -Q(s_t,a_t))^2 \bigg]
\end{align}
where $V_\theta(s_t)$ is the state value output by the critic network and $Q(s_t,a_t)$ is the state action value computed based on the observations throughout a trajectory, as defined in Eq.(~\ref{eqn:Q-function}).

Following the program flow of training described in the main text, with these formulas provided for various objective functions and advantage function, PPO can be trained accordingly. As noted in the main text, in our constructed model our task includes both discrete and continuous actions, so we use the HPPO algorithm \cite{2019arXiv190301344F} to learn the hybrid action space. H-PPO works similarly to conventional PPO, except that the action space is now hierarchically structured, consisted of multiple actors representing policies for various discrete and continuous actions. The actions in our model are also hierarchical, in the sense that certain discrete action (such as which particle to be modified) must necessarily be accompanied by continuous actions that parameterize it (such as the mass and coupling of that particle). In addition to the multiple-actor networks, there is one global critic network that updates the policy parameter $\theta$ of all sub-actor networks. 

To describe the action space in a mathematical way, we follow the method presented in \cite{2019arXiv190301344F}: we have a finite set of discrete action $\mathcal{A}_d = \{a_1, a_2, \cdots , a_k\}$ and for each action $a \in \mathcal{A}_d$, there are multiple sets of real-valued continuous parameters $x^i \in \mathcal{X}_a^i $ associated with $a$ to be specified. A complete action $(a,x^1,x^2,\cdots,x^i)$ is then composed of a discrete action and continuous parameters to be executed with that action. The whole hybrid action space $\mathcal{A}$ then takes the form:
\begin{align}
    \mathcal{A} = \bigcup_{\substack{a\in \mathcal{A}_d \\i \in \{1,2,\dots,n\} }}\{(a,x^1,\cdots,x^n)| x^i \in \mathcal{X}_a^i\}.
\end{align}
To deal with this action space, there are two actors: A discrete actor, which will learn a policy over the discrete actions $\mathcal{A}_d$, and a continuous actor, which will learn policies over all the continuous parameters in $\mathcal{X}_a^i$ for each discrete action $a$. The two actors' objective functions are evaluated separately according to the PPO objective function following Eq.~(\ref{eq:PPO-obj}), and added together to compute the full H-PPO objective function. Crucially, in H-PPO the network will generate probability distributions over \emph{all} the parameters for all possible actions, even those which aren't ultimately chosen by the top-level discrete action. When computing the objective function, however, only those outputs which are actually used in parameterizing the action at each step are included.
Our action space has a tree structure that describes multiple layers of actions. For example, as discussed in the main text, if we select that a particle's parameters should be modified, we would then have to select which particle should be modified, which parameter of that particle should be modified, and finally what modification we should make to that parameter. In this hierarchical action space, the hybrid actor-critic architecture contains multiple actor networks and one critic network. There is one actor network for each discrete or continuous action-selection sub-problems. The critic network follows the same virtue as the one in PPO. Each of the actor networks generates either a stochastic discrete policy or a stochastic continuous policy, and these actors are updated separately as independent policies using PPO during training. 

 With the hybrid actor-critic architecture, H-PPO uses PPO as the policy optimization method for all of its discrete policy $\pi_{\theta_d}$ and continuous policy $\pi_{\theta_c}$. The policies are updated separately and independently for each actor by maximizing their individual clipped surrogate objective function. For each actor with discrete policy $\pi_{\theta_d}^i$, with $i$ denotes the $i$-th actor, the objective takes the form
\begin{align}
    L_{PPO,d}^{i}(\theta_{d,i}) = \hat{\mathbb{E}}_t \bigg[ \textrm{min}( r_t^{d,i}(\theta_{d,i}) \hat{A}_t,  \textrm{clip}(r_t^{d,i}(\theta_{d,i}), 1-\epsilon, 1 + \epsilon) \hat{A}_t) \bigg],\\
    r_t(\theta_{d,i}) = \frac{\pi^i_{\theta_{d,i}} (a_t | s_t)}{ \pi^i_{\theta_{d,i},\textrm{old}} (a_t | s_t)}.
\end{align}
Whereas for each actor $i$ with continuous policy $\pi_{\theta_c}^i$, the objective function is:
\begin{align}
    L_{PPO,c}^{i}(\theta_{c,i}) = \hat{\mathbb{E}}_t \bigg[ \textrm{min}( r_t^{c,i}(\theta_{c,i}) \hat{A}_t,  \textrm{clip}(r_t^{c,i}(\theta_{c,i}), 1-\epsilon, 1 + \epsilon) \hat{A}_t) \bigg],\\
    r_t(\theta_{c,i}) = \frac{\pi^i_{\theta_{c,i}} (x^i_t | s_t)}{ \pi^i_{\theta_{c,i},\textrm{old}} (x^i_t | s_t)}.
\end{align}
Here we can see that even if multiple discrete and continuous policies work together to specify a complete action, their objective functions are independent and their policies are viewed as separate distributions.

\section{Additional Experimental Parameters}\label{appendix:experiment-details}

In this Appendix, we summarize a number of details of our experimental procedure that are not essential to understanding the main text, but in the interest of completeness must be reported.

To begin, several additional specifications must be made in order to precisely describe the architecture of the neural network. First, we must specify the operation that we employ for our graph message-passing layers. Here, we select the graph convolution proposed in \cite{2018arXiv181002244M}, which mimics the form of our simple linear example message-passing layer given in Eq.(\ref{eq:message-passing}) in Section \ref{sec:graph}, with no edge features. To accelerate training, we include a GraphNorm layer after each message-passing layer, with the GraphNorm operation defined in \cite{2020arXiv200903294C}, ReLU activation is then applied to the nodes after the GraphNorm layer to introduce nonlinearities. To accommodate our graphs' heterogeneous structure, we aggregate the results of convolutions with different node types by summing each output before applying the activation function.

For the multilayer perceptron (MLP) neural networks appearing in our agent's structure, we limit ourselves to stacks of conventional dense layers interspersed with batch normalization and ReLU nonlinearities-- in keeping with the default parameters of the MLP class in Pytorch Geometric. Finally, we specify our neurons completely by defining that all hidden layers will have 64 output channels (up to slight modifications from concatenating vectors with additional state information or feature vectors from skip layers).

\begin{figure}
    \centering
    \begin{tikzpicture}
        \node at (0,0) [input, fill=blue!20] (inp-id) {Input};
        \node at (0, -2) [graph, fill=red!20] (emb-id) {Embedding};
        \draw[->] (inp-id) -- (emb-id);
        \node at (-2, -4) [graph, fill=red!20] (choice-id) {Node Choice};
        \draw[->] (emb-id) -- (choice-id);
        \node at (-4, -3.5) [left, mlp, fill=gray!20] (rem-choice-id) {Removal Probability};
        \node at (-4, -4.5) [left, mlp, fill=gray!20] (mod-choice-id) {Modification Probability};
        \draw[->] (choice-id) -- (rem-choice-id);
        \draw[->] (choice-id) -- (mod-choice-id);
        \node at (2, -4) [graph, fill=red!20] (mod-id) {Modification};
        \draw[->] (emb-id) -- (mod-id);
        \node at (4, -3) [right, mlp, fill=gray!20] (mod-sel-id) {Exponential or Continuous};
        \node at (4, -4) [right, mlp, fill=gray!20] (mod-part-id) {Particle Modifications};
        \node at (4, -5) [right, mlp, fill=gray!20] (mod-coup-id) {Coupling Modifications};
        \draw[->] (mod-id) -- (mod-sel-id);
        \draw[->] (mod-id) -- (mod-part-id);
        \draw[->] (mod-id) -- (mod-coup-id);
        \node at (0, -6) [pooling, fill=green!20] (pool-id) {Pooling};
        \draw[->] (emb-id) -- (pool-id);
        \node at (-2, -8) [left, mlp, fill=gray!20] (add-id) {Added Particles};
        \node at (0, -8) [mlp, fill=gray!20] (crit-id) {Critic};
        \node at (2, -8) [right, mlp, fill=gray!20] (mas-id) {Master Action};
        \draw[->] (pool-id) -- (add-id);
        \draw[->] (pool-id) -- (crit-id);
        \draw[->] (pool-id) -- (mas-id);
    \end{tikzpicture}
    \caption{A diagram depicting the organization of various subnetworks in the agent, and how an input (circle) is processed. In this diagram, rounded rectangles represent stacks of message-passing layers, while sharp-edged rectangles denote multilayer perceptrons. Multilayer perceptrons directly connected to message-passing layers are applied to all nodes within the output graph, while those attached to the diamond-shaped ``pooling'' node are applied only to the finite-dimensional representation of the graph as a whole, obtained from pooling the feature vectors of all nodes. }
    \label{fig:agent-diagram}
\end{figure}
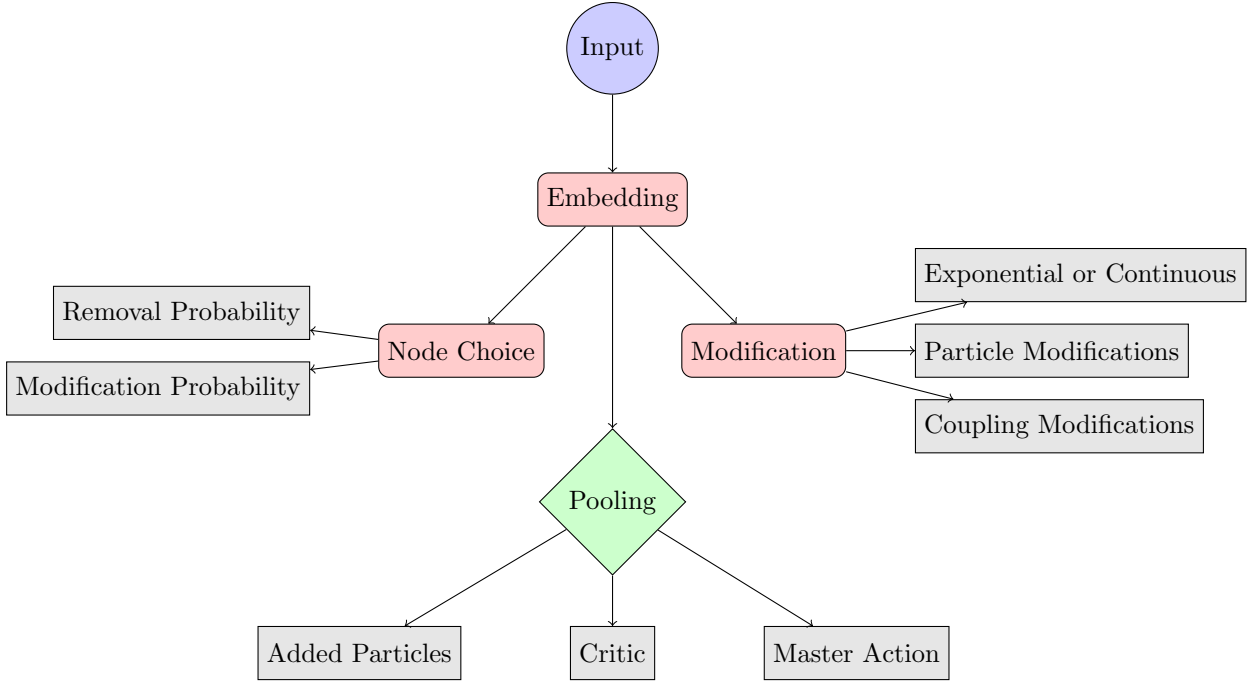

We must also provide some further specification of how our agent is constructed from the modules described in Figures \ref{fig:node-action} and \ref{fig:graph-action}. In Figure \ref{fig:agent-diagram}, we depict a diagram of the organization of the different stacks of message-passing layers and MLP networks that compose the agent in its entirety. To complete this description, we summarize the function of each network depicted in the diagram, by its label:

\begin{itemize}
    \item \textbf{Embedding:} A stack of 4 message-passing layers which will take the underlying state graph and learn a representation which is shared among all subsequent agent modules.
    \item \textbf{Node Choice:} A stack of 3 message-passing layers which performs further transformations of the graph after the embedding layers, the output of which is used to generate probabilities that a given node will be either removed from the graph or modified, in the event that the corresponding top-level action is selected by the agent.
    \item \textbf{Modification:} A stack of 3 message-passing layers which performs further transformations of the graph after the embedding layers, the output of which is used to generate probability distributions over the parameters governing modifications that the agent might make to particles or couplings in the model.
    \item \textbf{Master Action:} A 3-layer multilayer perceptron that will output log-probabilities for different top-level actions that the agent can take. In graph representation A, these are: Add a particle, remove a particle, modify a particle's features, or modify a Yukawa coupling's features. In graph representation B, these are: Add a vector-like doublet, add a vector-like singlet, remove a vector-like doublet, remove a vector-like singlet, modify a doublet's feature vector, modify a singlet's feature vector, modify a Yukawa doublet-singlet Yukawa coupling node, modify a doublet-doublet Yukawa coupling node, and modify a singlet-singlet Yukawa coupling node.
    \item \textbf{Critic:} A 3-layer multilayer perceptron that will output an estimate for the state value function, as described in Appendix \ref{appendix:PPO}.
    \item \textbf{Added Particles:} A 3-layer multilayer perceptron, the outputs of which govern the probability distribution for parameters of a newly-added particle in the model. In the case of graph representation A, this consists of four pairs of means and variances parameterizing Gaussian distributions over the four continuous features of a particle, followed by a series of outputs which will represent log-probabilities for the different discrete particle parameters, such as the electroweak representation, dark charge, and the exponential parameters of the various Yukawa couplings in scientific notation. In graph representation B, the outputs are similar, except that there is no discrete parameter for determining the electroweak representation of the new particle, and a full set of parameters for new doublets and singlets are considered separately.
    \item \textbf{Removal Probability:} A 3-layer multilayer perceptron that will output the log-probability that a given node should be removed from the model, if the top-level action for particle removal is selected. Note that the agent can only directly remove particles, not couplings, from the model, so this neural network is only applied to particle nodes. In representation A, this element is a single multi-layer perceptron, while in representation B there are two different networks, one for electroweak doublets and the other for electroweak singlets.
    \item \textbf{Modification \; Probability:} A 3-layer multilayer perceptron that will output the log-probability that a given node should have its feature vector modified, if the top-level action for particle or coupling removal is selected. In representation A, this element consists of two networks-- one for particles and one for couplings, but in representation B, this element consists of five networks, for doublet particles, singlet particles, doublet-singlet Yukawa couplings, doublet-doublet Yukawa couplings, and singlet-singlet Yukawa couplings.
    \item \textbf{Exponential or Continuous:} A 3-layer multilayer perceptron that will output the log-probability that if a given node will have its feature vector modified, the agent will modify the $O(1)$ continuous coefficients of the model's feature vectors, instead of their discrete exponential parameters in scientific notation-- in short, the agent must decide whether to modify the precise value or the order of magnitude of the parameters that are represented in scientific notation in the model state. The number of different networks composing this element are the same in the two graph representations as the Modification Probability element: Two for A, and five for B.
    \item \textbf{Particle Modifications:} A 3-layer multilayer perceptron that will output the parameters governing probability distributions over particle modification actions. In graph representation A, these outputs consist of four sets of means and variances for the four different continuous features of a particle, log-probabilities governing which continuous parameter should be modified, and finally log-probabilities for each possible discrete modification of the exponential parameters of the particle Yukawa couplings. In graph representation B, the construction is nearly identical, but there are separate outputs for the parameters governing doublet and singlet particles.
    \item \textbf{Coupling Modifications:} A 3-layer multilayer perceptron that will output the parameters governing probability distributions over coupling modification actions. In graph representation A, these outputs consist of two sets of means and variances for the two different continuous features of a coupling node, log-probabilities governing which continuous parameter should be modified, and finally log-probabilities for each possible discrete modification of the exponential parameters of the Yukawa coupling features. In graph representation B, the construction is nearly identical, but there are separate outputs for the parameters governing doublet-singlet, doublet-doublet, and singlet-singlet Yukawa coupling nodes.
\end{itemize}

We conclude this Appendix with a summary of several hyperparameters in our reinforcement learning scan that are not essential for following our analysis, but we should specify for completeness. These are:

\begin{itemize}
    \item \textbf{The Discount Factor:} We specify the discount factor $\gamma$, appearing in our $Q$-value definition in Eq.(\ref{eqn:Q-function}), as $\gamma = 0.999$.
    \item \textbf{Generalized Advantage Estimation:} To reduce variance, we employ generalized advantage estimation (GAE), described in \cite{2015arXiv150602438S}. Following the notation of that work, our parameter $\lambda$ (as defined there) is set at 0.9.
    \item \textbf{Learning Rate:} For our neural network training, we employ an Adam optimizer with a learning rate of $10^{-3}$.
\end{itemize}

\section{Formulas for Calculation of Observables}\label{appendix:Observables}

In this Appendix, we discuss the computation of the various precision observables listed in Section \ref{sec:observables} in greater detail. Our numerical analysis of a given model of the class described in Section \ref{sec:model} begins with the creation of the model's charged-lepton mass matrix $\mathcal{M}$, initially computed from the parameters in the action of Eq.(\ref{eq:model-action}). This mass matrix takes the form,
\begin{align}\label{eq:model-mass-mat}
    &\mathcal{M} \sim \begin{pmatrix}
        \mathbf{y}_{SM} v & \mathbf{0} & \boldsymbol{\lambda}^\pm_L v_D & \mathbf{y}^0_E v & \mathbf{0}\\
        \mathbf{y}^0_L v & \mathbf{M}^{L,0} & \boldsymbol{\lambda}^{0 \pm}_L v_D & \mathbf{y}^0_{LE} v & \mathbf{0}\\
        \mathbf{0} & \boldsymbol{\lambda}^{\pm 0}_L v_D & \mathbf{M}^{L, \pm} & \mathbf{0} & \mathbf{y}^{\pm}_{LE} v\\
        \mathbf{0} & \mathbf{y}^0_{EL} v & \mathbf{0} & \mathbf{M}^{E,0} & \boldsymbol{\lambda}^{0 \pm}_E v_D\\
        \boldsymbol{\lambda}^{\pm}_{E} v_D & \mathbf{0} & \mathbf{y}^{\pm}_{EL} v & \boldsymbol{\lambda}^{\pm 0}_E v_D & \mathbf{M}^{E,\pm}
    \end{pmatrix},\\
    &\mathbf{y}_{SM} = diag(y_e, y_\mu, y_\tau) \nonumber,
\end{align}
where we use the bolded versions of the symbols in Eq.(\ref{eq:model-action}) to denote the corresponding matrices, and  suppress factors of $1/\sqrt{2}$ associated with each factor of $v_D$ and $v$ for clarity. When generating $\mathcal{M}$ from a set of input vector-like lepton masses and couplings, we specify the SM lepton Yukawa couplings $y_{e,\mu,\tau}$ such that we recreate the experimentally observed lepton masses $(m_{e}, m_{\mu}, m_{\tau}) \approx (511 \; \textrm{keV}, 105 \; \textrm{MeV}, 1.777 \; \textrm{GeV})$ using numerical root-finding with the eigenvalues of the full model's mass matrix.

Using the mass matrix $\mathcal{M}$, we then numerically find the mass eigenstates and eigenvalues for each charged lepton in the model, which we can use to bi-diagonalize $\mathcal{M}$ in the usual manner, that is, identify unitary matrices $U_L$ and $U_R$ such that
\begin{align}
    \mathcal{M}_D = U_L^\dagger \mathcal{M} U_R,
\end{align}
where $\mathcal{M}_D$ is a diagonal matrix. Using $U_{L,R}$, we can then compute the tree-level couplings featuring the charged-lepton mass eigenstates.\footnote{Strictly speaking, computing tree-level $W$ boson couplings requires finding the neutral lepton mass eigenstates as well-- however the neutrino rotation matrices relevant to our analysis here will only be nontrivial when at least one dark-charged electroweak doublet (that is, $L_\pm$) is included in the model, at which point processes associated with the sub-GeV scale dark Higgs and dark photon will dominate new physics contributions over any processes associated with $W$ boson couplings (namely, adjustments to the electroweak precision parameters and the contribution of $W$ boson loops to anomalous magnetic moments and lepton flavor violating couplings). So for simplicity, we shall assume that the the neutral lepton mass matrix is already diagonal, introducing mild numerical inaccuracies into some of our numerical computations in scenarios in which these computations are already subdominant contributions to our results.} To standardize the notation we shall use throughout this Appendix, we write these mass-eigenstate interactions which will be relevant for computing our physical observables in Eq.(\ref{eq:num-couplings}). Denoting the vector of electrically charged lepton mass eigenstates (including both predominantly SM and predominantly vector-like leptons) as $\vec{l}$ and the vector of electrically neutral lepton mass eigenstates as $\vec{\nu}$, we write
\begin{align}\label{eq:num-couplings}
    \mathcal{L} &\supset \overline{l}_i \slashed{Z} [(\mathbf{g}^{Z}_L)_{i j} P_L + (\mathbf{g}^{Z}_R)_{i j} P_R)] l_j + \overline{l}_i \slashed{A}_D [(\mathbf{g}^{D}_L)_{i j} P_L + (\mathbf{g}^{D}_R)_{i j} P_R)] l_j \\
    &- \frac{h}{\sqrt{2}} \overline{l}_i (\mathbf{y}_h)_{i j} l_j - \frac{h_D}{\sqrt{2}} \overline{l}_i (\mathbf{y}_D)_{i j} l_j + \overline{\nu}_i \slashed{W}^+ [(\mathbf{g}^{W}_L)_{i j} P_L + (\mathbf{g}^{W}_R)_{i j} P_R] l_j + h.c. \nonumber
\end{align}
where $A_D$ and $h_D$ refer to the dark photon and dark Higgs, while $h$, $Z$, and $W$ refer to the SM Higgs, $Z$, and $W$ bosons, respectively. In Table \ref{tab:inputs}, we provide the numerical parameters we used in our likelihood calculations for the experiments of Section \ref{sec:experiments}.

\begin{table}[t!]
    \centering
    \begin{tabular}{| c | c | c |}
    \hline
    Input & Definition & Value\\
    \hline
    $M_Z$ & $Z$ boson mass & $91.1875 \; \textrm{GeV}$\\
    \hline
    $G_F$ & Fermi constant & $1.16637 \times 10^{-5} \; \textrm{GeV}^{-2}$\\
    \hline
    $\alpha_{\textrm{em}}$ & Fine structure constant & $7.755 \times 10^{-3}$\\
    \hline
    $m_e$ & electron mass & $511 \; \textrm{keV}$\\
    \hline
    $m_\mu$ & muon mass & $105 \; \textrm{MeV}$\\
    \hline
    $m_\tau$ & $\tau$ mass & $1.777 \; \textrm{GeV}$\\
    \hline
    $g_D$ & Dark $U(1)$ gauge coupling & 0.3\\
    \hline
    $v_D$ & Dark Higgs vev & $1.0 \; \textrm{GeV}$\\
    \hline
    $m_{hD}$ & Dark Higgs mass & $1.0 \; \textrm{GeV}$\\
    \hline
    \end{tabular}
    \caption{The parameters that are inserted into our numerical calculations. The values for $M_Z$, $G_F$, and $\alpha_{\textrm{em}}$ are extracted from \cite{Workman:2022ynf}, while the precise values of $v_D$, $g_D$, and $m_{hD}$ will have a negligible effect on any numerical results and are specified merely for definiteness.}
    \label{tab:inputs}
\end{table}
We begin our discussion of specific observable calculations with the electroweak precision observables $m_W$, $\Gamma(W \rightarrow l \nu)$, $R_l$, $A_l$, and $A_{FB}^{0,l}$. Our treatment resembles that of \cite{Crivellin:2020ebi}, with the exception that our calculations are done using the tree-level modifications of the relevant SM couplings, as computed in the procedure outlined above, rather than expressing the vector-like lepton contributions in terms of effective dimension-6 operators. The electroweak observables are sensitive to tree-level modifications of couplings to $W$ and $Z$ bosons to SM particles-- for clarity, we write these coupling modifications as
\begin{align}\label{eq:EWPO-coupling-mods}
    (\delta \mathbf{g}^{Z,W}_{L})_{i j} = (\mathbf{g}^{Z,W}_{L})_{ij}/ g^{Z,W}_{L,SM} - \delta_{ij}, \; \; (\delta \mathbf{g}^{Z}_{R})_{i j} = (\mathbf{g}^{Z}_{R})_{ij}/ g^{Z}_{R,SM} - \delta_{ij},\\
    g^{Z}_{L,SM} \equiv \frac{g_2}{c_W} \bigg( -\frac{1}{2} + s_W^2 \bigg), \;\; g^{Z}_{R,SM} \equiv \frac{g_2}{c_W} s_W^2, \;\; g^{W}_{L, SM} \equiv \frac{g_2}{\sqrt{2}}, \nonumber
\end{align}
where $g_2$ is the gauge coupling constant for the $SU(2)_L$ gauge group of the Standard Model and $c_W$ and $s_W$ represent the cosine and sine of the usual Weinberg angle, respectively.
The indices $i$ and $j$ are limited to only include the three lightest electrically charged or neutral fermions, corresponding to the SM-like charged leptons and neutrinos, since these couplings are the only ones that are relevant for the electroweak precision observables.
In the electroweak fit, we take
\begin{align}
    M_Z = 91.1875 \; \textrm{GeV} \nonumber\\
    G_F = 1.16637 \times 10^{-5} \; \textrm{GeV}^{-2}\\
    \alpha_{\textrm{em}} = 7.755 \times 10^{-3} \nonumber
\end{align}
as inputs, and use the coupling modifications of Eq.(\ref{eq:EWPO-coupling-mods}) to compute changes in the relationship between these quantities and the electroweak observables listed in Table \ref{tab:observables}. We begin by noting that the Fermi constant is measured using the partial width of the process $\mu \rightarrow e \nu \nu$, which will be subject to modifications to the $W$ boson couplings to the electron and muon from mixing with vector-like leptons. Hence, the measured Fermi constant $G_F$ and the constant appearing in the Lagrangian (which we denote by $G_F^0$) will differ by a factor of
\begin{align}
    G_F = G_F^0 (1 + \delta G_F), \;\; \delta G_F \equiv \delta (\mathbf{g}^W_L)_{e e} + \delta (\mathbf{g}^W_L)_{\mu \mu}.
\end{align}
In turn, this leads to corrections to the predicted $W$ boson mass, as well as further modification to the effective $Z$ boson couplings beyond what appears in $\delta \mathbf{g}^{Z}_{L,R}$. We find that to leading order, the $W$ boson mass is corrected as
\begin{align}\label{eq:mw-correction}
    M_W \approx M_{W,SM} \bigg( 1 - \frac{s_W^2}{2 (c_W^2 - s_W^2)} \delta G_F \bigg),
\end{align}
while the leptonic asymmetry parameters $A_f$ and $A_{FB}^{(0,f)}$, where $f = e,\mu,\tau$, undergo the corrections
\begin{align}\label{eq:Af-correction}
    A_f \approx A_{f,SM} \bigg[ 1 + \delta A_f \bigg], \;\; \delta A_f = \bigg( A_{f,SM}- \frac{1}{A_{f,SM}} \bigg) \bigg(\delta (\mathbf{g}^{Z}_L)_{ff} - \delta (\mathbf{g}^{Z}_R)_{ff} - \frac{c_W^2}{(c_W^2 - s_W^2)^2} \delta G_F \bigg),\\
    A_{FB}^{(0,f)} \approx A_{FB,SM}^{(0,f} (1 + \delta A_{FB}^{(0,f)}), \;\; \delta A_{FB}^{(0,f)} = (1 + \delta A_f + \delta A_e).
\end{align}
The observable $R_f$, defined as the ratio of the hadronic partial width of the $Z$ to the leptonic partial width to lepton species $f$, undergoes two corrections. One is from the indirect modification of $\delta G_F$ to the hadronic partial decay width, which we denote using the variable $\delta R_{f}^{(h)}$, and another is from the modification to the leptonic decay width, denoted as $\delta R_{f}^{(f)}$:
\begin{align}\label{eq:Rl-correction}
    &R_f = R_{f,SM} (1 + \delta R_{f}^{(h)} + \delta R_{f}^{(f)}), \nonumber\\
    &\delta R_{f}^{(h)} \equiv \bigg( \frac{6 g^{d}_R (g^d_L + g^d_R) + 4 g^u_R (g^u_L + g^u_R)}{3 ((g^d_L)^2 + (g^d_R)^2) + 2 ((g^u_L)^2 + (g^u_R)^2)} \bigg) \frac{(g^l_R - 2 g^l_L)}{2 g^l_L} \delta G_F,\\
    &\delta R_{f}^{(f)} \equiv - \frac{2}{(g^l_L)^2 + (g^l_R)^2} \bigg( (g^l_L)^2 \delta (\mathbf{g}^Z_L)_{ff} + (g^l_R)^2 \delta (\mathbf{g}^Z_R)_{ff} + \frac{g^l_R}{2 g^l_L} (g^l_L + g^l_R)(g^l_R - 2 g^l_L) \delta G_F \bigg), \nonumber\\
    &g^{u,d,l}_L \equiv (T_3)_{u,d,l} - Q_{u,d,l} s_W^2, \nonumber\\
\end{align}
where $T_3$ denotes the weak isospin quantum number and $Q$ denotes the electric charge (in units of $|e|$), and the index $l$ refers to the (universally shared) quantum numbers for any of the three generations of charged leptons.

The final electroweak precision observables that we consider are the branching fractions of the $W$ to each different SM lepton generation. All partial $W$ widths are rescaled uniformly due to the correction to $G_F$, leaving (up to flavor-specific corrections to the total width that will be subleading, due to the fact that each $W \rightarrow l \overline{\nu}$ branching fraction is only $\sim 10\%$)
\begin{align}\label{eq:BR-W-correction}
    BR(W \rightarrow f \overline{\nu}) \approx BR(W \rightarrow f \overline{\nu})_{SM} (1 + 2 \delta (\mathbf{g}^Z_L)_{ff}).
\end{align}
Using the tree-level proportional corrections given in Eqs.(\ref{eq:mw-correction}-\ref{eq:BR-W-correction}), we can then approximate a model's predictions for each observable by extracting the best-fit SM predictions from the electroweak precision fit of \cite{Workman:2022ynf}.

Apart from the electroweak precision data, the final tree-level effect that we compute when estimating model likelihood stems from the adjustment to the $\mu$ and $\tau$ lepton Yukawa couplings from their SM expectations, given as $\kappa_\mu$ and $\kappa_\tau$ in the rescaling formalism of \cite{LHCHiggsCrossSectionWorkingGroup:2013rie}, where $\kappa_{\tau,\mu} = 1$ in the Standard Model. In our construction, the SM lepton Yukawa couplings $y_{\mu,\nu,\tau}$ are selected in order to recreate the observed charged lepton masses precisely at tree level, resulting in a small correction to the SM expectation of $y_{f,SM} = \sqrt{2} m_f / v$. We therefore determine the observables $\kappa_{\mu,\tau}$ as
\begin{align}
    \kappa_{\mu,\tau} = \frac{y_{\mu, \tau} v}{\sqrt{2} m_{\mu,\tau}}.
\end{align}

The next observables that we consider are the lepton flavor violating decays $\mu \rightarrow e \gamma$, $\tau \rightarrow \mu \gamma$, and $\tau \rightarrow e \gamma$, as well as $Z$-mediated $\mu-e$ conversion in gold nuclei. In the case of flavor violating $l \rightarrow l' \gamma$ decays, we extract the relevant expressions from \cite{Lavoura:2003xp}. The flavor violating decay rates are then determined by computing the effects of $W$, $Z$, $h$, $A_D$, and $h_D$ loops with internal vector-like lepton lines. If vector-like leptons with nonvanishing dark charge are present, the contributions arising from dark Higgs and dark photon loops are overwhelmingly dominant, due to their smaller mass and that the $W$, $Z$, and $h$ couplings between SM and vector-like leptons will feature suppression by a small mixing angle between SM and BSM particles, which $h_D$ and the dark photon's longitudinal mode don't share. Meanwhile, we can extract the $Z$ boson contribution to $\mu-e$ conversion in the nucleus from the expression in \cite{Crivellin:2020ebi}.

The final observables that we consider in our results are the BSM corrections to the anomalous magnetic moments of the muon and the electron. These corrections emerge at the one-loop level from interactions between the SM leptons and vector-like fermions facilitated by the dark photon, dark Higgs, and electroweak gauge bosons. We extract expressions for numerical fit from the generic-case calculation done in \cite{Leveille:1977rc}. Much as in the case of the lepton flavor violating decays we have considered earlier, for models with any dark-charged vector-like leptons the dark photon and dark Higgs contributions will dominate the electroweak ones.

%----------------------------------------------------------------
% References
%----------------------------------------------------------------
%\clearpage
\setlength{\bibsep}{3pt}
\bibliographystyle{JHEP}
\bibliography{main}

\end{document}